\documentclass[a4paper,10.5 pt]{article}

\usepackage[T1]{fontenc}
\usepackage[latin1]{inputenc}
\usepackage[font=small]{caption}
\usepackage{latexsym} 
\usepackage{amssymb}
\usepackage{hyperref}
\usepackage{rotating}
\usepackage{fancyvrb}
\usepackage{animate}
 \usepackage{listings}
\usepackage{pdflscape}
\usepackage{animate}
\usepackage{media9}
 \usepackage{graphicx}
\usepackage[T1]{fontenc}
\usepackage[latin1]{inputenc}
\usepackage[font=small]{caption}
\usepackage{latexsym} 
\usepackage{amssymb}
\usepackage{hyperref}
\usepackage{rotating}
\usepackage{fancyvrb}
\usepackage{animate}
\usepackage{listings}
\usepackage{geometry}
\usepackage{pdflscape}
\usepackage{animate}
\usepackage{media9}
\usepackage{graphicx}
\usepackage{color}
\usepackage{colortbl}
\usepackage{xcolor}
\definecolor{GGG}{RGB}{213,230,255}
\definecolor{GG}{RGB}{220,220,220}
\definecolor{diag}{RGB}{255,255,255}
\definecolor{var}{RGB}{234,137,154}
\usepackage[english]{babel}
\usepackage{amsmath}
\usepackage{array,longtable}
\usepackage{amsthm}
\usepackage{multirow}
\usepackage{slashbox,booktabs}
\usepackage{tikz}
\usepackage{algorithmic}
\usepackage{longtable}
\usepackage{arydshln}
\usepackage{setspace}
\usetikzlibrary{spy}

\usepackage[first=0,last=9]{lcg}

\usepackage[T1]{fontenc}
\usepackage[latin1]{inputenc}
\usepackage[english]{babel}
\usepackage{latexsym} 
\usepackage{amssymb}
\usepackage{amsmath}
\usepackage{array,longtable}
\usepackage{bbding} 
\usepackage{mathpazo} 

\usepackage{amsthm}
\usepackage{xcolor}
\usepackage{multirow}
\usepackage{slashbox,booktabs,amsmath}

\usepackage{tikz}
\usepackage{algorithmic}
\usepackage{longtable}
\usepackage{arydshln}

\newcolumntype{P}[1]{>{\centering\arraybackslash}p{#1}}

\usepackage{setspace}
\allowdisplaybreaks

\usepackage{cite}

\usepackage{graphicx}
\usepackage{tikz}
\usetikzlibrary{spy}

\newlength\imagewidth
\newlength\imagescale

\theoremstyle{plain}

\newtheorem{definition}{Definition}[section]

\theoremstyle{definition}

\newtheorem*{theorem*}{Theorem}

\theoremstyle{remark}

\newcommand{\reviewtimetoday}[2]{
}
\reviewtimetoday{\today}{Draft Version v.0.1}

\title{Nonparametric estimation of directional highest density regions}
\author{Paula Saavedra-Nieves\and Rosa M. Crujeiras}

\date{
	Department of Statistics, Mathematical Analysis and Optimization\\
	Universidade de Santiago de Compostela\\
	\texttt{\{paula.saavedra, rosa.crujeiras\}@usc.es}\\[2ex]%
}
\begin{document}
	\maketitle
	\begin{abstract}
		Reconstruction of sets from a random sample of points intimately related to them is the goal of  set estimation theory. Within this context, a particular problem is the one related with the reconstruction of density level sets and specifically, those ones with a high probability content, namely highest density regions.
		
		We define highest density regions for directional data and provide a plug-in estimator, based on kernel smoothing. A suitable bootstrap bandwidth selector is provided for the practical implementation of the proposal. An extensive simulation study shows the performance of the plug-in estimator proposed with the bootstrap bandwidth selector and with other bandwidth selectors specifically designed for circular and spherical kernel density estimation. The methodology is applied to analyze two real data sets in animal orientation and seismology.\\
		
	\end{abstract}
	
	\tableofcontents
	
	\section{Introduction}
	\label{sec:intro}
	Set estimation is focused on the reconstruction of a set (or the approximation of any of its characteristic features such as its boundary or its volume) from a random sample of points. One of the specific topics in this area is concerned with the estimation of sets directly related to density functions such as level sets. Mathematically, for a given level $t>0$, the goal is to reconstruct the unknown set 
	\begin{equation}\label{lgt}
		G_g(t)=\{x\in \mathbb{R}^d:g(x)\geq t\}
	\end{equation}
	from random sample of points of a density function $g$ on $\mathbb{R}^d$. This topic has received considerable attention in the statistical literature, specially since the notion of population clusters was established in \cite{har} as the connected components of the set in (\ref{lgt}). This cluster definition relies clearly on the user-specified level $t$, so for addressing this problem, an algorithm for estimating the smallest level with more than a single connected component was proposed in \cite{ste}. Furthermore, interesting applications of this clustering approach have emerged into different fields such as astronomy in \cite{jan}; cytometry in \cite{roe}; detection of mine fields in \cite{huo}; detection of outliers  in \cite{gar} or \cite{mar} and quality control in \cite{dev}, \cite{bai1} and \cite{bai2}. For a general review on clustering, see \cite{Anderberg}, \cite{everitt}, \cite{cuefra} and \cite{rin}.

The rationale for establishing this definition of cluster is quite related to the notion of mode. In fact, several cluster algorithms are based on the detection of modes (see, for example, \cite{silverman}) noting that the number of modes (local maxima of f) is not usually smaller than the number of clusters.
Nevertheless, the concept of cluster is easier to handle, since it has a global and geometrical nature, whereas the local maxima depend on analytical properties. There exist several works in literature dealing with the issue of inference on the number of modes with an approach based on density estimates (see \cite{silverman1981}, \cite{mammen1992}, \cite{minnote}, \cite{mammen1995}, \cite{hall} and \cite{fraiman}), but restricted up to dimension two. More recent contributions on this perspective are analyzed in \cite{azza}, \cite{stue} and \cite{burman}. There are also contributions from a testing perspective, with extensions for circular data (see \cite{Ame1} and \cite{Ame2}, and references therein).

The number of clusters is a basic feature for a statistical population. However, the problem of its estimation is not always taken into account in cluster analysis where it is usually chosen by the practitioner as a first step. Since the number of clusters is equal to the number of connected components of a level set, a very natural estimator for this populational parameter is the number of the connected components of the level set reconstruction. This perspective that solves the problem of selecting this unknown population parameter is considered, for instance, in \cite{cuefebr}, \cite{cueff} and \cite{biau}.

	
	Although initially established for a density supported on an Euclidean space such as in equation (\ref{lgt}), some generalizations of level set estimation theory have been already introduced in the literature. For instance, estimation of level sets for general functions (not necessarily a density) is considered in \cite{cuevasManteiga}. As an illustration, the authors show a first approach for the estimation of density level sets from data on a sphere. More recently, the reconstruction of density level sets on manifolds is studied in \cite{cholaq}. Through some simulations, the behavior of the proposed method is analyzed on the torus and on the sphere. Despite the introduction of these specific extensions, the general problem of reconstructing density level sets from random samples in the directional setting (that is, on a $d$-dimensional sphere) has not been formally established yet.
	
	Unfortunately, for most practical purposes, the specific value of the level $t$ in (\ref{Gtest}) is fully unknown by the practitioner. In addition, areas of the distribution support where $f$ is close to zero (\textit{non-effective support}) are usually of limited interest for applications. Therefore, in this work, we aim to introduce an alternative definition of directional level set where the practitioner establishes the probability content instead of the level $t$. This type of regions is known as highest density regions (HDRs) in the Euclidean space, so an appropriate definition of HDRs for the directional setting as well as a procedure for their estimation in practice (based on plug-in ideas considering a kernel density estimator) is presented in this work. For the practical computation of the proposed plug-in estimator, a bootstrap bandwidth designed for reconstructing directional HDRs is also introduced. Its performance is analyzed through an extensive simulation study and compared with other bandwidth selectors specifically devised for density estimation.

	One may argue that such an absence of a general and effective proposal for directional level set estimation may be due to a lack of practical interest, but this is far from the truth, so let us present two application examples that motivate the developments in this work. The first one concerns a problem from animal orientation studies and the second one is related to earthquakes occurrences.\\
	
	\subsection{Some motivating examples}
	\label{intro:data}
	\emph{Animal orientation example.} Behavioral plasticity is considered by biologists as a feature of adaptation to changing beach environments. In particular, orientation is an adaptation characteristic that can not be modified by a single factor. Nonetheless, experts found some regularities in the orientation of sandhoppers and other animals from beach environments by changing one factor at a time under other controlled conditions.  
	
	\begin{figure}[h!]\centering
		\hspace{-.6cm}\begin{picture}(0,160)	
		\put(-162,-23.3){\includegraphics[scale=0.161]{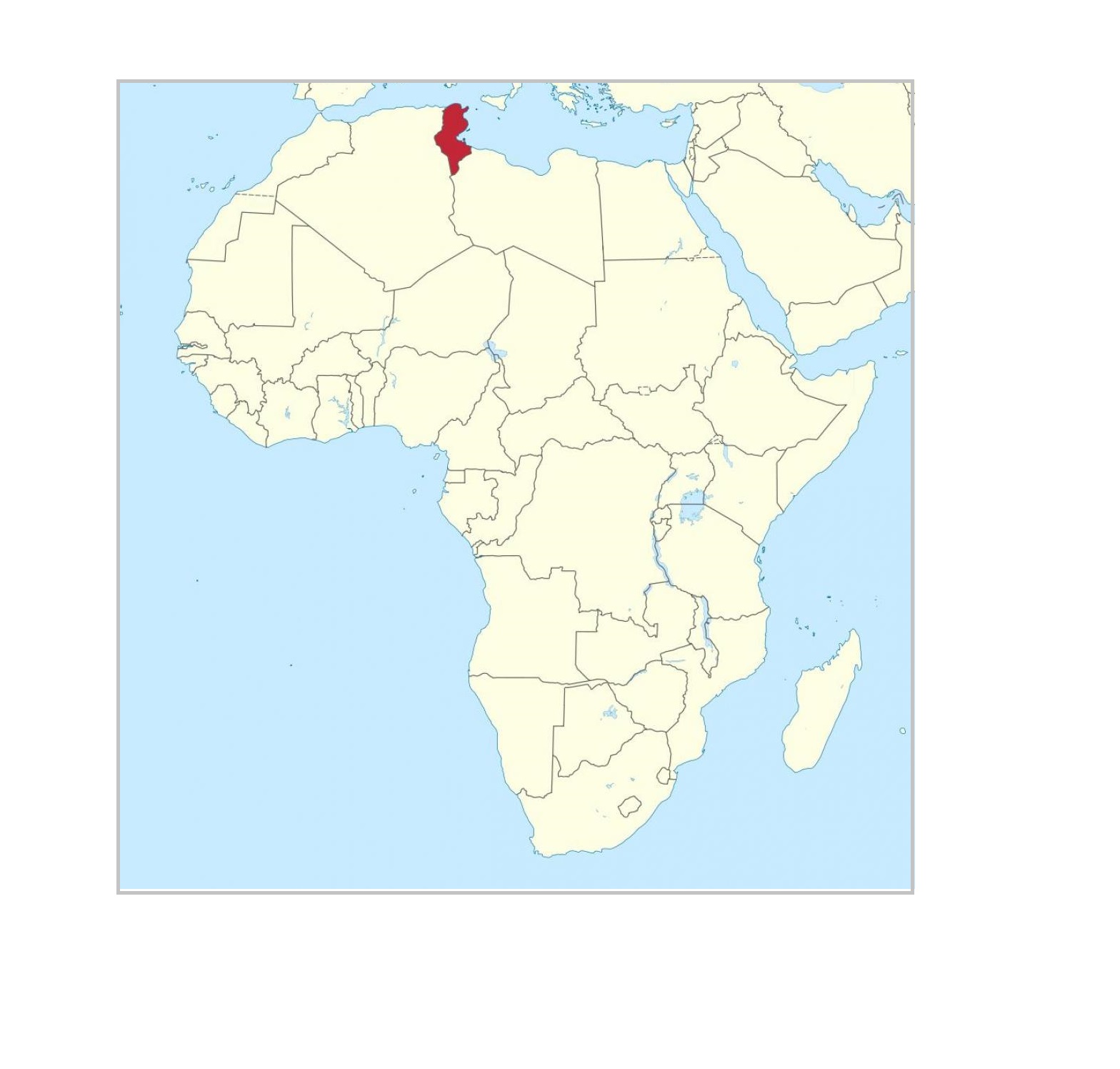}}
		\put(-3.8,100.1){ \includegraphics[scale=0.317]{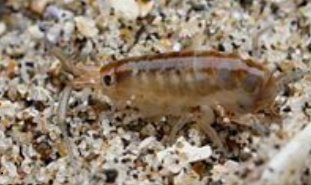}}
		\put(78,100){\includegraphics[scale=0.32]{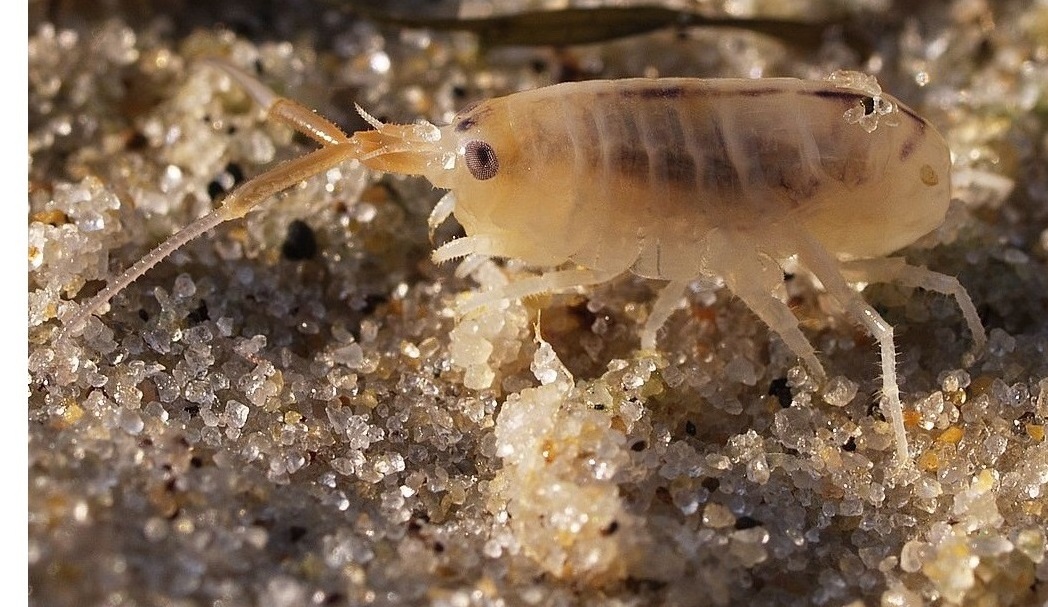}}
		\put(-2.2,5){\includegraphics[scale=0.318]{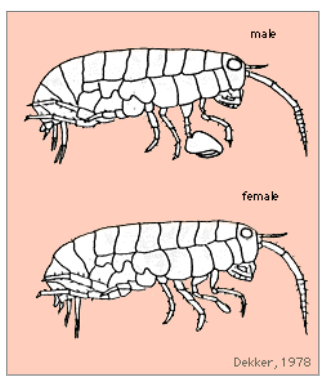}\hspace{.1cm}\includegraphics[scale=0.318]{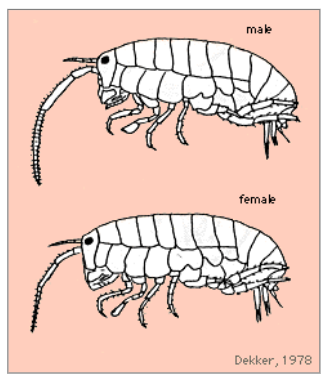}}
		\put(-84,140){\circle*{2}}
		\put(-81,139){\tiny{Zouara beach}}
		\end{picture}
		\caption{Geographical location of Zouara beach (right). Talorchestia brito (center) and Talitrus saltator (left). }\label{sandhop}
	\end{figure}
	
	For instance, the orientation of two sandhoppers species (Talitrus saltator and Talorchestia brito) is analyzed in \cite{scapini2002}. The experiment was carried out on the exposed non-tidal sand of Zouara beach located in the Tunisian northwestern coast. Both species are shown in Figure \ref{sandhop}. Bottom pictures can be found in \cite{dekker1978}. Apart from the specie and the orientation angles, this dataset contains information about other variables such as sex (male, female), month (April, October) and moment of the day (morning, afternoon and noon) when the experiment was done. We refer to \cite{scapini2002} and \cite{scapini2003} for further details on the dataset and the experimental design.
	
\begin{figure}[h!]\centering 
		$\hspace{-.55cm}$\includegraphics[scale=0.43]{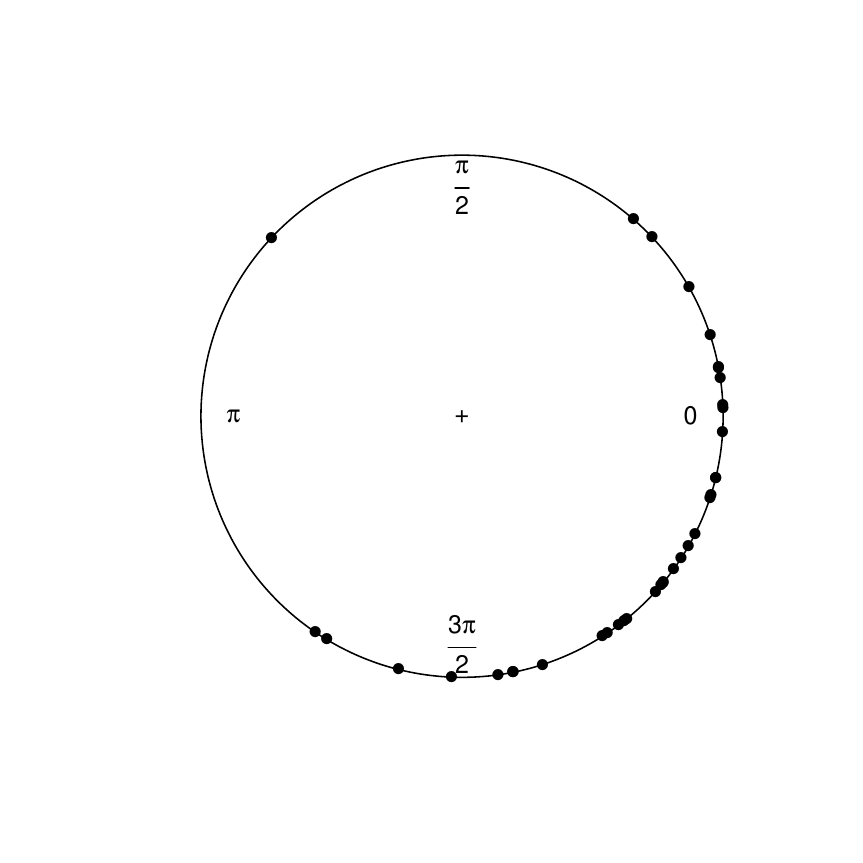}
		\includegraphics[scale=0.43]{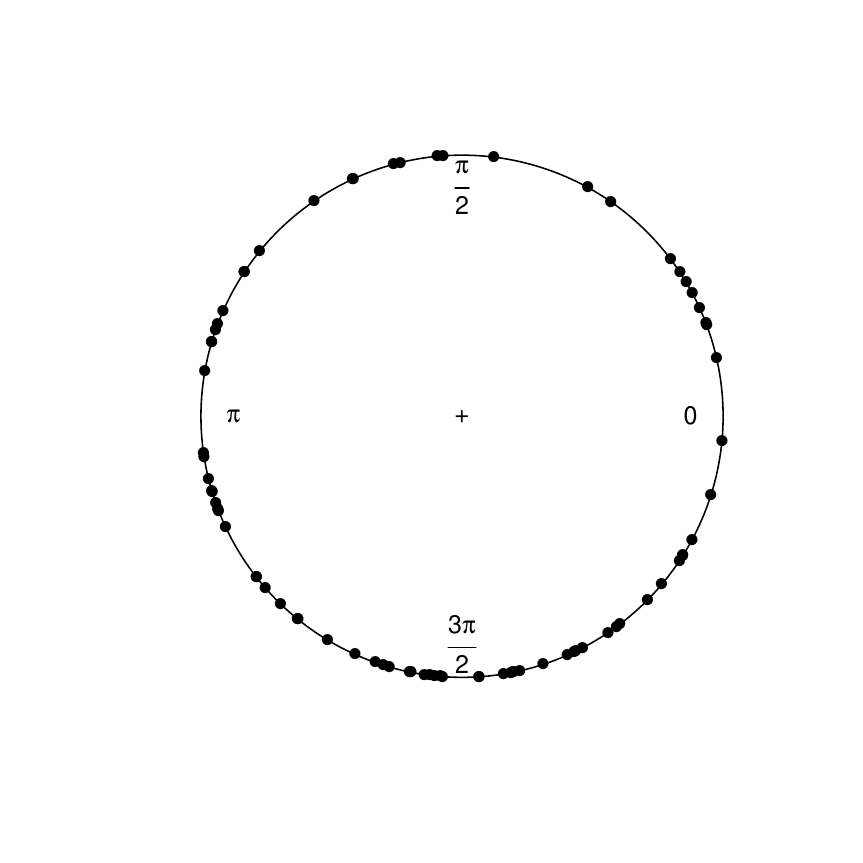}\vspace{-1cm}\\
		\caption{Orientation data (slightly jittered) corresponding to males of the specie Talitrus saltator registered in the noon in April (left) and October (right).}\label{intro0}
	\end{figure}

	Comparing the two species through regresion procedures, \cite{scapini2002} conclude that Talitrus saltator showed more differentiated orientations, depending on the time of day, period of the year and sex, with respect to Talorchestia brito. Moreover, it seems that Talitrus saltator shows a higher flexibility (variation) of orientation than Talorchestia brito under the same environmental conditions, supporting the hypothesis that the former has a higher level of terrestrialization. As an illustration, Figure \ref{intro0} (left panel) shows the $36$ orientation points (slightly jittered) corresponding to males of the specie Talitrus saltator measurements during the noon in April. It also contains the $77$ angles (slightly jittered) when the measures are taken in  October (Figure~\ref{intro0}, right panel). Differences in the distribution on the circle of these two samples can be easily observed. Therefore, the month of the year seems to play a significant role in sandhoppers behavior. In particular, two clusters for October measurements can be detected around the angle $\pi$ but they are not present for the April sample. Similar comments could be done for the situation registered around the angles $\pi/2$.  Therefore, cluster identification under the established conditions can be considered as an useful alternative to analyze sandhoppers orientation. 

\begin{figure}[h!]
		\begin{picture}(-200,430)
		\put(-5,270){\includegraphics[scale=0.7]{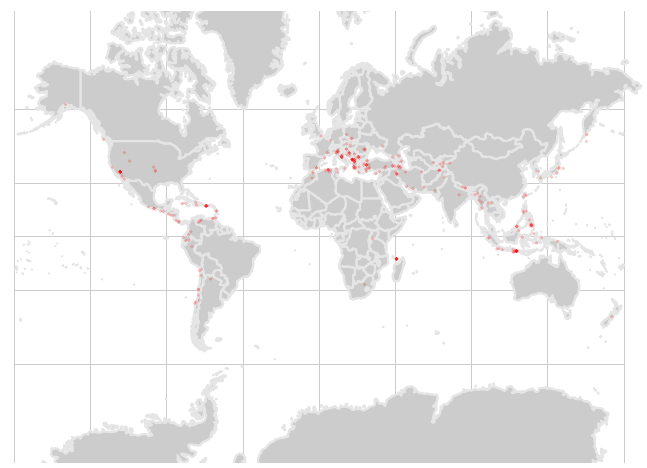}}
		\put(205,215){\includegraphics[scale=0.5]{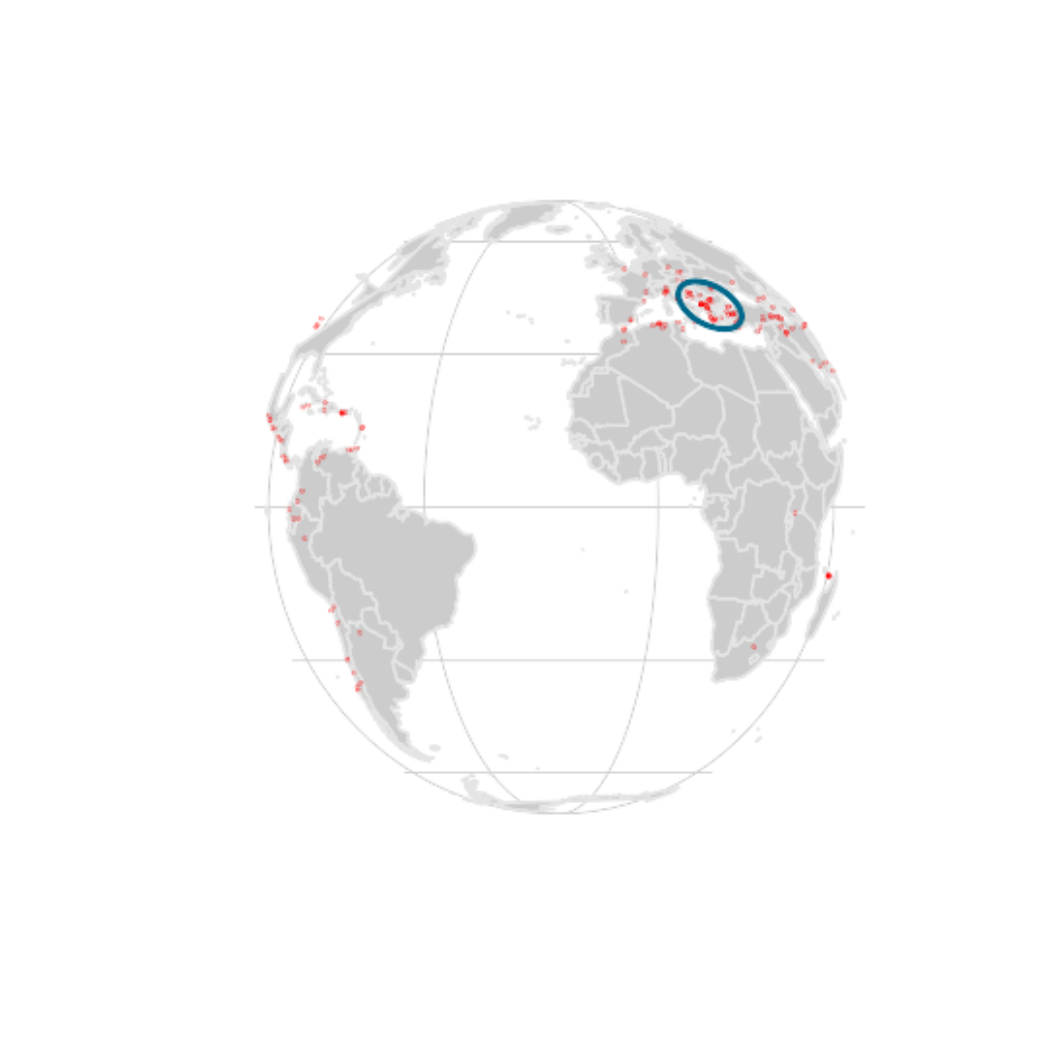}}
		\end{picture}  \vspace{-9.2cm}
		\caption{Distribution of earthquakes around the world between October 2004 and April 2020 (left). Density level set contour obtained from the sample of world earthquakes registered between October 2004 and April 2020 (right).}\label{earthquakes0}
	\end{figure}

	\emph{Earthquakes occurrences.} The European-Mediterranean Seismological Centre (EMSC)\footnote{European-Mediterranean Seismological Centre: \url{www.emsc-csem.org}.} is a non-governmental and non-profit organisation that has been established in 1975 at the request of the European Seismological Commission. Since the European-Mediterranean region has suffered  several destructive earthquakes, there was a need for a scientific organisation to be in charge of the determination, as quickly as possible (within one hour of the earthquake occurrence), of the characteristics of such earthquakes. These predictions are based on the seismological data received from more than 65 national seismological agencies, mostly in the Euro-Med region. Figure \ref{earthquakes0} (left) shows the geographical coordinates (red points), downloaded from EMSC website, of a total of $272$ medium and strong world earthquakes registered between 1$^{th}$ October 2004 and 9$^{th}$ April 2020. The magnitude of all these events is at least $2.5$ degrees on the Richter scale. Of course, these planar points correspond to spherical coordinates on Earth. Due to the important damages that earthquakes cause, cluster detection could be useful to identify, from a real dataset, where earthquakes are specially likely. This information is key for decision-making, for example, to update construction codes guaranteeing a better building seismic-resistance. An interactive representation of the sphere can be seen in Appendix \ref{AppendixA}.
	
	\subsection{Paper organization}
	This paper is organized as follows. Section 2 contains some background on level set estimation in the Euclidean setting, extending the definition for directional data and proposing a plug-in estimator. HDRs are the topic of Section 3, where a proper definition, jointly with a plug-in estimator are presented. This plug-in estimator is based on a directional kernel density estimator, which requires a smoothing parameter (bandwidth) for practical implementation. An appropriate bootstrap bandwidth selector is also introduced in this section. Section 4 presents an extensive simulation study illustrating the performance of the plug-in estimator for the HDRs (for circular and spherical domains) with the proposed bandwidth selector, comparing the results with those provided when other directional bandwidth selector criteria are considered. The proposed methodology is applied to the two real data examples presented in the Introduction. Finally, some conclusions and ideas for further research are presented in Section 6. This work is completed with some supplementary material. Appendix \ref{AppendixB} includes further information on the datasets. Appendix \ref{AppendixC} specifies the parameters taken for the construction of the  spherical densities in the simulation study. Appendix \ref{AppendixA} collects the description of the bandwidth selectors considered in the simulation study.

	\section{Some background on level sets}
	\label{sec:spher-sets}
	The specific problem of reconstructing density level sets in the directional setting is addressed in this section: a definition of directional level set is provided jointly with a plug-in estimator. Based on the real data and simulated examples, some discussion about how to measure the estimation error is also included.
	
	\subsection{On directional level sets}
	Consider a random vector $X$ taking values on a $d$-dimensional unit sphere $S^{d-1}$ with density $f$. Given a level $t>0$, the directional level set is defined as:
	\begin{equation}\label{G(t)}
		G_f(t)=\{x\in S^{d-1}:f(x)\geq t\}.
	\end{equation}
	Note that each $x\in S^{d-1}$ fully characterizes a point in $\theta\in [0,2\pi)^{d-1}$. Therefore, definition in (\ref{G(t)}) could have been also equivalently established as a subset of points in $[0,2\pi)^{d-1}$.
	
	\begin{figure}[h!]
		\begin{picture}(-200,410)
		\put(-70,200){\includegraphics[scale=0.6]{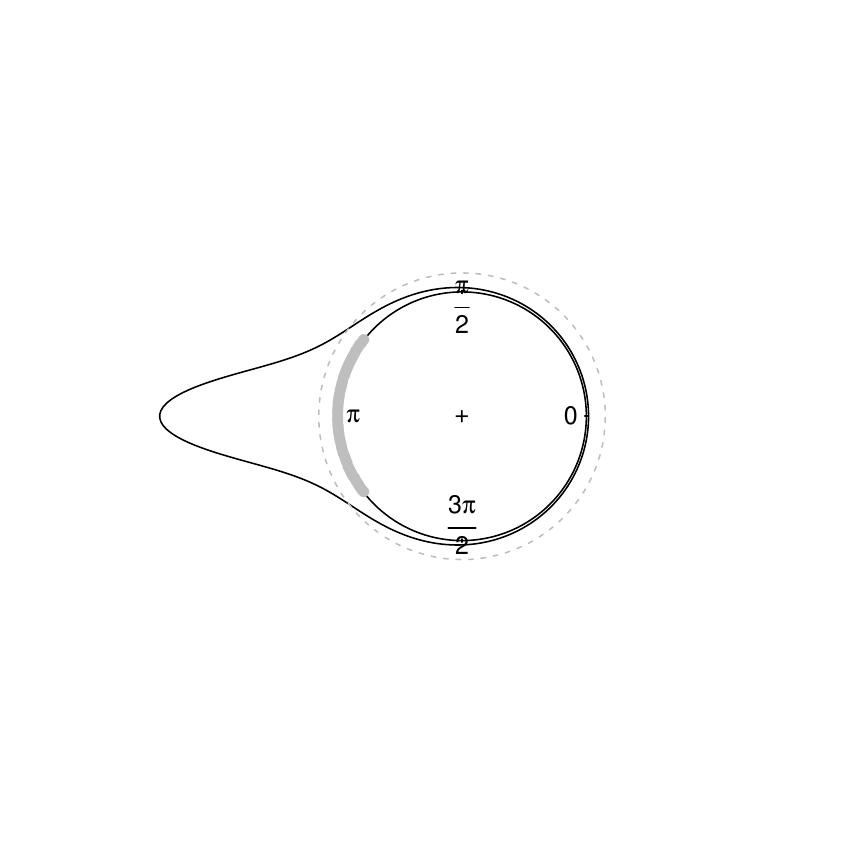}}
		\put(70,200){\includegraphics[scale=0.6]{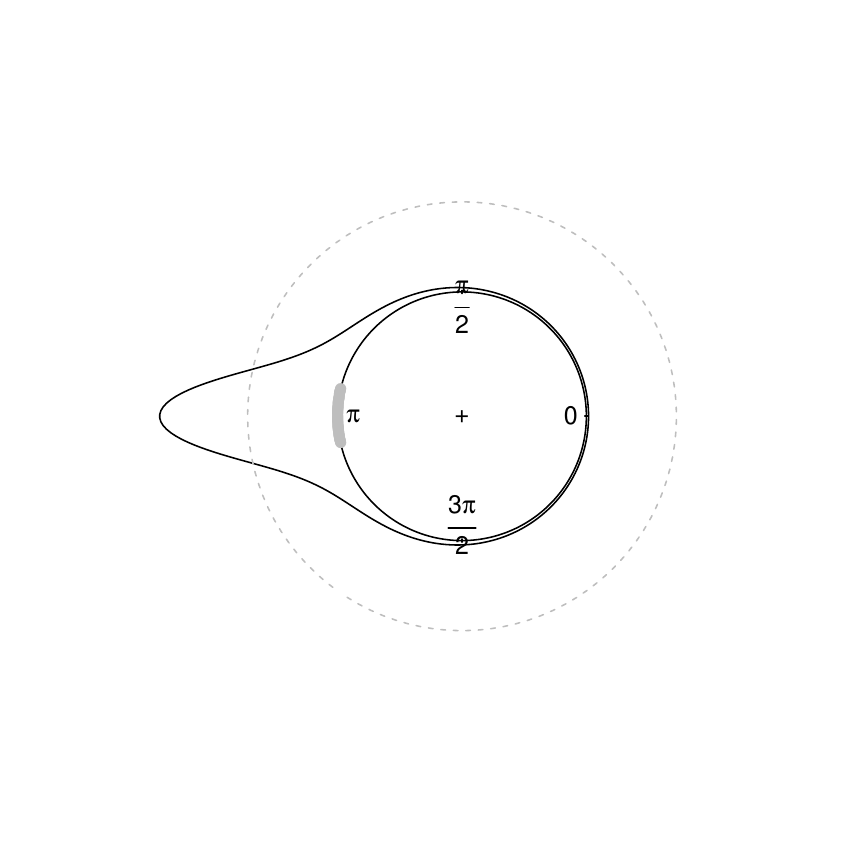}}
		\put(230,200){\includegraphics[scale=0.6]{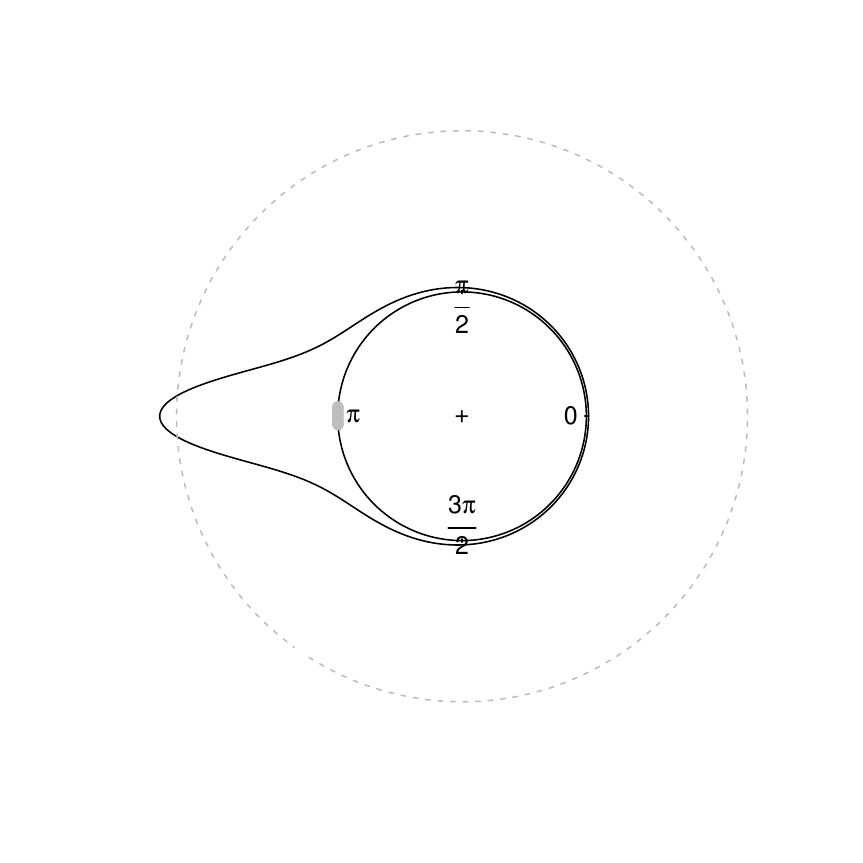}}
		
		\put(-70,60){\includegraphics[scale=0.6]{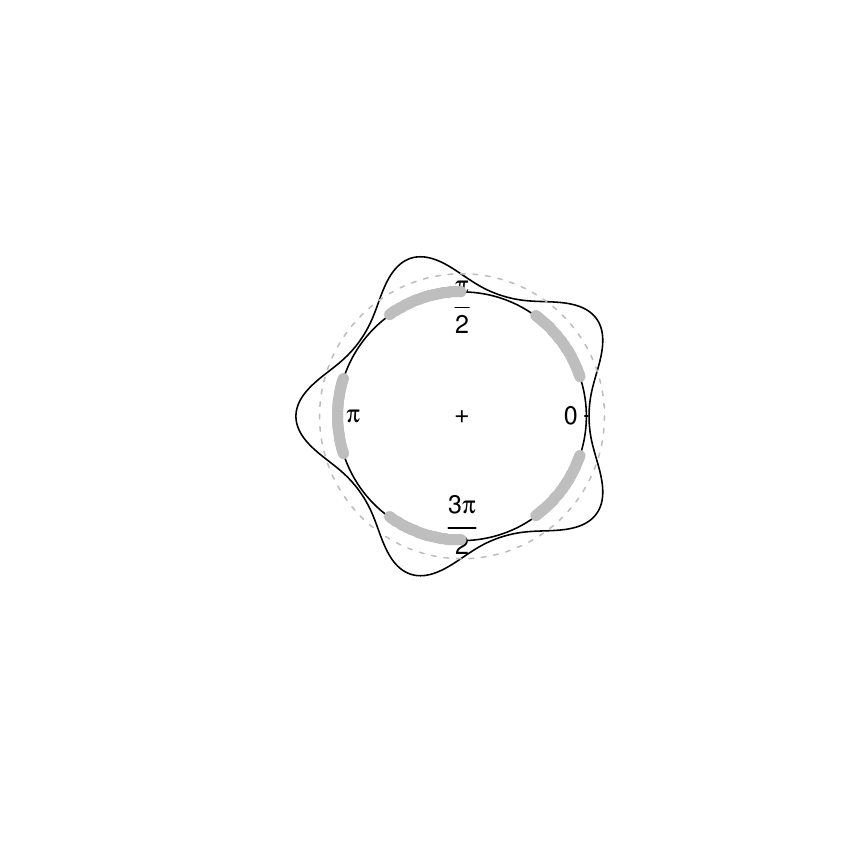}}
		\put(70,60){\includegraphics[scale=0.6]{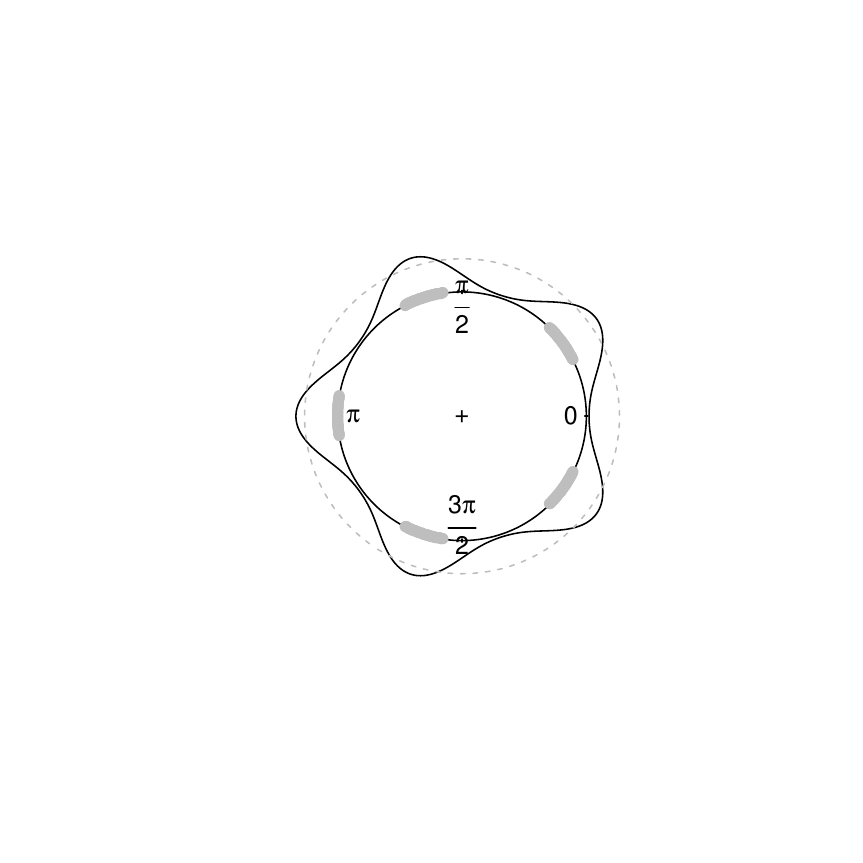}}
		\put(230,60){\includegraphics[scale=0.6]{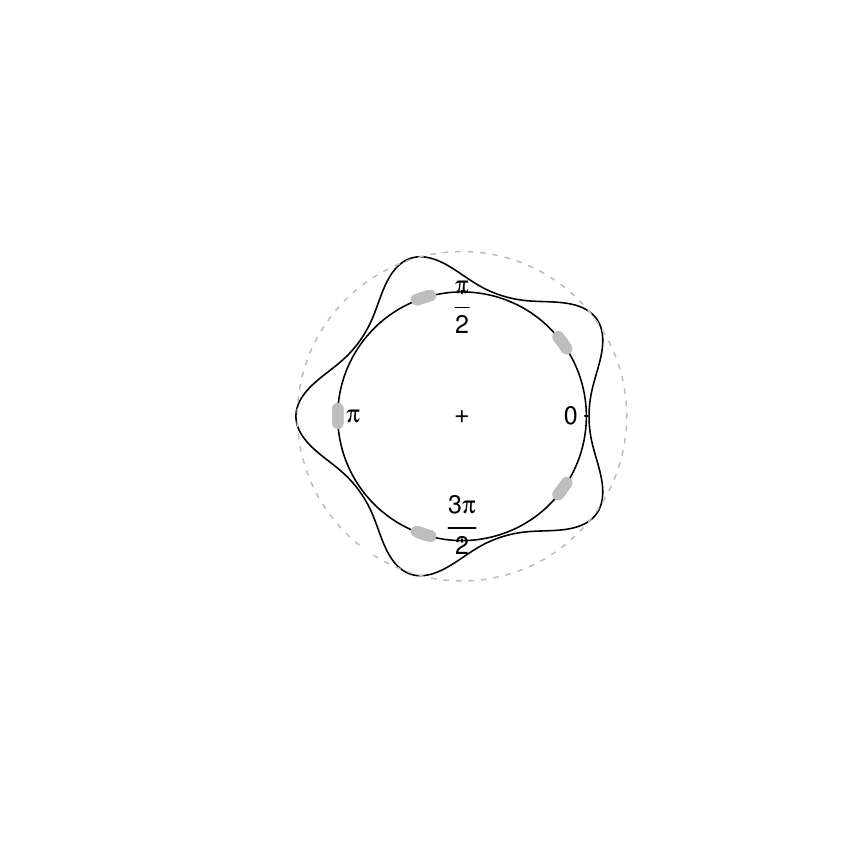}}
		
		\put(-70,-50){\includegraphics[scale=0.6]{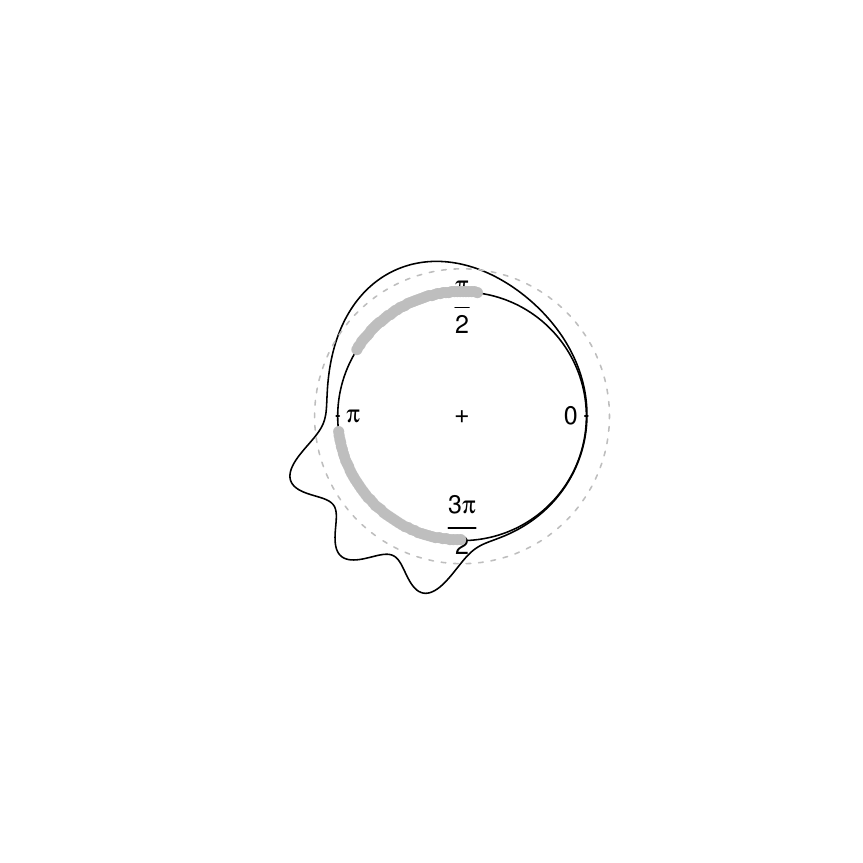}}
		\put(70,-50){\includegraphics[scale=0.6]{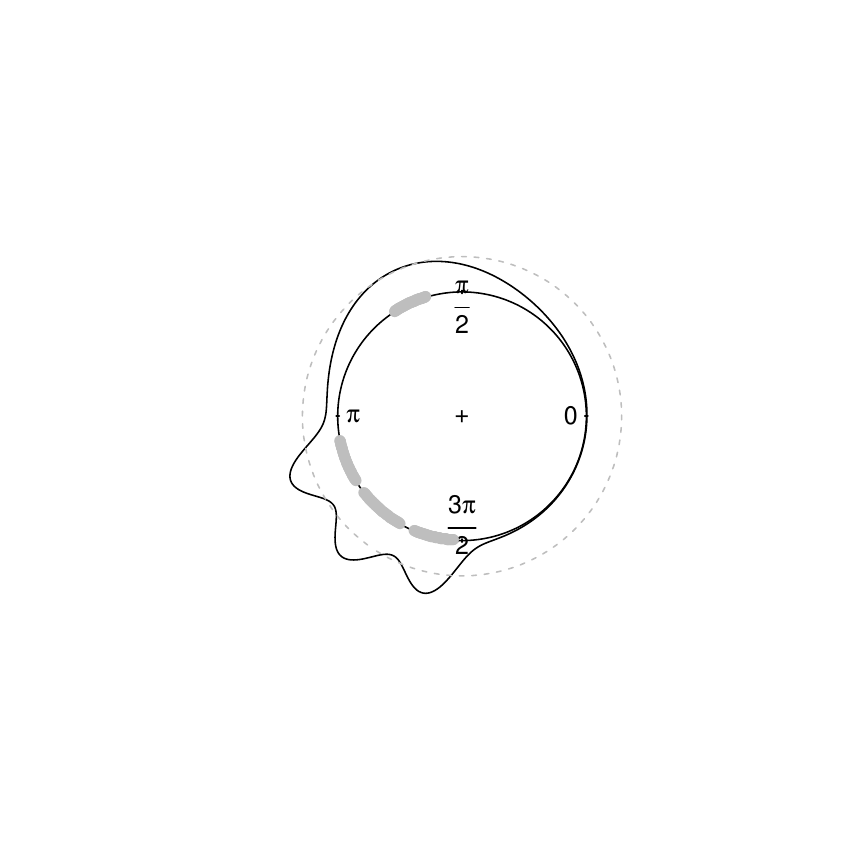}}
		\put(230,-50){\includegraphics[scale=0.6]{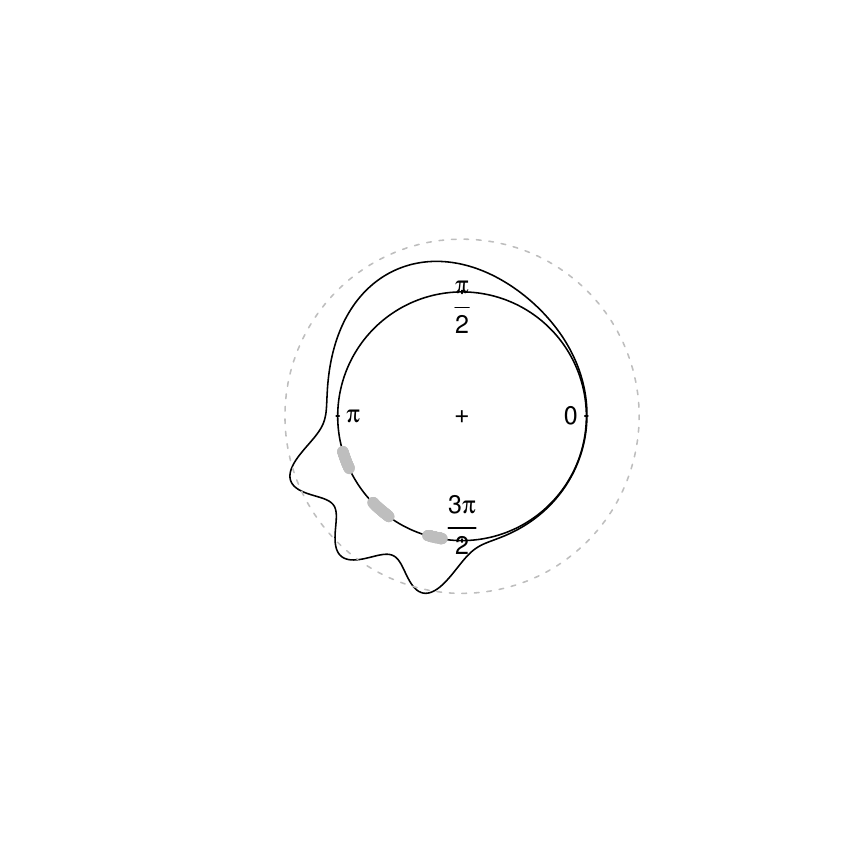}}
		\end{picture}  \vspace{-.9cm}
		\caption{For thee different circular densities, $G_f(t)$ for $t=t_1$ (first column), $t=t_2$ (second column) and $t=t_3$ (third column) verifying $0<t_1<t_2<t_3$. Equivalently, $L(f_\tau)$ for $\tau=0.2$ (first column), $\tau=0.5$ (second column) and $\tau=0.8$ (third column).}\label{ltau}
	\end{figure}
	
	The nature of different level sets is shown in Figure \ref{ltau}, wich represents $G_f(t)$ in grey color for three different circular densities and three different values of the level $t$. The threshold $t$ is represented through a dotted grey line. Note that, if large values of $t$ are considered (bottom row in Figure~\ref{ltau}), $G_f(t)$ coincides with the greatest modes. However, for small values of $t$, the level set $G_f(t)$ is virtually equal to the support of the distribution. 
	
	It is important to noticed that, following \cite{har}, the concept of cluster in directional setting can be established as the connected components of the level set $G_f(t)$. With this view in mind, note that the density represented in the second row of Figure \ref{ltau} presents four connected components for all of the considered values for $t$, determining four population clusters.

	Plug-in estimation is the most natural and common choice for reconstructing density level sets in the Euclidean space. A review of other existing estimation alternatives can be seen in \cite{rosaa}. Plug-in methods are devised to reconstruct (\ref{lgt}) as 
	$$\hat{G}_g(t)=\{x\in \mathbb{R}^d:g_n(x)\geq t\}$$
	where $g_n$ usually denotes the classical kernel estimator for euclidean data (see \cite{parzen} and \cite{rosenblatt}). This methodology, which has received considerable attention (see, for instance,\cite{r4}, \cite{bai3}, \cite{mas}, \cite{rig}, \cite{rig2}, \cite{pol2} or \cite{chen}) can be easily generalized to the directional setting. Given a random sample $\mathcal{X}_n=\{X_1,\cdots,X_n\}\in S^{d-1}$ of the unknown directional density $f$, $G_f(t)$ in (\ref{G(t)}) can be reconstructed as
	\begin{equation}\label{Gtest}
		\hat{G}_f(t)=\{x\in S^{d-1}:f_n(x)\geq t\}
	\end{equation}
	where $f_n$ denotes a nonparametric directional density estimator. Following the ideas of the classical linear (for real-valued random variables) kernel estimator, a kernel estimator on $S^{d-1}$ is provided in \cite{bai}. Strong pointwise consistency, uniform consistency, and $L_1-$norm consistency of the estimator are proved. Almost simultaneously, a similar kernel density estimation procedure also on $S^{d-1}$ is presented in \cite{hall87}. Some of the results in \cite{hall87} are later extended in \cite{kle3}. Following \cite{bai}, from a random sample on a $d$-dimensional sphere, $\mathcal{X}_n$, the directional kernel density estimator at a point $x\in S^{d-1}$ is defined as
	\begin{equation}
		\label{estimacionnucleo}
		f_n(x)= \frac{1}{n}  \sum_{i=1}^n K_{vM}(x;X_i;1/h^2),
	\end{equation}where  $1/h^2 > 0$ is concentration parameter and $K_{vM}$ denotes the von Mises-Fisher kernel density. The von Mises-Fisher distribution plays the role of the normal distribution in directional setting (\cite{Mardia}). Formally, its density function can be written as
	$$
	K_{vM}(x;\mu;\kappa)=C_{d}(\kappa)\exp\{\kappa x^T\mu\},\quad\mbox{with }\quad C_{d}(\kappa)=\frac{\kappa^{\frac{d-1}{2}}}{(2\pi)^{\frac{d+1}{2}} \mathcal{I}_{\frac{d-1}{2}} (\kappa)  }
	$$
	where $\mu\in S^{d-1}$ is the directional mean, $\kappa>0$ the concentration parameter around the mean, $T$ stands for the transpose operator and $\mathcal{I}_p$ is the modified Bessel function of order $p$, given by
	$$\mathcal{I}_p(z)=\frac{(\frac{z}{2})^p}{ \pi^{1/2}\Gamma(p+1/2)      } \int_{-1}^{1}(1-t^2)^{p-1/2}e^{zt}dt$$ where $\Gamma(p)=\int_{0}^\infty x^{p-1}e^{-x} dx,$ with $p>-1$.

	Note that the kernel estimator in (\ref{estimacionnucleo}) can be viewed as a mixture of von Mises-Fisher. Furthermore, the concentration parameter $1/h^2$ plays an analogous role to the bandwidth in the Euclidean case. For small values of $1/h^2$, the density estimator is oversmoothed. The opposite effect is obtained as $1/h^2$ increases: with a large value of $1/h^2$, the estimator is clearly undersmoothing the underlying target density. Hence, the choice of $h$ is a crucial issue. For simplicity, in what follows, we refer to $h$ as bandwidth parameter. Several approaches for selecting $h$ in practice, in circular and even directional settings, have been proposed in the literature. However, no one of these existing proposals was designed focusing on the problem of directional level set estimation. For real-valued random variables, this problem was already widely treated. See, for instance, \cite{bai2}, \cite{sin}, \cite{sam}, \cite{Qiao} and \cite{doss}.

	\subsection{Estimation error}
\begin{figure}[h!]
		\begin{picture}(-100,400)
		\put(-40,200){\includegraphics[scale=0.6]{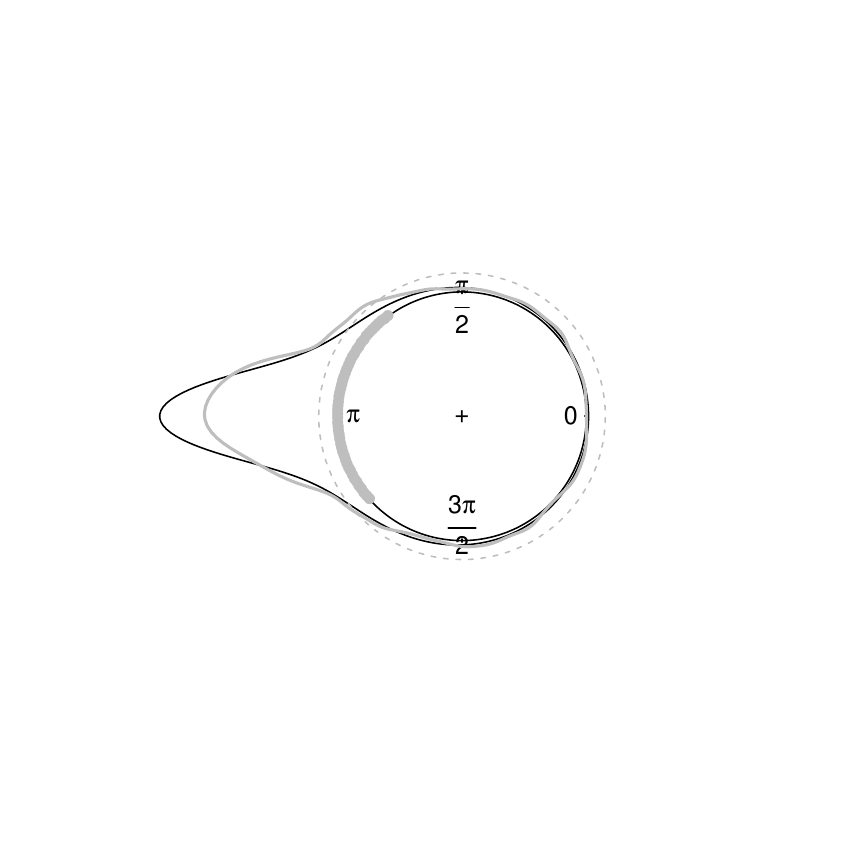}}
		\put(90,200){\includegraphics[scale=0.6]{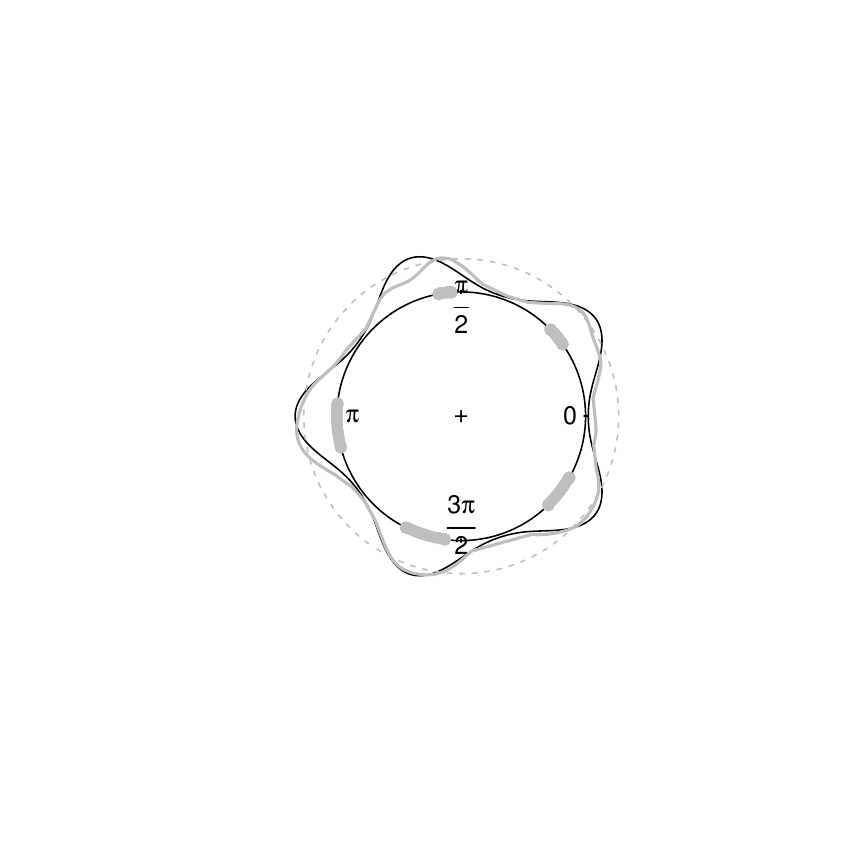}}
		\put(230,200){\includegraphics[scale=0.6]{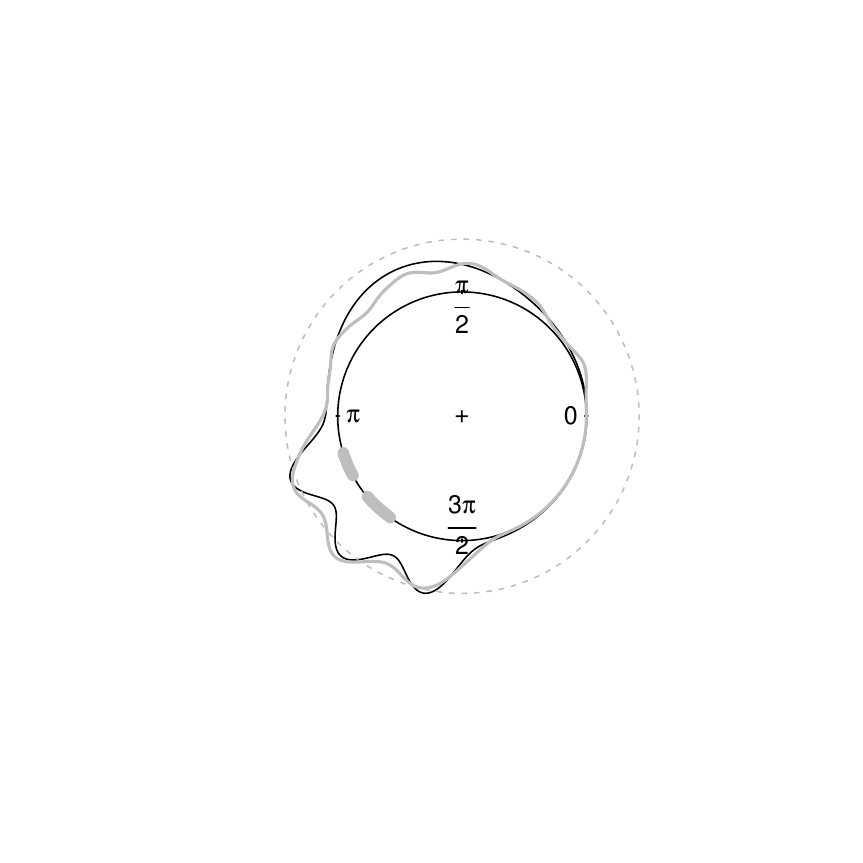}}
		\end{picture}\vspace{-10.3cm}\\
		\caption{Plug-in density level sets $\hat{G}_f(t)$ from $\mathcal{X}_{250}$ for three different circular densities with $t_1$ (first column), $t_2$ (second column) and $t_3$ (third column) verifying $0<t_1<t_2<t_3$. }\label{ltau2}
	\end{figure}

	Figure \ref{ltau2} shows three plug-in estimators $\hat{G}_f(t)$ for models (black colour) and levels $t_i$, $i=1,2,3$ (dotted grey line) considered in Figure \ref{ltau}. Kernel density estimators (grey color) in (\ref{estimacionnucleo}) have been determined from samples of size $250$ considering the proposal in \cite{oli1} as smoothing parameter. Note that $\hat{G}_f(t_3)$ in third column presents two connected components. However, Figure \ref{ltau} shows that the theoretical level set $G_f(t_3)$ has exactly three. Therefore, the estimation error is considerable and distances between sets should be used to measure it. 
	
	Since $(S^1,d_E)$ is a metric space when $d_E$ denotes the metric induced by the Euclidean norm $\|\cdot\|$ in $S^1$, it is possible to write $$d_E^2(x,y)=(x-y)^T(x-y)=x^Tx+y^Ty-2x^Ty=2(1-x^Ty)\mbox{ for }x,y\in S^1.$$Let us recall that, if $A$ and $B$ are non-empty compact sets in $(S^1,d_E)$, the Hausdorff distance between $A$ and $B$ is established as follows
	$$d_H(A,B)=\max\left\{\sup_{x\in A}d_E\left(\{x\},B\right),\sup_{y\in B}d_E\left(\{y\},A\right)\right\}$$where $d_E(\{x\},B)=\inf_{y\in B}\{d_E(x,y)\}$.
	The metric $d_H$ is not completely successful in detecting differences in shape properties. In other
	words, two sets can be very close in $d_H$ and still show quite different shapes. This typically happens where the boundaries $\partial A$ and $\partial B$ are far apart, no matter the proximity of $A$ and $B$. So a natural way to reinforce the notion of visual proximity between two sets provided by Hausdorff distance is to account also for the proximity of the respective boundaries. In particular, this error criterion is considered in \cite{cuevasManteiga} in order to establish the consistency in the sphere of the plug-in estimator defined in (\ref{Gtest}).

\begin{figure}[h!]\centering 
		\vspace{-3.5cm}
		$\hspace{-1.8cm}$\includegraphics[scale=.68]{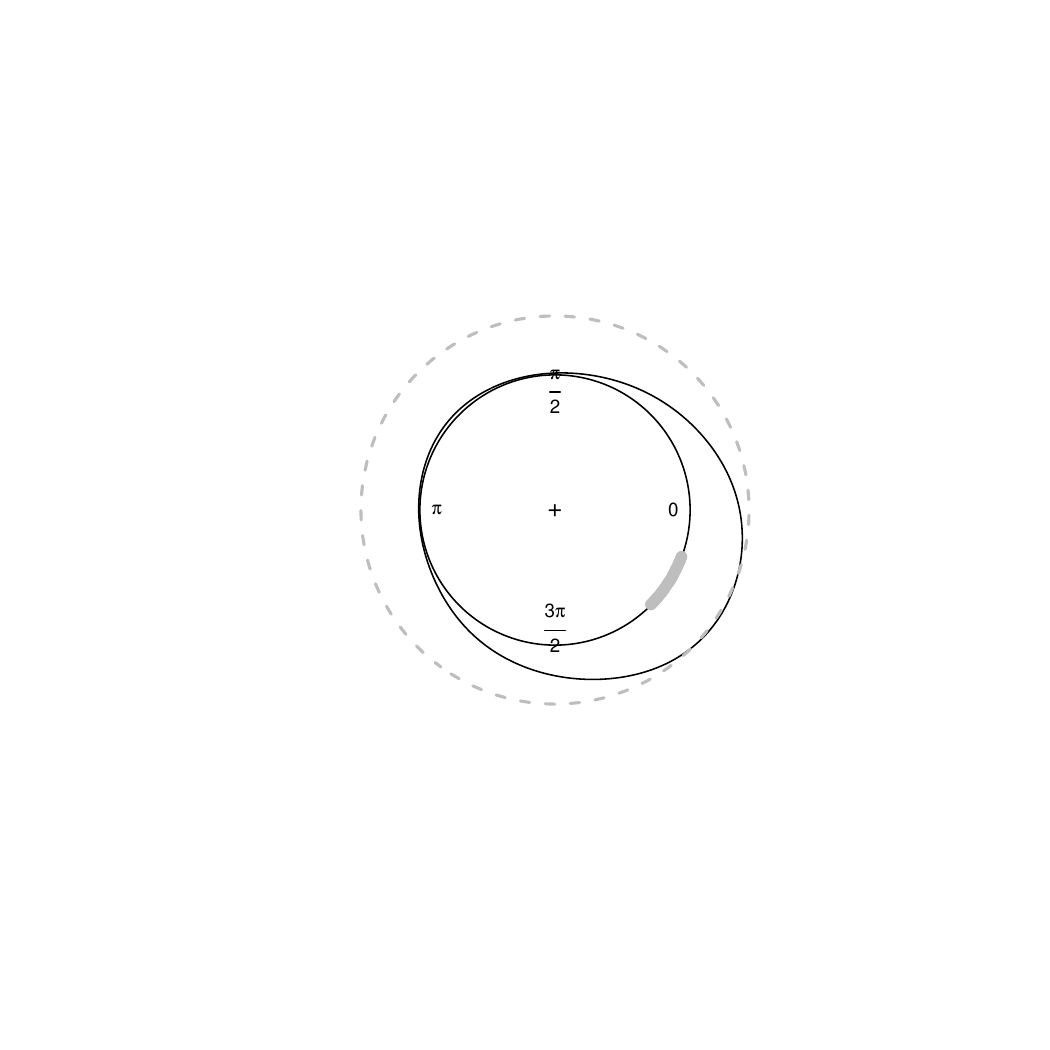}$\hspace{-6cm}$\includegraphics[scale=.68]{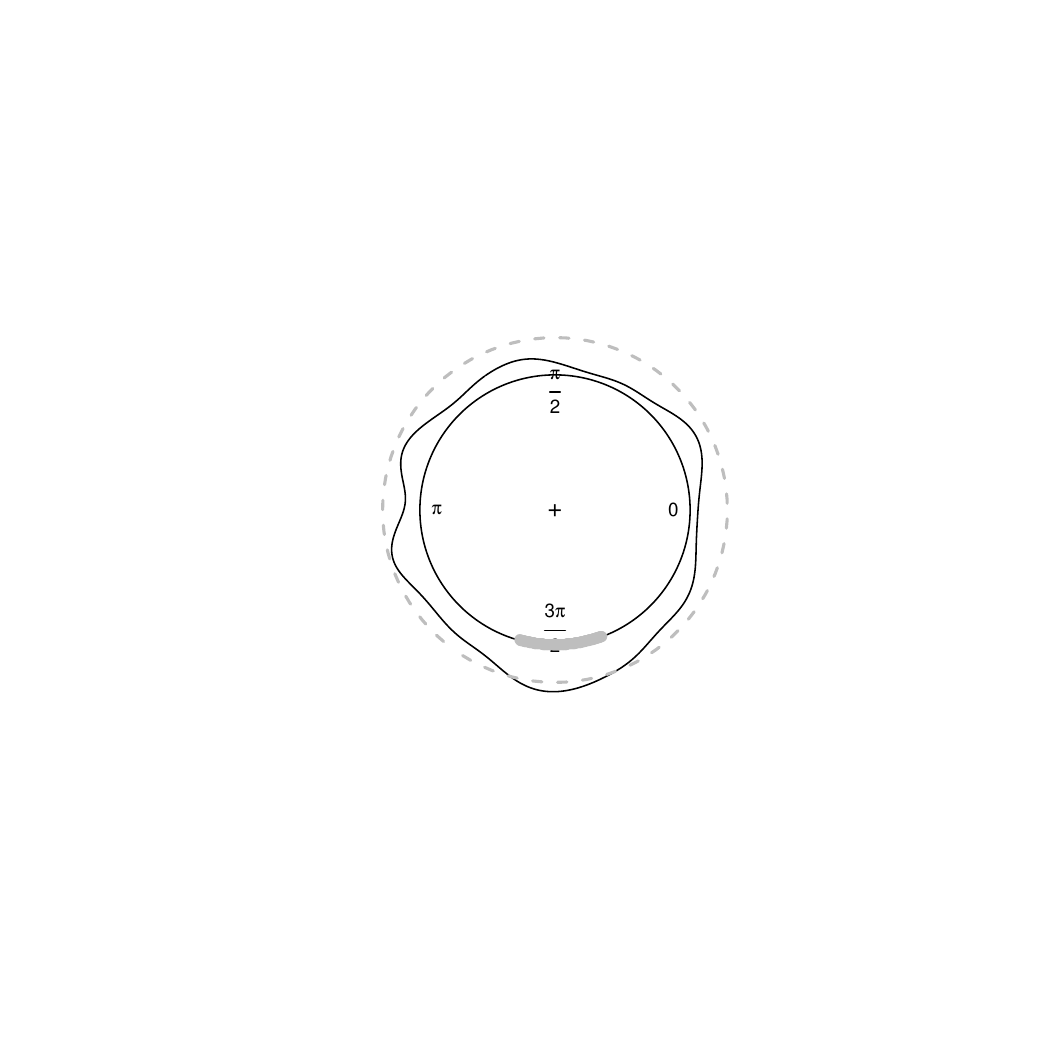}\vspace{-3.5cm}\\
		\caption{Plug-in density level sets $\hat{G}_f(t)$ obtained from the orientation samples corresponding males of the specie Talitrus saltator registered in the noon in April (left) and October (right).}\label{intro1}
	\end{figure}

	For instance, for the sandhoppers example, Figure \ref{intro1} shows the plug-in estimators obtained for the two samples of sandhoppers represented in Figure \ref{intro0}. Note that the value of the level $t$ considered is large enough in order to detect the greatest modes of the two sample distributions corresponding to April and October samples. These results allow us to confirm the differences between the two populations. The largest cluster of April orientations is located around the angle $7\pi/4$. However, the pattern observed for October registries is completely different. Although an only cluster is identified around the angle $3\pi/2$, if the level $t$ decreases slightly two additional groups can be detected around the angles $3\pi/4$ and $5\pi/4$, respectively. 
	
	Regarding the earthquakes illustration, Figure \ref{earthquakes0} (right) shows the plug-in density level set contour in blue obtained from the selected sample of world earthquakes considered. Chosing a convenient value of the level $t$, the greatest mode of sample distribution is identified in the Southeast of Europe. Countries such as Italy, Greece or Turkey (located withint this cluster) are clearly risky areas.

	\section{HDRs in the directional setting}
	\label{sec:HDR}
	As noted in the  Itroduction, the level $t$ is usually unknown and, for practical purposes, the interest usually focus on the effective support reconstruction for the density $f$ considering a fixed probability content. Figure \ref{hdr3333} (top) shows four different 50\% circular regions (regions containing 50\% of the probability, empirically approximated) for the kernel density estimator $f_n$ represented in grey. Although all of them have probability content equal to 50\%, they are completely different. Therefore, it is obvious that there exists an infinite number of ways to choose a region with given coverage probability. Depending on the specific problem, a certain region may be selected but in a general scenario, it may not be clear which region must be chosen. The same happens for real-valued random variables, and \cite{hyn} suggests that HDRs are the best subset to summarize a probability distribution. The concept of HDRs will be extended to the directional setting in what follows. 
	
	The usual purpose in summarizing a probability distribution by a region of the sample space is to delineate a comparatively small set which contains most of the probability, although the density may be nonzero over infinite regions of the sample space. Therefore, as in the linear case, it is necessary to decide what properties the region has to verify. The following conditions are natural:
	\begin{enumerate}
		\item[(C1)] The region should occupy the smallest possible volume in the sample space.
		\item[(C2)] Every point inside the region should have probability density at least as large as every point outside the region.
	\end{enumerate}
	
	Following \cite{box}, conditions (C1) and (C2 are equivalent and lead to regions called HDRs. Definition \ref{hdrr} formalizes this concept in the directional context taking into account the second criterion.
	
	\begin{definition}\label{hdrr}
		Let $f$ be a directional density function on $S^{d-1}$ of a random vector $X$. Given $\tau\in(0,1)$, the $100(1 - \tau)$\% HDR is the subset
		\begin{equation}\label{conjuntonivel2}
			L(f_\tau)=\{x\in S^{d-1}:f(x)\geq f_\tau\}
		\end{equation}where $f_\tau$ can be seen as the largest constant such that 
		\begin{equation}\label{umbral}
			\mathbb{P}(X\in L(f_\tau))\geq 1-\tau
		\end{equation}with respect to the distribution induced by $f$.\end{definition} According to \cite{pol3} and \cite{gar} in the Euclidean context, $L(f_\tau)$ is the minimum volume level set with probability content at least $(1-\tau)$. Figure \ref{ltau} shows the HDR $L(f_\tau)$ in grey for three different circular densities and three different values of $\tau$. The threshold $f_\tau$ is represented through a dotted grey line. Note that, if large values of $\tau$ are considered, $L(f_\tau)$ is equal to the greatest modes and, therefore, the most differentiated clusters can be easily identified. However, for small values of $\tau$, $L(f_\tau)$ is almost equal to the support of the distribution.

	\subsection{Plug-in estimation of directional HDRs}
	\label{sec:HDR-plugin}
	
	The first step to reconstruct the HDR established in Definition \ref{hdrr} for a given $\tau\in (0,1)$ is to estimate the threshold $f_\tau$. As in the Euclidean case, numerical integration methods could be also used in the directional setting in order to approximate its value. However, when the dimension increases, the computational cost becomes a major issue due to the complexity of the numerical integration algorithms considered on high diemsnional spaces. An alternative approach reducing the computational cost is described next.
	
	As before, let $X$ be a random vector with directional density $f$ and let $Y = f(X)$ be the random vector obtained by transforming $X$ by its own density function. Since $\mathbb{P}(f(X)\geq f_\tau)=1-\tau$, $f_{\tau}$ is exactly the $\tau-$ quantile of $Y$. Following \cite{hyn} in the linear case, $f_{\tau}$ can be estimated as a sample quantile from a set of independent and identically distributed random vectors with the same distribution as $Y$.

	In particular, if $\mathcal X_n=\{X_1,\cdots,X_{n}\}$ denotes a set of independent observations in $S^{d-1}$ from a density $f$. Then, $\{f(X_1), \cdots, f(X_n)\}$ is a set of independent observations from the distribution of $Y$. Let $f_{(j)}$ be the $j$-th largest value of $\{f(X_i)\}_{i=1}^n$ so that $f_{(j)}$ is the $(j/n)$ sample quantile of $Y$. We shall use $f_{(j)}$ as an estimate of $f_\tau$. Specifically, we choose $\hat{f}_\tau = f_{(j)}$ where $j = \lfloor \tau n\rfloor$. Then, $\hat{f}_\tau$ converges to $f_\tau$ as $n$ tends to $\infty$, and therefore $L(\hat{f}_\tau)$ converges to $L(f_\tau)$ as $n$ tends to $\infty$.
	
	Of course, if $f$ is a known function, the observations can be generated pseudorandomly and the estimation of $f_\tau$ could be made arbitrarily accurate by increasing $n$. In practice, as for density level set estimation, $f$ is often unknown. In this case, we have as only information a random sample of points  $\mathcal{X}_n$ from an unknown density $f$. From this sample, we propose first to determine the kernel estimator $f_n$ established in (\ref{estimacionnucleo}). If $n$ is large enough, we propose to calculate the set $\{f_n(X_1),\cdots,f_n(X_n)\}$ in order to estimate $f$ empirically. If $n$ is moderate, it may be preferable to generate observations $\mathcal X_n=\{X_l,\cdots,X_N\}$ of large size $N$ from $f_n$. For small values of $n$ it may not be possible to get a reasonable density estimate. Besides, with few observations and no prior knowledge of the underlying density, there seems little point in attempting to summarize the sample space. See \cite{wand1995} for some discussion on the number of observations needed for a reasonable linear density estimate. Note that the problem here is not with the density quantile algorithm (that give results to an arbitrary degree of accuracy given a density), but with estimating the density from insufficient data. 
	
	Once the threshold $f_\tau$ is estimated, plug-in methods reconstruct the $100(1 - \tau)$\% HDR $L(f_\tau)$  in (\ref{conjuntonivel2}) as
	\begin{equation}\label{Ltauest}
		\hat{L}(\hat{f}_\tau)=\{x\in S^{d-1}:f_n(x)\geq \hat{f}_\tau\}.
	\end{equation}

\begin{figure}[h!]
		\hspace{-1.6cm}	\begin{picture}(100,400) 
		\put(-40,200){\includegraphics[scale=0.6]{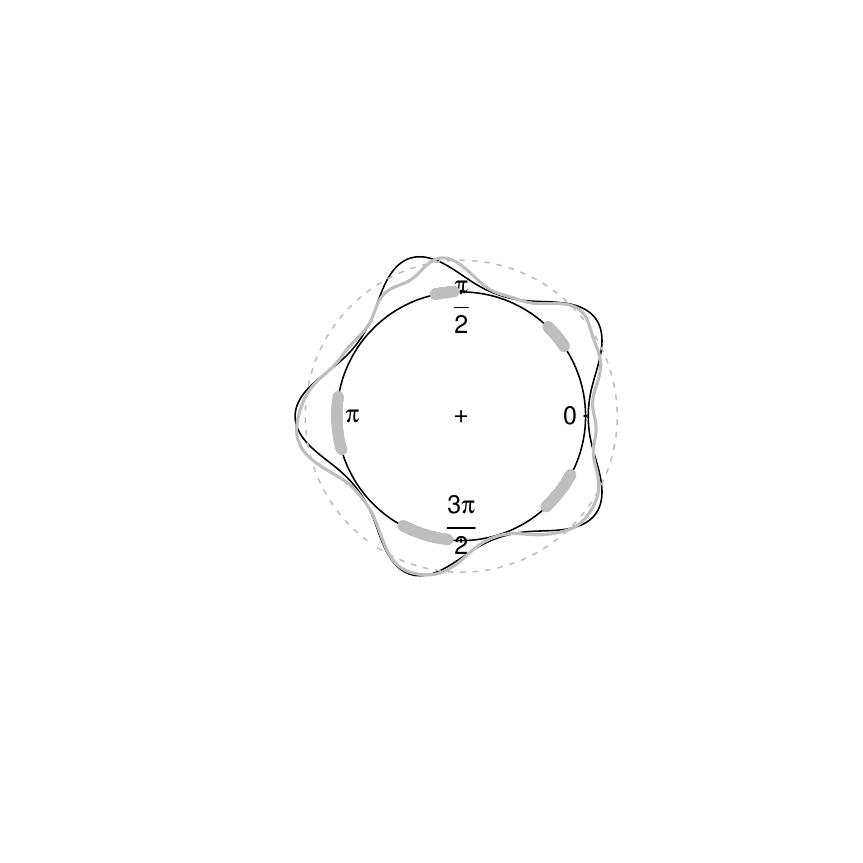}}
		\put(80,200){\includegraphics[scale=0.6]{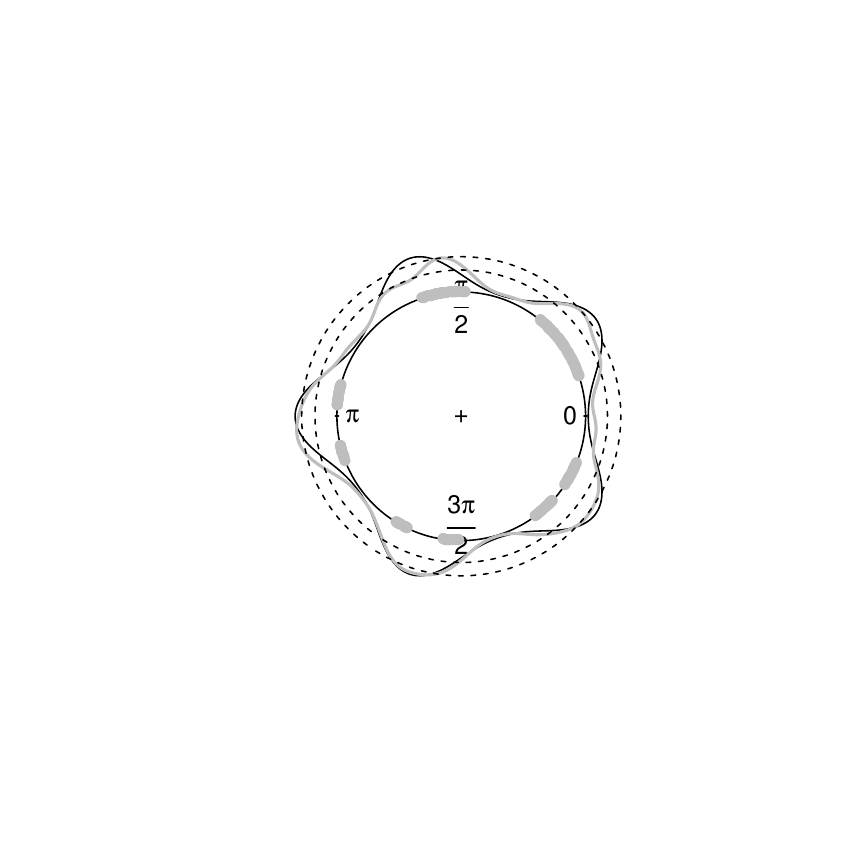}}
		\put(200,200){\includegraphics[scale=0.6]{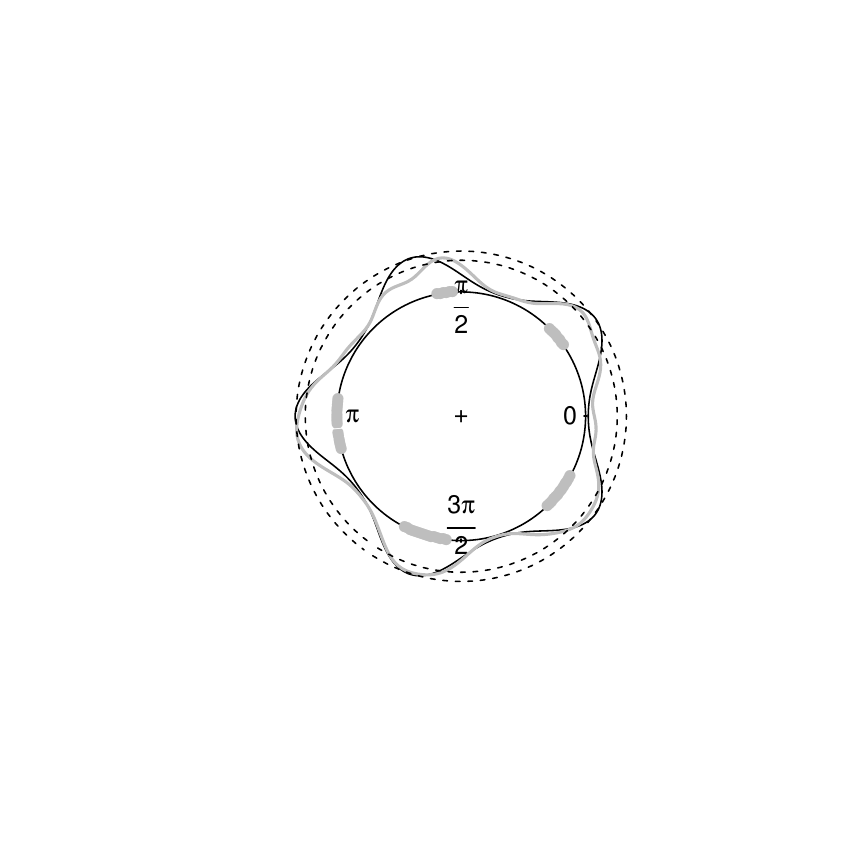}}
		\put(320,200){\includegraphics[scale=0.6]{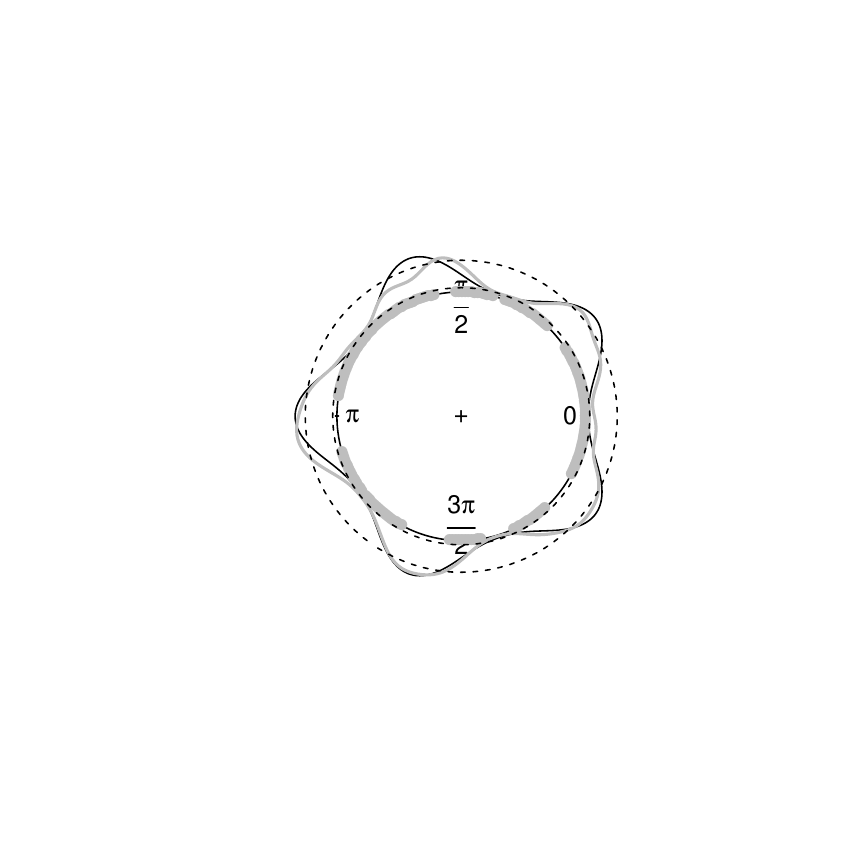}}
		
		\put(20,160){\includegraphics[scale=0.3]{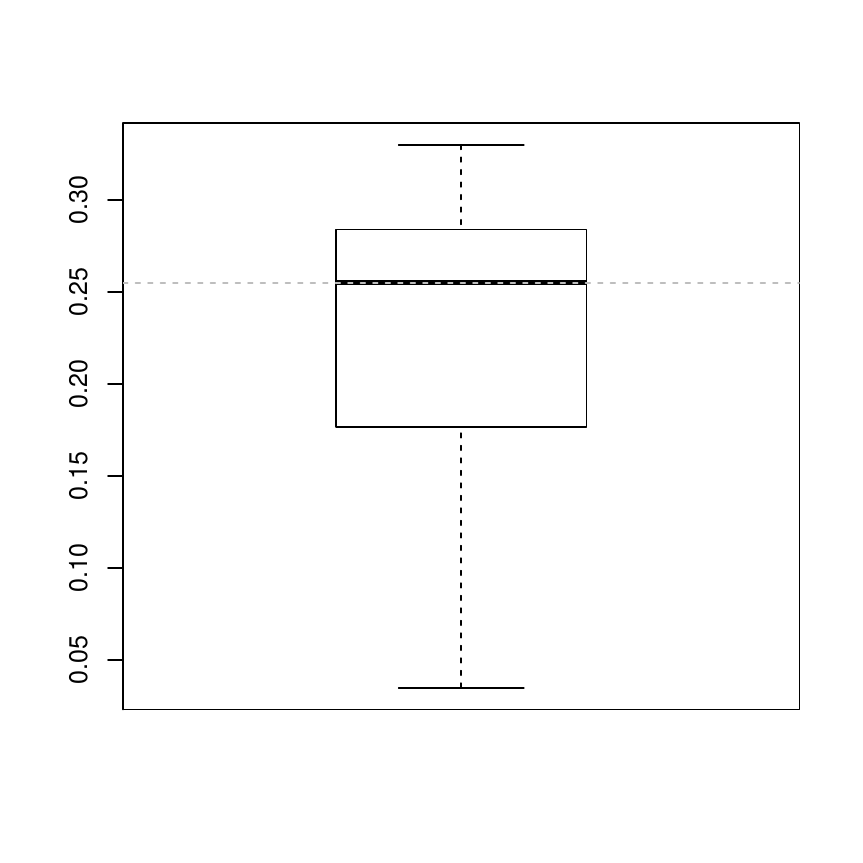}}
		\put(140,160){\includegraphics[scale=0.3]{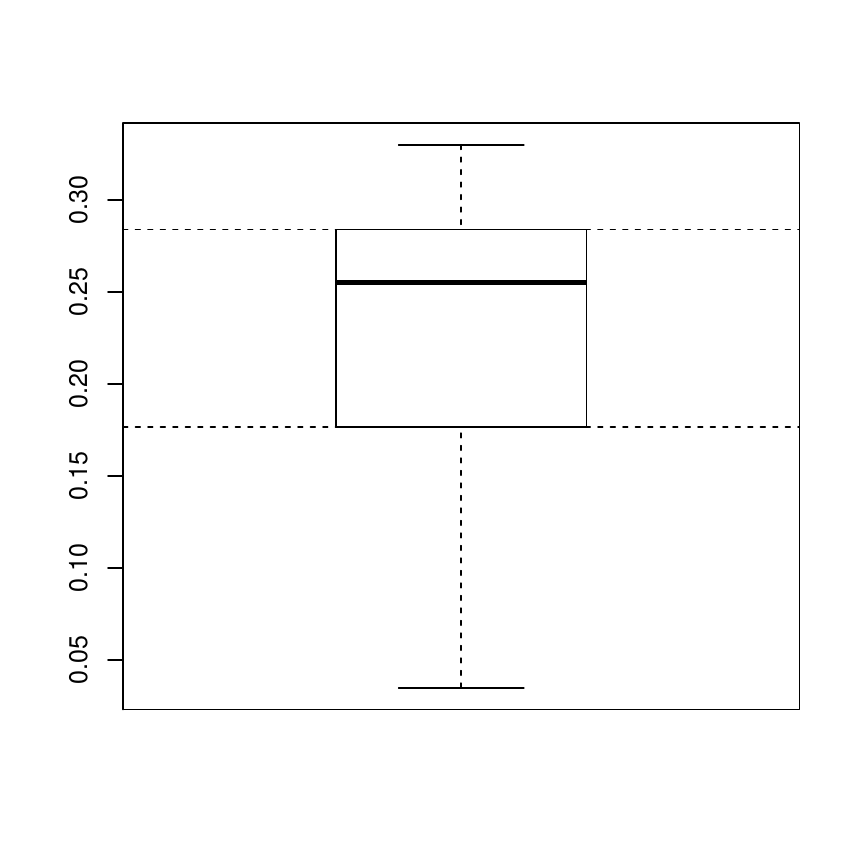}}
		\put(260,160){\includegraphics[scale=0.3]{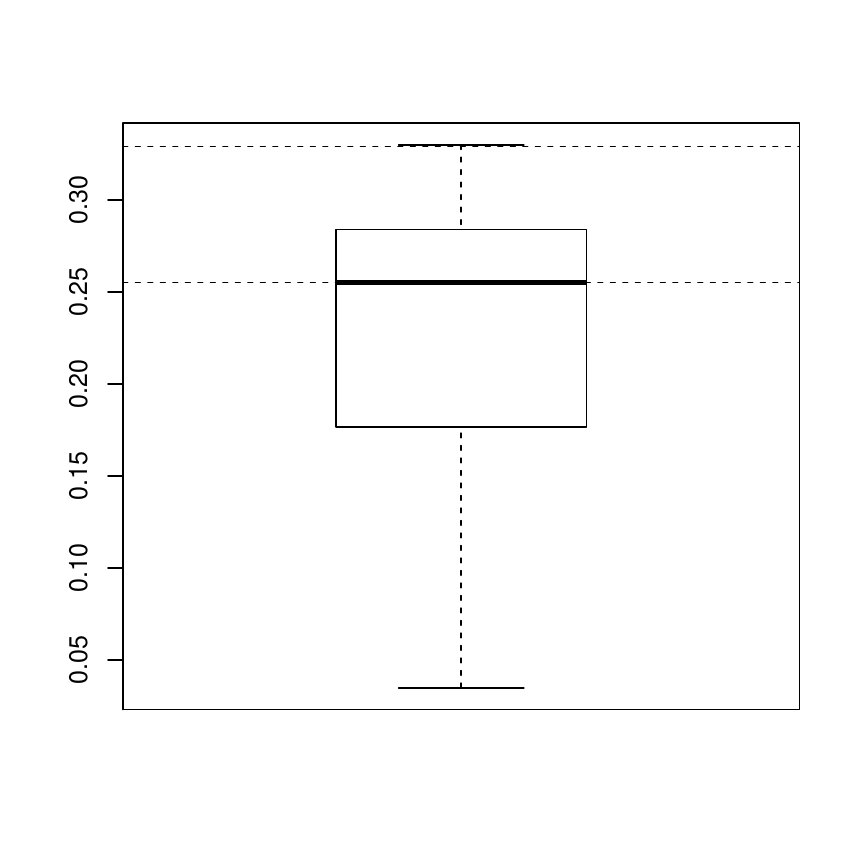}}
		\put(380,160){\includegraphics[scale=0.3]{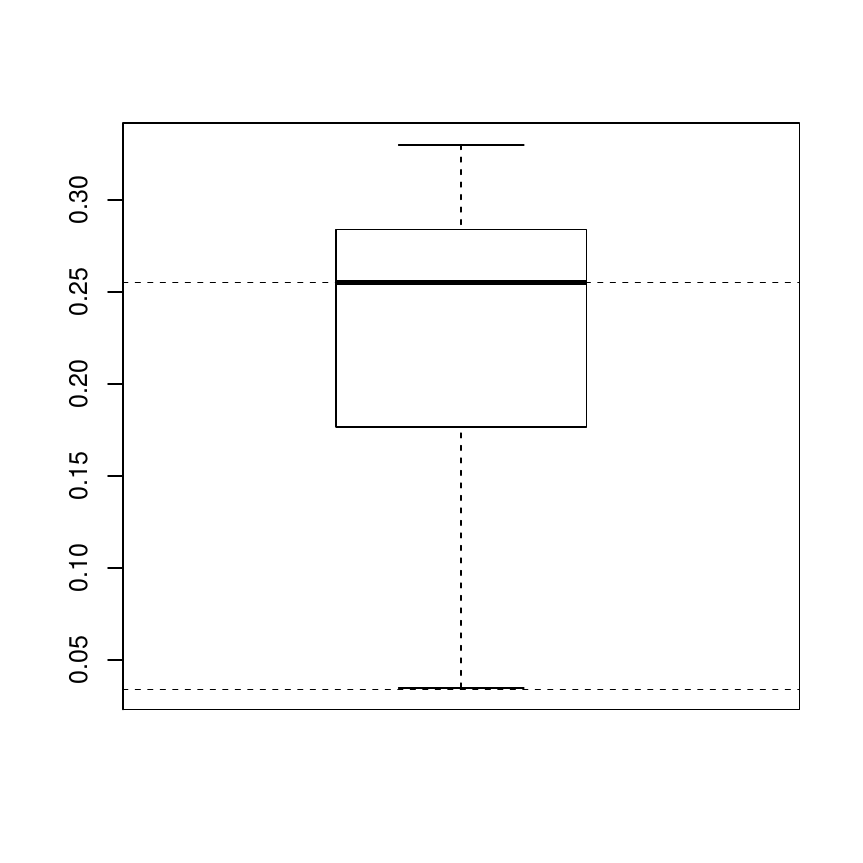}}
		\end{picture}
		\vspace*{-6.5cm}
		\caption{50\% circular regions obtained from the circular kernel estimator $ f_n$ (Grey color) obtained from a sample $\mathcal{X}_{250}$. Boxplots of $\{f_n(X_1),\cdots,f_n(X_{250})\}$ and quantiles (dotted lines) that determine the 50\% regions (bottom).}\label{hdr3333}
	\end{figure}

	Figure \ref{hdr3333} shows the circular kernel estimator $f_n$ (grey color) calculated from a sample  $\mathcal{X}_{250}$ generated from the second model (black color) in Figure \ref{ltau} and different empirically approximated 50\% circular regions (grey color, top). The boxplot of the transformed values denoted by $\{f_n(X_1),\cdots,f_n(X_{250})\}$ is also shown (bottom). The dotted lines represent the quantiles that determine the corresponding 50\% (probability coverage) circular region. Note that only the estimated HDR (left), $\hat{L}(\hat{f}_\tau)$, is able to show the existence of the four existing modes.

	\subsection{A suitable bootstrap bandwidth selector}
	\label{sec:HDR-boot}
	The construction of the kernel density estimator in (\ref{estimacionnucleo}) requires an appropriate selection of $h$. Although there exist several proposals in the literature for this task, none of them has been specifically designed for reconstructing a directional level set. This is the goal of this section. The previous bandwidth selectors (designed for density estimation) are briefly introduced in Appendix \ref{AppendixA}, as supplementary information for the simulation study.
	
	A bootstrap bandwidth selector focused on the problem of reconstructing density level sets is introduced in what follows. The idea is to use an error criterion that quantifies the differences between the theoretical set and its reconstructions instead of measuring the accuracy of kernel estimators. In the real-valued setting, these ideas are also considered in \cite{sam} for proposing one of the first bandwidth selectors in level set estimation setting.
	
	In the directional case, the closed expression of  $d_H(\partial L(f_\tau),\partial\hat{L}(\hat{f}_\tau))$ is not known. However, it could be estimated through a bootstrap procedure. Therefore, a new bandwidth selector can be established  as
	\begin{equation}
		h_{1}=\arg \min_{h>0}\mathbb{E}_B\left[d_H(\partial L^{*}(\hat{f}_\tau^*),\partial \hat{L}(\hat{f}_\tau))\right]
		\label{eq:h1}
	\end{equation}
	where $\mathbb{E}_B$ denotes the bootstrap expectation with respect to random samples $\mathcal X_n=\{X_1^*,\cdots,X_n^*\}$ generated from the directional kernel $f_n$ that, of course, is heavily dependent on a pilot bandwidth.
	
	This bandwidth selector is specifically designed for HDRs estimation, but it may be argued that a plug-in estimator may be computed just taking a kernel density estimator with a suitable bandwidth for reproducing the curve, that is, minimizing some global error on the curve estimate. As mentioned before, there are other approaches for selecting the bandwidth parameter in (\ref{estimacionnucleo}), such as the circular rule-of-thumb by \cite{Taylor} ($h_2$ in this work) or the improved version by \cite{oli1} (namely $h_3$). Cross-validation methods (likelihood $h_4$ and least squares $h_5$) were introduced by \cite{hall87} whereas a bootstrap bandwidth ($h_6$) was presented by \cite{dimarzio2011}. From the previous proposals, cross-validation bandwidth selectors can be applied for data on a sphere $S^{d-1}$ for any $d$. For spherical data, a plug-in bandwidth selector was also introduced by \cite{eduardo2013} ($h_7$).
	
\begin{figure}[h!] \vspace{-2.5cm}
		$ \hspace{1.5cm}$	\begin{picture}(0,0)
		\put(-75,-200){\includegraphics[scale=.6]{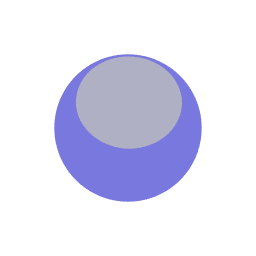}}
		\put(17,-200){\includegraphics[scale=.6]{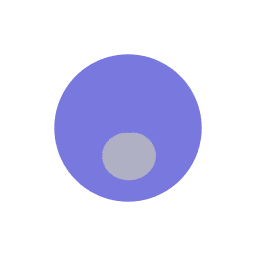}}
		\put(110,-200){\includegraphics[scale=.6]{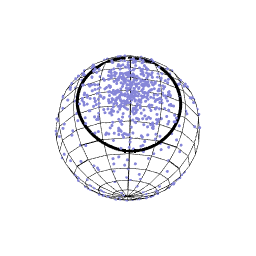}}
		\put(205,-200){\includegraphics[scale=.6]{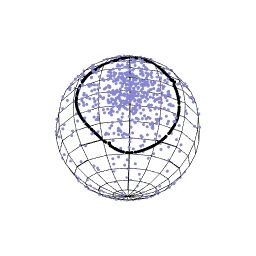}}
		\put(300,-200){\includegraphics[scale=.6]{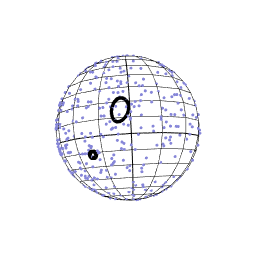}}
		\end{picture}
		\vspace*{6.5cm}
		\caption{Theoretical HDR for model S3 when $\tau=0.5$ (first and second columns). Sample of size $n=1000$ of model 3 (blue color) and corresponding plug-in level set estimators (black color) when $\tau=0.5$ considering $h_7$ (third column) and $h_5$ (fourth and fifth columns) as smoothing parameters. Note that the last two columns show two views of the sphere.}\label{xustificacionngrande}
	\end{figure}

	Figure \ref{xustificacionngrande} shows the theoretical HDR for model S3 (see Section 4.2) when $\tau=0.5$ (first and second columns). Moreover, the plug-in level set estimator $\hat{L}(\hat{f}_\tau)$ obtained from a sample of size $n=1000$ of this density and considering $h_7$ when $\tau=0.5$ is also represented (third column). Note that, for this sample size, only the largest mode is detected. In this particular case, the Hausdorff error is smaller if the level set is reconstructed from $h_5$  (fourth and fifth columns). A relevant issue appears when $h_1$ is estimated from imprecise level set estimators such as the obtained one from $h_7$. Remember that the minimization procedure considered for determining $h_1$ involves the boundary of the set $\hat{L}(\hat{f}_\tau)$. If this set is poorly approximated the resulting bandwidth surely will not provide competitive results. Therefore, largest sample sizes will be considered in this section for avoiding this problem. Additionally, the bandwidth $h_5$ will be used as pilot in order to determine the set $\hat{L}(\hat{f}_\tau)$.

	\section{Simulation study}
	\label{sec:simulation}
	The performance of different bandwidth selectors (our proposal in Section \ref{sec:HDR} and other selectors for density estimation described in Appendix \ref{AppendixA}) is checked through a simulation study. Circular and spherical HDRs are estimated considering the plug-in methods that arise of the consideration of these bandwidths parameters. The code for computing $h_1$ can be obtained from the authors upon request. All the rest bandwidths are implemented in the R packages \texttt{NPCirc}\footnote{\url{https://CRAN.R-project.org/package=NPCirc}} and \texttt{Directional}\footnote{\url{https://CRAN.R-project.org/package=Directional}}.  Sections \ref{circularsimus} and \ref{sphericalsimus} contain the results obtained in circular and  spherical settings, respectively.

	
	\subsection{Circular level set estimation}
	\label{circularsimus}
\begin{figure}[h!] \vspace{-5cm}
		$\hspace{.8cm}$	\begin{picture}(0,0)
		
		\put(60,-160){\textbf{C1 }\textbf{(5)}}
		\put(188,-160){ \textbf{C2 }\textbf{(6)}}
		\put(300,-160){ \textbf{C3 }\textbf{(7)}}

		\put(59,-318){\textbf{C4 } \textbf{(8)}} 
		\put(190,-318){\textbf{C5 }\textbf{(10)}} 
		\put(305,-318){\textbf{C6 }\textbf{(11)}} 
		
		\put(57,-441){ \textbf{C7 } \textbf{(16)}} 
		\put(193,-441){\textbf{C8 } \textbf{(19)}}  
		\put(300,-441){ \textbf{C9 } \textbf{(20)}}  


		\put(-40,-340){\includegraphics[scale=.5]{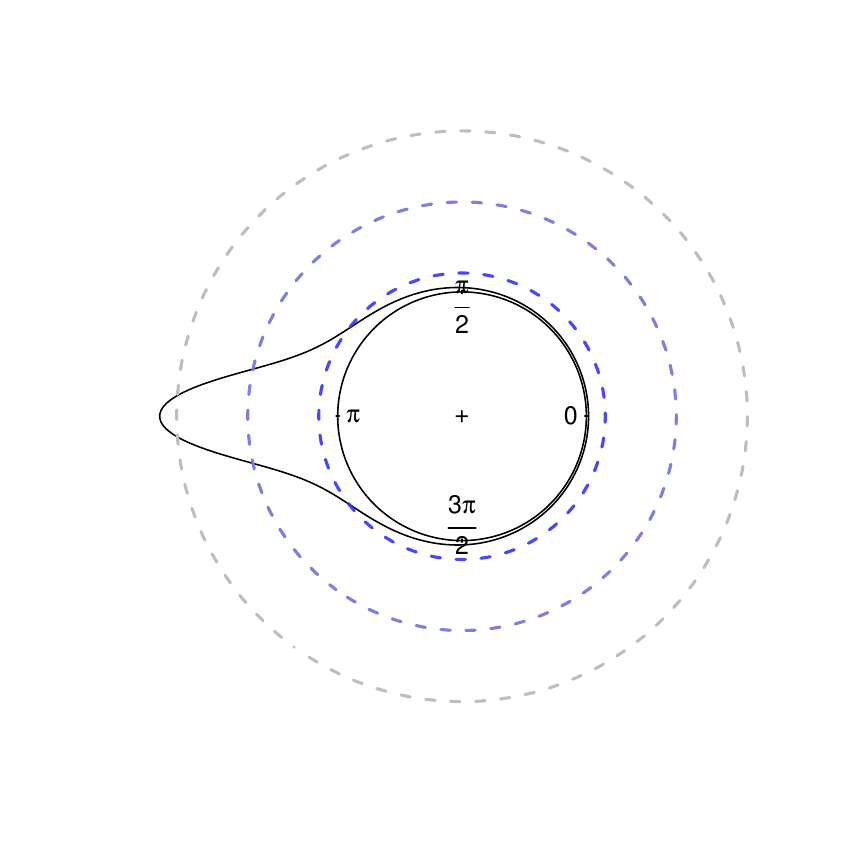}}
		\put(95,-340){\includegraphics[scale=.5]{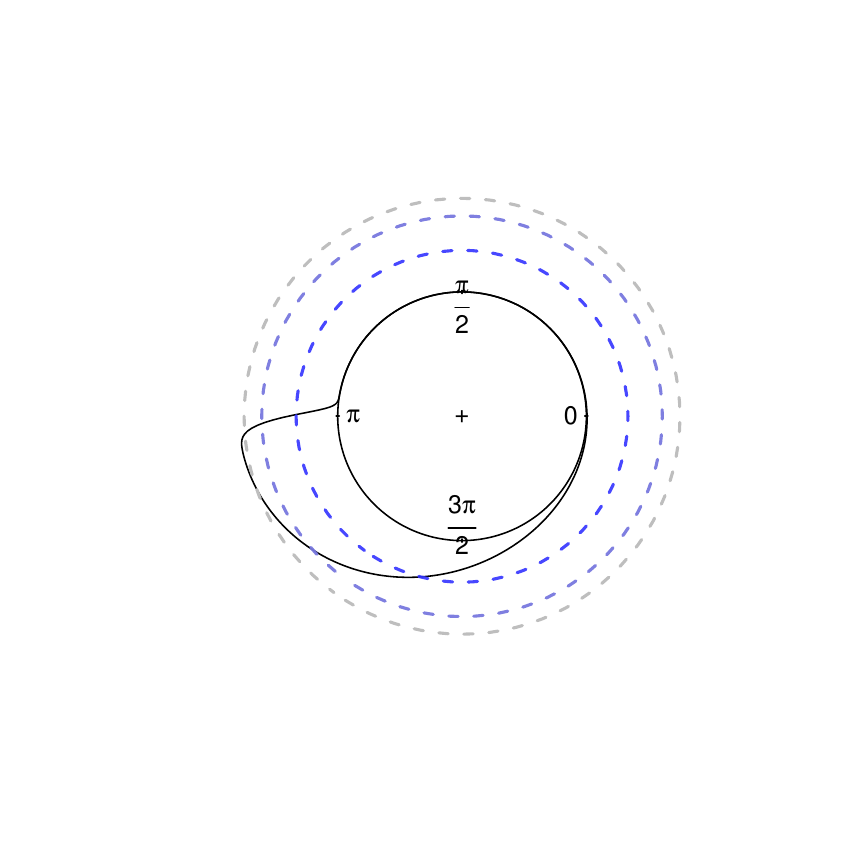}}
		\put(205,-340){\includegraphics[scale=.5]{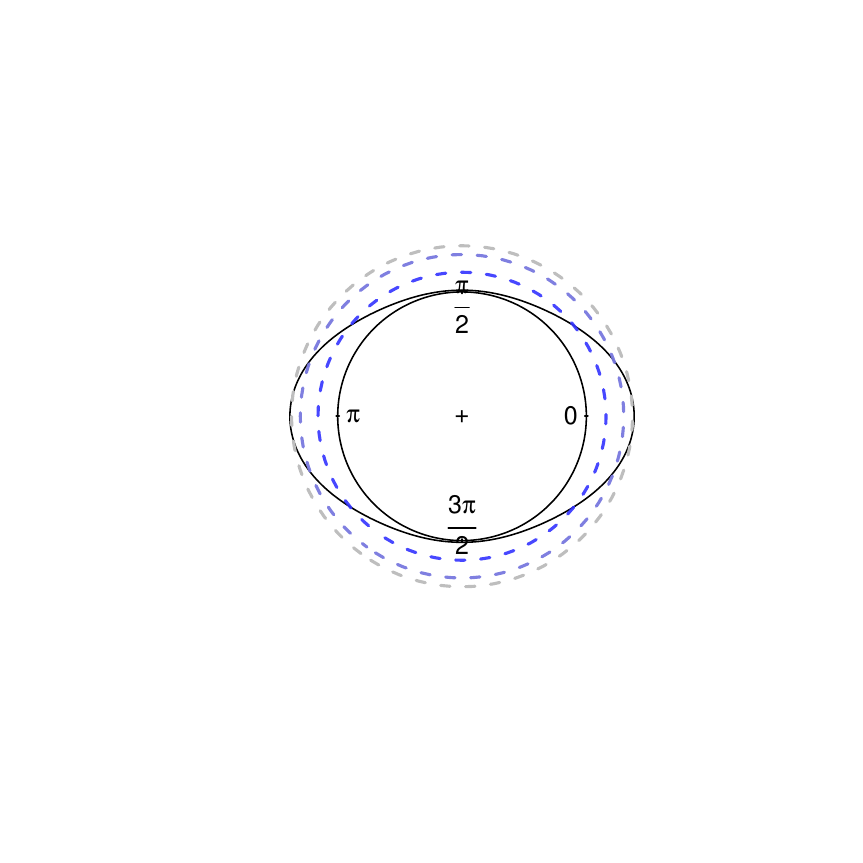}}
		
		\put(-40,-480){\includegraphics[scale=.5]{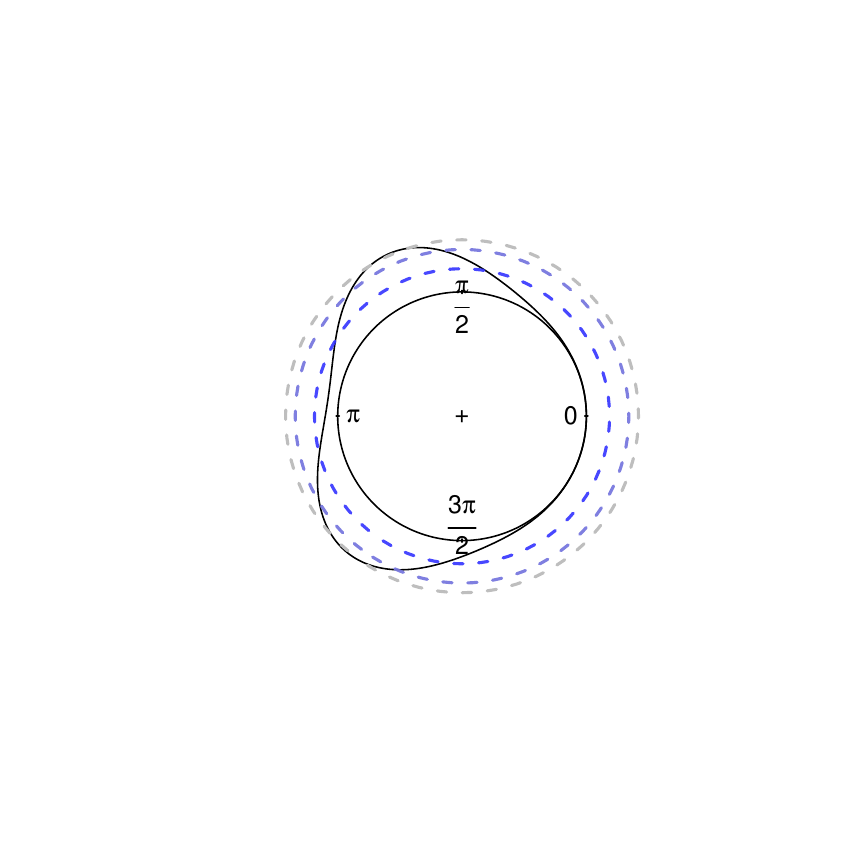}}	
		\put(95,-480){\includegraphics[scale=.5]{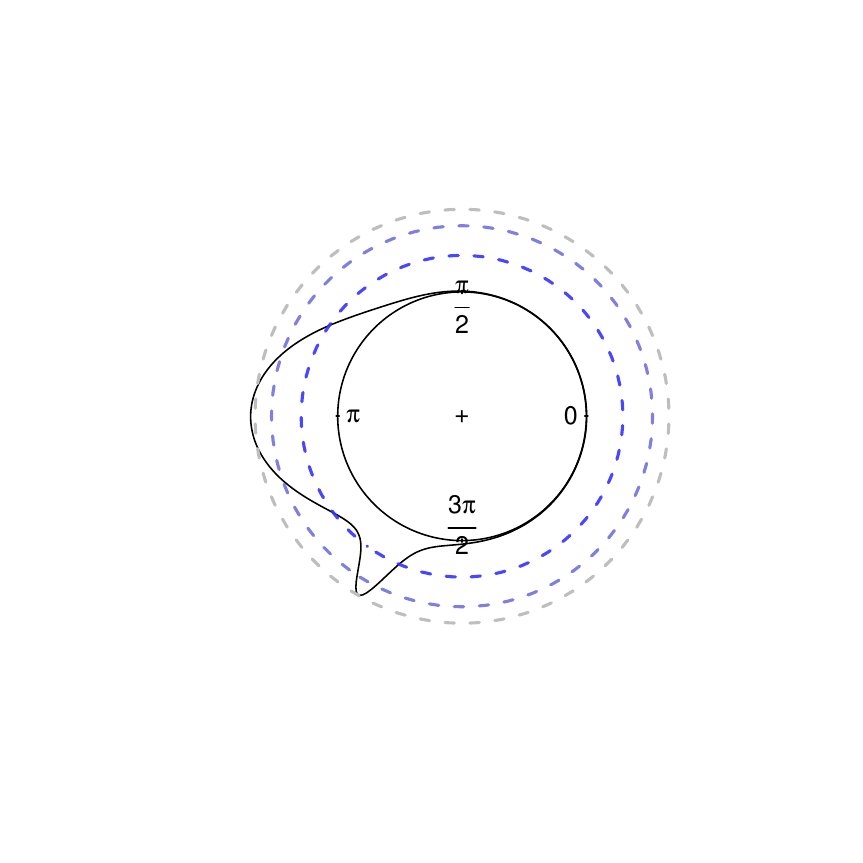}}
		\put(205,-480){\includegraphics[scale=.5]{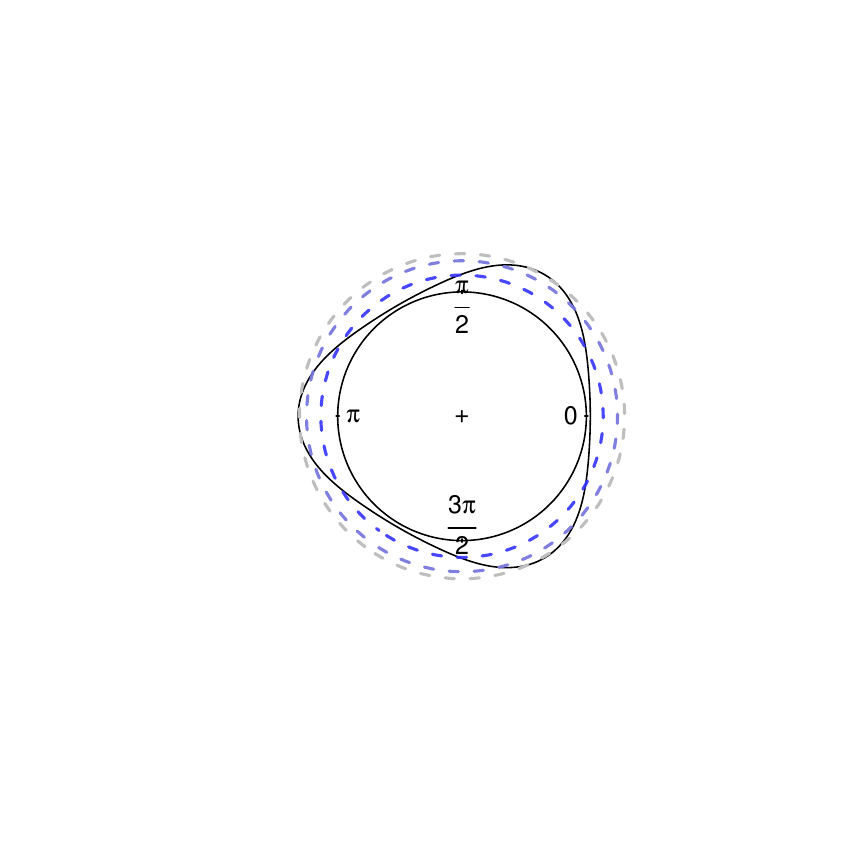}}

		\put(-40,-600){\includegraphics[scale=.5]{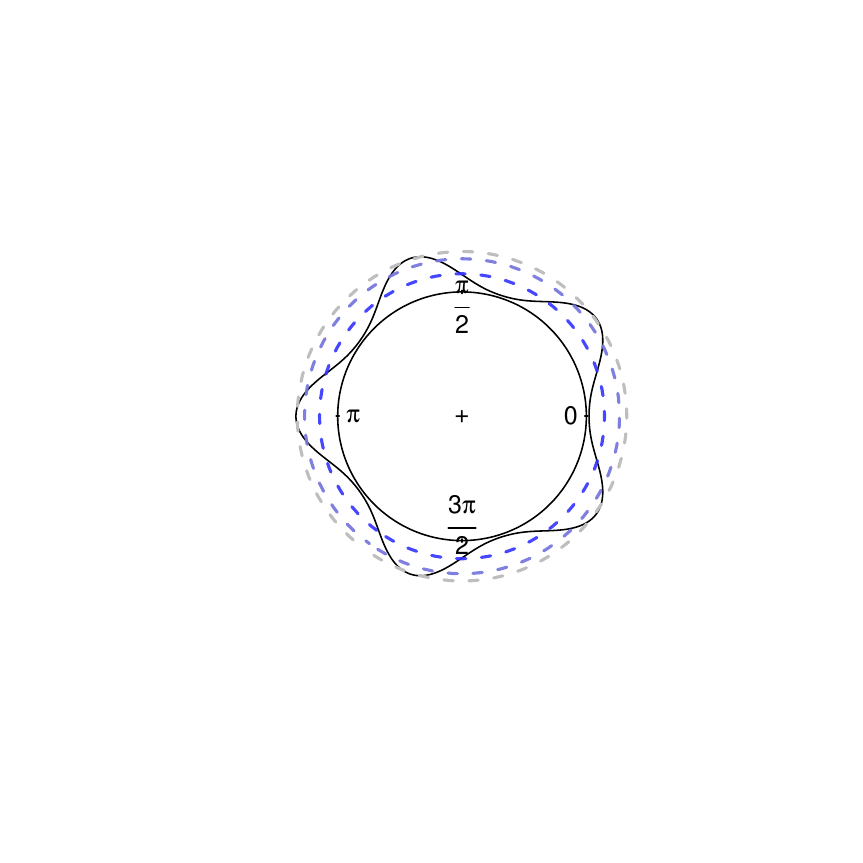}}
		\put(95,-600){\includegraphics[scale=.5]{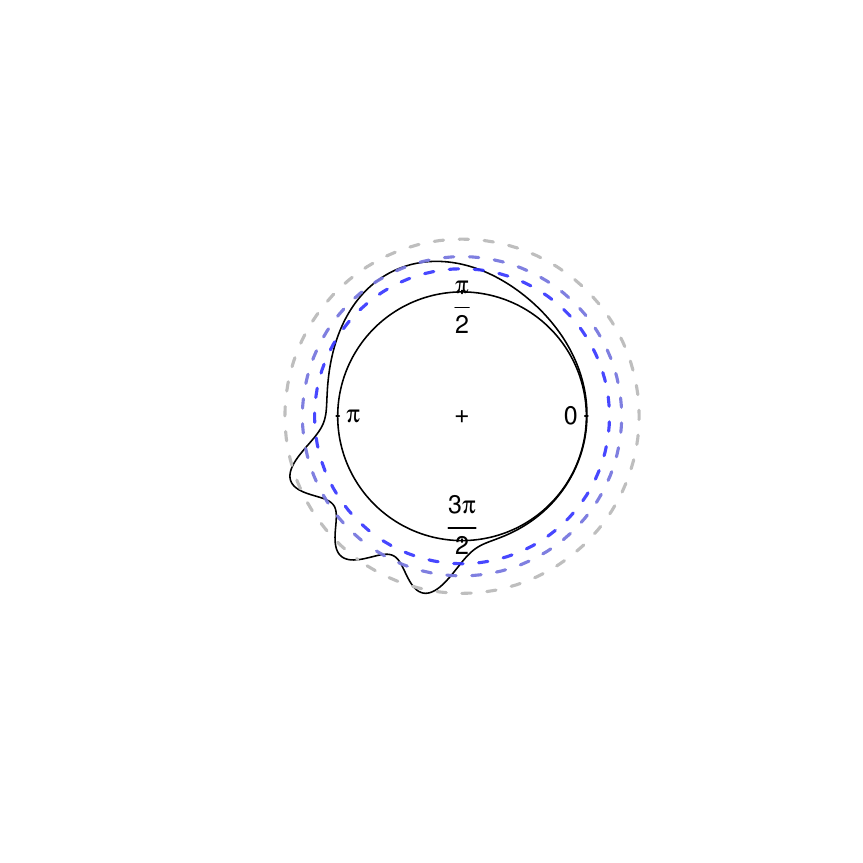}}
		\put(205,-600){\includegraphics[scale=.5]{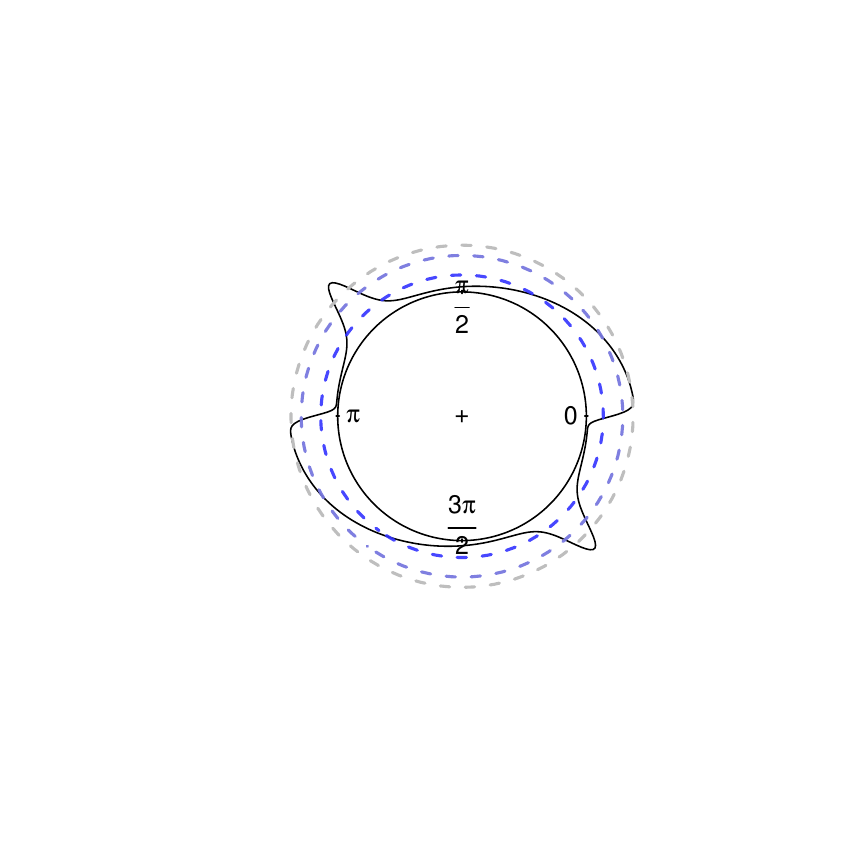}}
		
		
		\end{picture}\vspace{20cm}
		\caption{Circular density models for simulations. Dotted circles represent the threshold $f_\tau$ when $\tau=0.2$, $\tau=0.5$ and $\tau=0.8$, respectively.}\label{circulardensities}
	\end{figure}
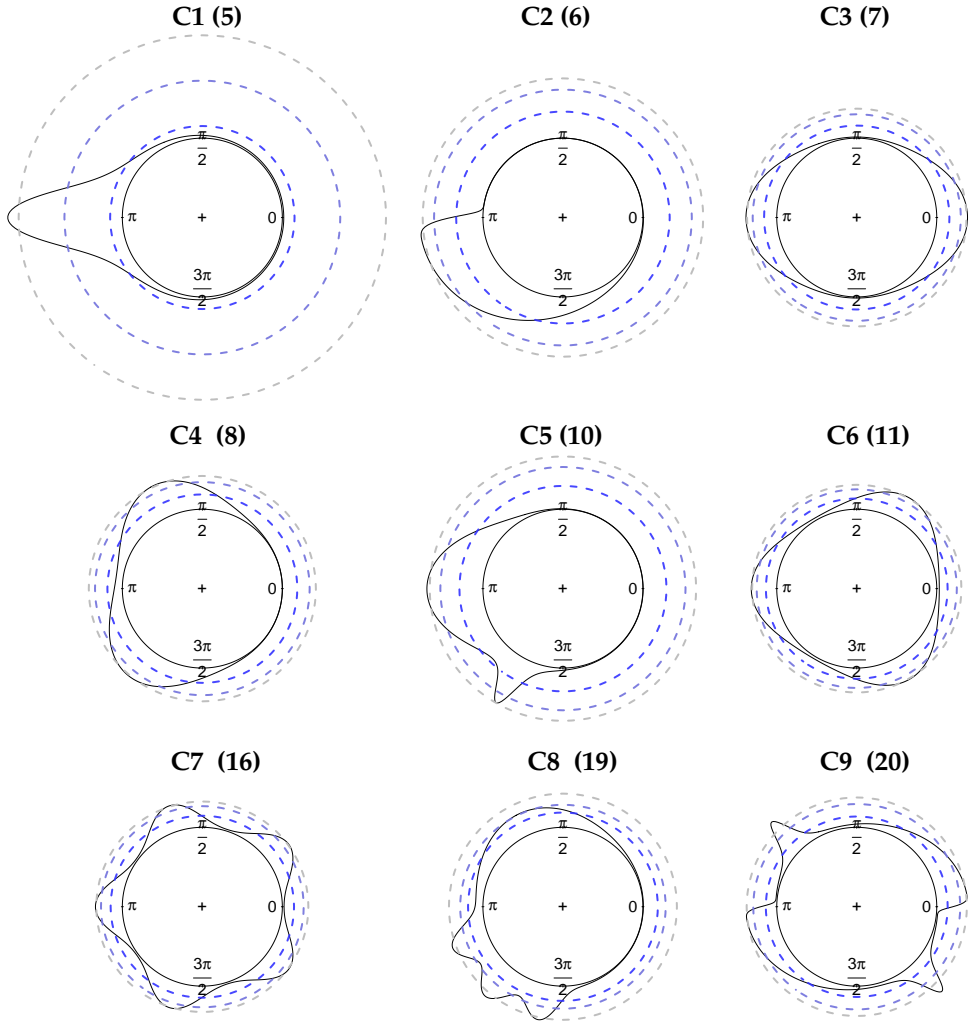

	A collection of 9 circular densities (models C1 to C9) have been considered in this simulation study. These models are mixture of different circular distributions and they correspond to densities 5, 6, 7, 8, 10, 11, 16, 19 and 20 fully described in \cite{oli2}. Figure \ref{circulardensities} shows these densities and the thresholds $f_\tau$ for $\tau=0.2$, $\tau=0.5$ and $\tau=0.8$ through dotted circles.
	
	A total of $250$ random samples of sizes $n=500$ and $n=1000$ were generated for each of these models. From each sample, circular HDRs are reconstructed for $\tau=0.2$, $\tau=0.5$ and $\tau=0.8$. The behavior of plug-in methods that emerge from the consideration of different bandwidth parameters will be checked. Note that for computing $h_1$, a pilot bandwidth is required. In this study, $h_3$ has been taken as a pilot, and $B=200$ resamples are considered for obtaining $h_1$.

	For each method and each sample, the estimation error is measured by computing the Hausdorff and Euclidean distances ($d_H$ and $d_E$) between the boundaries of estimated level set and the frontier of theoretical set. Note that the Euclidean distance is not as informative as the Hausdorff criterion to detect differences between sets. Therefore, just results for $d_H$ are shown. As a reference, note that the maximum value of both criteria in $S^1$ is $2$. This upper bound coincides exactly with the length of the diameter of the circle. 
	
	Tables \ref{meanstau02n500} and \ref{meanstau02n1000} show the means and the standard deviations of the $250$ estimation errors obtained when $\tau=0.2$ from samples of sizes $n=500$ and $n=1000$, respectively. Blue cells corresponds to the lowest mean errors obtained for each density. Taking into account the variety of models considered, exhibiting different features, it is not surprising that all of the bandwidth selectors are the best ones for some model, showing $h_1$ a competitive behavior in all cases. This feature is also illustrated in Figure \ref{vioplot_cir_1}. Note that $h_2$ presents a poor behavior for models C3, C6, C7, C8 and C9, and $h_6$ performance is also unsatisfactory for models C2 and C9, although it improves with sample size. 
	
	Similar comments can be made for $\tau=0.5$ (see Tables \ref{meanstau05n500} and \ref{meanstau05n1000}, for $n=500$ and $n=1000$, respectively), although in this case, $h_1$ (being a competitive selector in all the scenarios) is the best one for models C3, C5, C6 and C8 (with $n=1000$).

	Tables \ref{meanstau08n500} and \ref{meanstau08n1000} contain the results obtained for $\tau=0.8$ when $n=500$ and $n=1000$, respectively. According to Table \ref{meanstau08n1000}, $h_1$ is the best selector for five models (C2, C6, C7, C8 and C9). It is clear that the new selector improves its results when large values of $\tau$ are considered and, therefore, largest modes are identified.

	\newpage
	\begin{landscape}
		\newgeometry{left=2cm,right=2cm,top=5.5cm,bottom=1cm} 
		\thispagestyle{empty}
		
		\begin{table}[h!]\centering
			\caption{Means (M) and standard deviations (SD) of $250$ errors in Hausdorff distance for $\tau=0.2$, $n=500$ and $B=200$.}\label{meanstau02n500}
			$\hspace{-6cm}$\begin{tabular}{ccccccccccccccccccc}
				\hline
				&\multicolumn{2}{c}{\textbf{C1}}&\multicolumn{2}{c}{\textbf{C2}}&\multicolumn{2}{c}{\textbf{C3}}&\multicolumn{2}{c}{\textbf{C4}}&\multicolumn{2}{c}{\textbf{C5}}&\multicolumn{2}{c}{\textbf{C6}}&\multicolumn{2}{c}{\textbf{C7}}&\multicolumn{2}{c}{\textbf{C8}}&\multicolumn{2}{c}{\textbf{C9}}\\
				&M& SD&M& SD&M& SD&M& SD&M& SD&M& SD&M& SD&M& SD&M& SD\\ 
				\hline	
				$h_1$&0.086&0.064&0.070&0.071&0.089&0.037&0.084&0.031&0.086&0.053&0.098&0.038&\cellcolor{GGG}0.081&0.026&0.154&0.098&0.145&0.065\\
				$h_2$&\cellcolor{GGG}0.067&0.039&\cellcolor{GGG}0.049&0.032&1.443&0.309&0.099&0.038&0.206&0.153&1.786&0.110&1.812&0.064&0.251&0.094&1.758&0.115\\
				$h_3$&0.094&0.060&0.065&0.059&0.090&0.036&0.085&0.031&\cellcolor{GGG}0.081&0.053&0.100&0.038&0.084&0.027&0.153&0.093&0.142&0.063\\
				$h_4$&0.075&0.042&0.051&0.036&0.090&0.036&0.085&0.031&0.097&0.072&0.099&0.038&\cellcolor{GGG}0.081&0.025&\cellcolor{GGG}0.128&0.067&\cellcolor{GGG}0.131&0.049\\
				$h_5$&0.075&0.041&0.051&0.036&0.091&0.036&0.084&0.031&0.190&0.150&0.100&0.038&\cellcolor{GGG}0.081&0.025&\cellcolor{GGG}0.128&0.067&\cellcolor{GGG}0.131&0.049\\
				$h_6$&0.093&0.058&\cellcolor{GGG}0.049&0.032&\cellcolor{GGG}0.087&0.033&\cellcolor{GGG}0.081&0.031&0.358&0.110&\cellcolor{GGG}0.097&0.037&0.082&0.026&0.156&0.091&1.686&0.372\\
				\hline			
			\end{tabular} 
		\end{table}

		\begin{table}[h!]\centering
			\caption{Means (M) and standard deviations (SD) of $250$ errors in Hausdorff distance for $\tau=0.2$, $n=1000$ and $B=200$.}\label{meanstau02n1000}
			$\hspace{-6cm}$\begin{tabular}{ccccccccccccccccccc}
				\hline
				&\multicolumn{2}{c}{\textbf{C1}}&\multicolumn{2}{c}{\textbf{C2}}&\multicolumn{2}{c}{\textbf{C3}}&\multicolumn{2}{c}{\textbf{C4}}&\multicolumn{2}{c}{\textbf{C5}}&\multicolumn{2}{c}{\textbf{C6}}&\multicolumn{2}{c}{\textbf{C7}}&\multicolumn{2}{c}{\textbf{C8}}&\multicolumn{2}{c}{\textbf{C9}}\\
				&M& SD&M& SD&M& SD&M& SD&M& SD&M& SD&M& SD&M& SD&M& SD\\
				\hline
				
				$h_1$&0.058&0.040&0.040&0.034&\cellcolor{GGG}0.061&0.025&0.058&0.024&\cellcolor{GGG}0.060&0.028&0.072&0.028&0.057&0.016&0.101&0.060&0.115&0.042\\
				$h_2$&\cellcolor{GGG}0.049&0.027&0.036&0.021&1.428&0.334&0.066&0.024&0.097&0.066&1.798&0.106&1.820&0.057&0.179&0.047&1.759&0.118\\
				$h_3$&0.059&0.039&0.037&0.028&0.063&0.024&0.058&0.025&0.061&0.024&0.073&0.028&0.057&0.016&0.097&0.060&\cellcolor{GGG}0.110&0.044\\
				$h_4$&0.053&0.028&0.036&0.021&0.063&0.024&0.057&0.024&0.077&0.037&0.072&0.028&\cellcolor{GGG}0.056&0.016&\cellcolor{GGG}0.091&0.043&0.114&0.041\\
				$h_5$&0.053&0.028&0.036&0.021&0.063&0.024&\cellcolor{GGG}0.056&0.023&0.083&0.052&0.072&0.028&\cellcolor{GGG}0.056&0.016&\cellcolor{GGG}0.091&0.043&0.114&0.041\\
				$h_6$&0.059&0.039&\cellcolor{GGG}0.035&0.020&\cellcolor{GGG}0.061&0.024&0.057&0.024&0.174&0.148&\cellcolor{GGG}0.071&0.028&0.057&0.016&0.107&0.038&0.111&0.042\\
				\hline	
			\end{tabular} 
		\end{table}  
		\begin{figure}[h!] 
			\begin{picture}(0,0)
			\put(-80,-110){\includegraphics[scale=0.35]{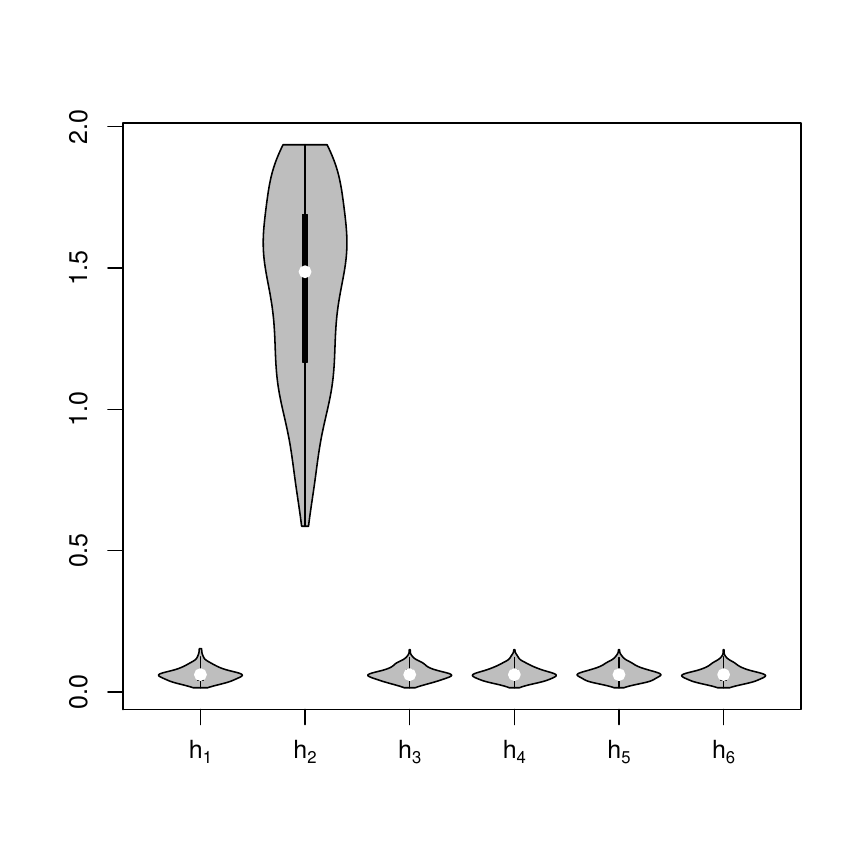}}
			\put(60,-110){\includegraphics[scale=0.35]{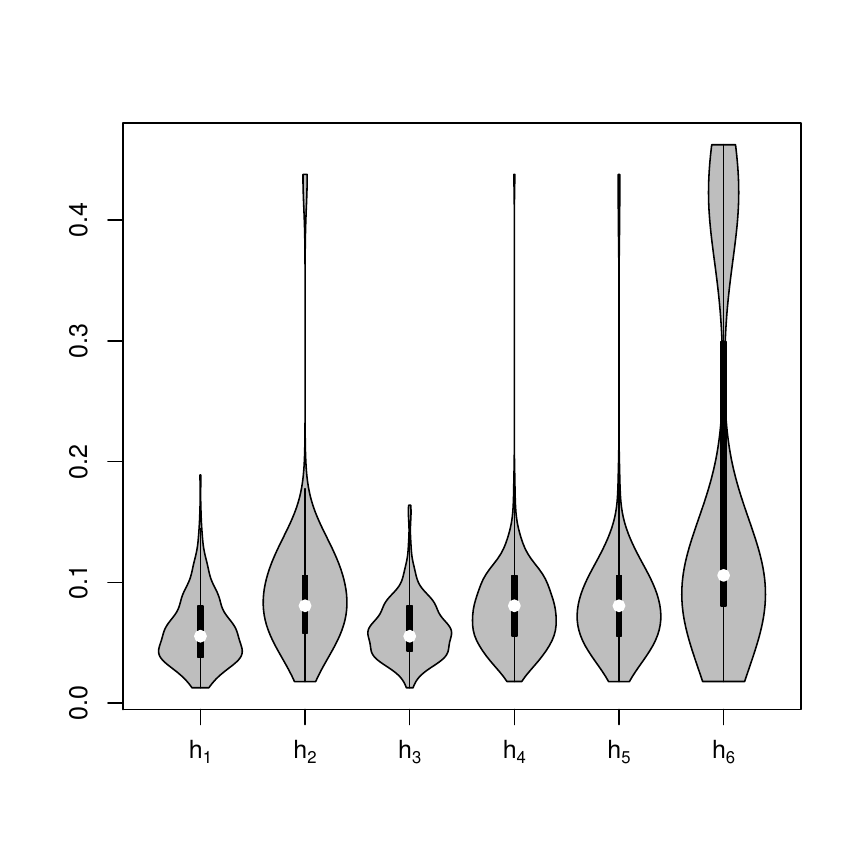}}
			\put(200,-110){\includegraphics[scale=0.35]{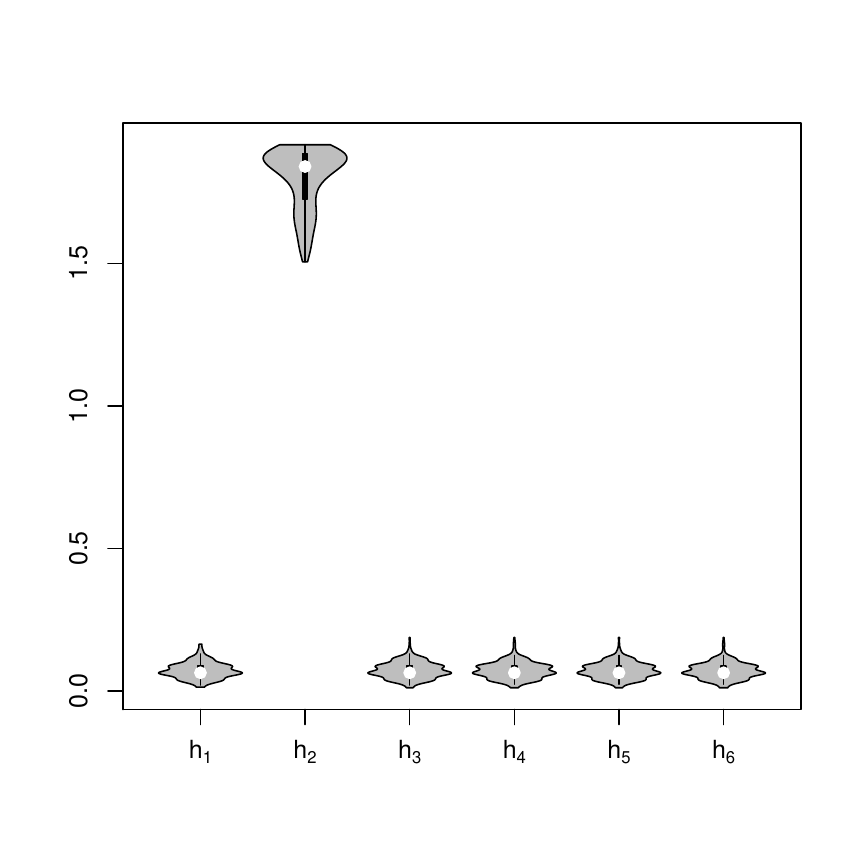}}
			\put(340,-110){\includegraphics[scale=0.35]{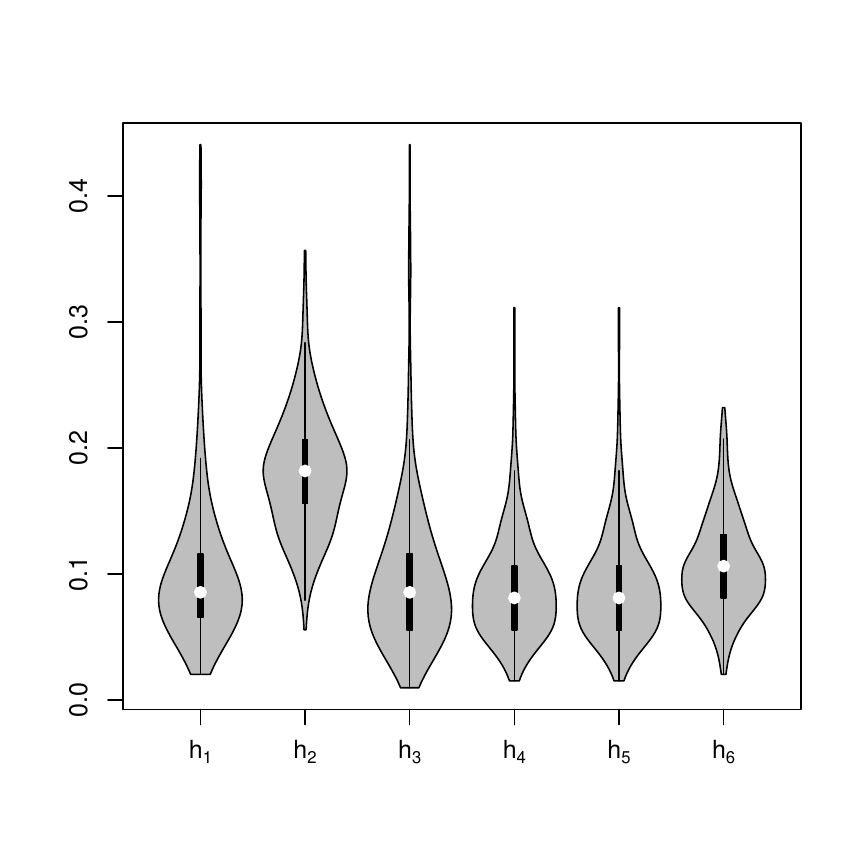}}
			\end{picture} \vspace{3.2cm}\\
			\caption{Violin plots of Hausdorff errors for models C3, C5, C6 and C8 for $\tau=0.2$ and $n=1000$. Note that due to the behaviour of $h_2$, the scale of these figures is different.}
			\label{vioplot_cir_1}
		\end{figure}
		\restoregeometry
	\end{landscape}
	\newpage
	\begin{landscape}
		\newgeometry{left=2cm,right=2cm,top=5.5cm,bottom=1cm} 
		\thispagestyle{empty}
		\begin{table}[h!]\centering
			\caption{Means (M) and standard deviations (SD) of $250$ errors in Hausdorff distance for $\tau=0.5$, $n=500$ and $B=200$.}\label{meanstau05n500}
			$\hspace{-6cm}$\begin{tabular}{ccccccccccccccccccc}
				\hline
				&\multicolumn{2}{c}{\textbf{C1}}&\multicolumn{2}{c}{\textbf{C2}}&\multicolumn{2}{c}{\textbf{C3}}&\multicolumn{2}{c}{\textbf{C4}}&\multicolumn{2}{c}{\textbf{C5}}&\multicolumn{2}{c}{\textbf{C6}}&\multicolumn{2}{c}{\textbf{C7}}&\multicolumn{2}{c}{\textbf{C8}}&\multicolumn{2}{c}{\textbf{C9}}\\
				&M& SD&M& SD&M& SD&M& SD&M& SD&M& SD&M& SD&M& SD&M& SD\\
				\hline
				$h_1$&0.027&0.015&0.120&0.092&\cellcolor{GGG}0.113&0.052&0.107&0.055&0.425&0.290&0.141&0.063&0.206&0.216&\cellcolor{GGG}0.504&0.409&0.243&0.211\\
				$h_2$&\cellcolor{GGG}0.026&0.014&0.103&0.042&1.303&0.372&0.104&0.047&0.655&0.103&1.427&0.085&1.313&0.060&0.777&0.385&1.327&0.200\\
				$h_3$&\cellcolor{GGG}0.026&0.015&0.100&0.088&0.120&0.054&0.112&0.056&\cellcolor{GGG}0.417&0.307&0.141&0.058&\cellcolor{GGG}0.192&0.221&0.554&0.407&\cellcolor{GGG}0.213&0.215\\
				$h_4$&\cellcolor{GGG}0.026&0.015&\cellcolor{GGG}0.090&0.054&0.123&0.060&0.112&0.056&0.588&0.219&0.142&0.062&0.221&0.270&0.665&0.407&0.403&0.363\\
				$h_5$&\cellcolor{GGG}0.026&0.015&0.091&0.054&0.122&0.059&0.108&0.054&0.633&0.154&0.143&0.063&0.221&0.270&0.667&0.404&0.403&0.363\\
				$h_6$&\cellcolor{GGG}0.026&0.015&0.104&0.051&\cellcolor{GGG}0.113&0.049&\cellcolor{GGG}0.103&0.048&0.659&0.054&\cellcolor{GGG}0.136&0.056&0.207&0.252&0.658&0.380&1.273&0.286\\
				
				\hline	
			\end{tabular}
		\end{table}
		\begin{table}[h!]\centering
			\caption{Means (M) and standard deviations (SD) of $250$ errors in Hausdorff distance for $\tau=0.5$, $n=1000$ and $B=200$.}\label{meanstau05n1000}
			$\hspace{-6cm}$\begin{tabular}{ccccccccccccccccccc}
				\hline
				&\multicolumn{2}{c}{\textbf{C1}}&\multicolumn{2}{c}{\textbf{C2}}&\multicolumn{2}{c}{\textbf{C3}}&\multicolumn{2}{c}{\textbf{C4}}&\multicolumn{2}{c}{\textbf{C5}}&\multicolumn{2}{c}{\textbf{C6}}&\multicolumn{2}{c}{\textbf{C7}}&\multicolumn{2}{c}{\textbf{C8}}&\multicolumn{2}{c}{\textbf{C9}}\\
				&M& SD&M& SD&M& SD&M& SD&M& SD&M& SD&M& SD&M& SD&M& SD\\

				\hline		
				
				$h_1$&0.019&0.011&0.071&0.050&\cellcolor{GGG}0.080&0.033&0.073&0.031&\cellcolor{GGG}0.410&0.307&\cellcolor{GGG}0.095&0.042&0.105&0.065&\cellcolor{GGG}0.488&0.421&0.133&0.105\\
				$h_2$&\cellcolor{GGG}0.018&0.010&0.086&0.030&1.262&0.340&\cellcolor{GGG}0.070&0.028&0.661&0.080&1.425&0.082&1.306&0.047&0.673&0.357&1.301&0.191\\
				$h_3$&0.019&0.010&\cellcolor{GGG}0.062&0.041&0.083&0.033&0.077&0.034&0.414&0.307&0.101&0.043&0.095&0.063&0.569&0.390&\cellcolor{GGG}0.117&0.098\\
				$h_4$&\cellcolor{GGG}0.018&0.010&0.078&0.031&0.084&0.038&0.076&0.033&0.628&0.159&0.099&0.040&0.104&0.116&0.650&0.377&0.297&0.319\\
				$h_5$&\cellcolor{GGG}0.018&0.010&0.079&0.031&0.085&0.038&0.075&0.034&0.631&0.153&0.099&0.041&0.104&0.116&0.650&0.377&0.297&0.319\\
				$h_6$&0.019&0.010&0.073&0.035&0.081&0.032&0.074&0.029&0.646&0.115&0.097&0.038&\cellcolor{GGG}0.092&0.036&0.624&0.357&0.168&0.210\\
				\hline
			\end{tabular}
		\end{table} 
		\begin{figure}[h!] 
			\begin{picture}(0,0)
			\put(-80,-110){\includegraphics[scale=0.35]{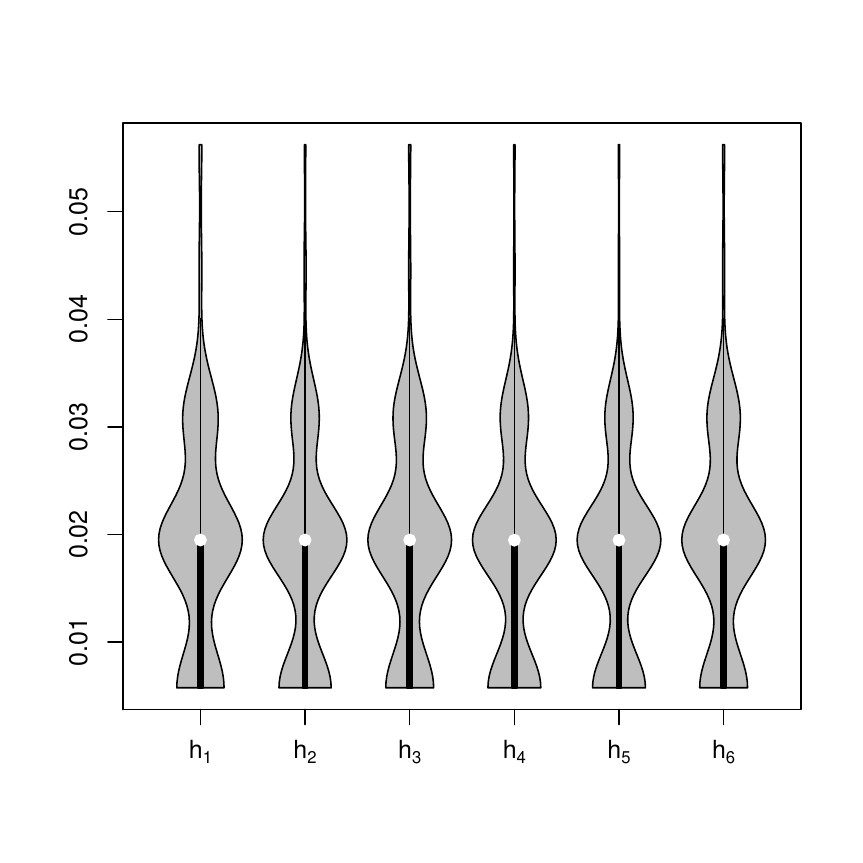}}
			\put(60,-110){\includegraphics[scale=0.35]{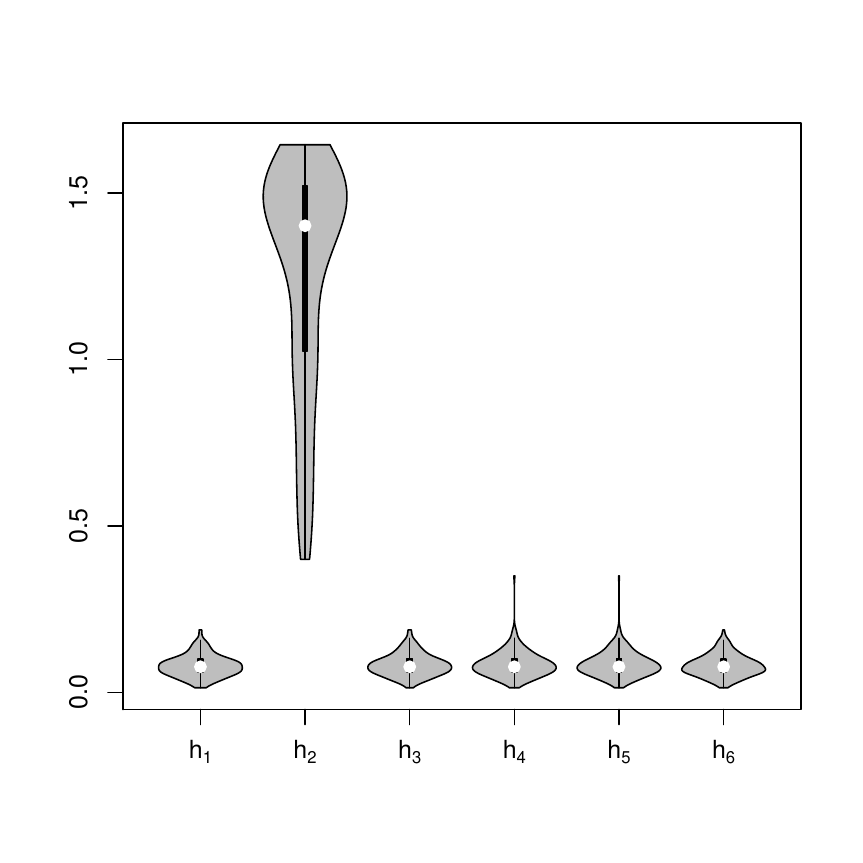}}
			\put(200,-110){\includegraphics[scale=0.35]{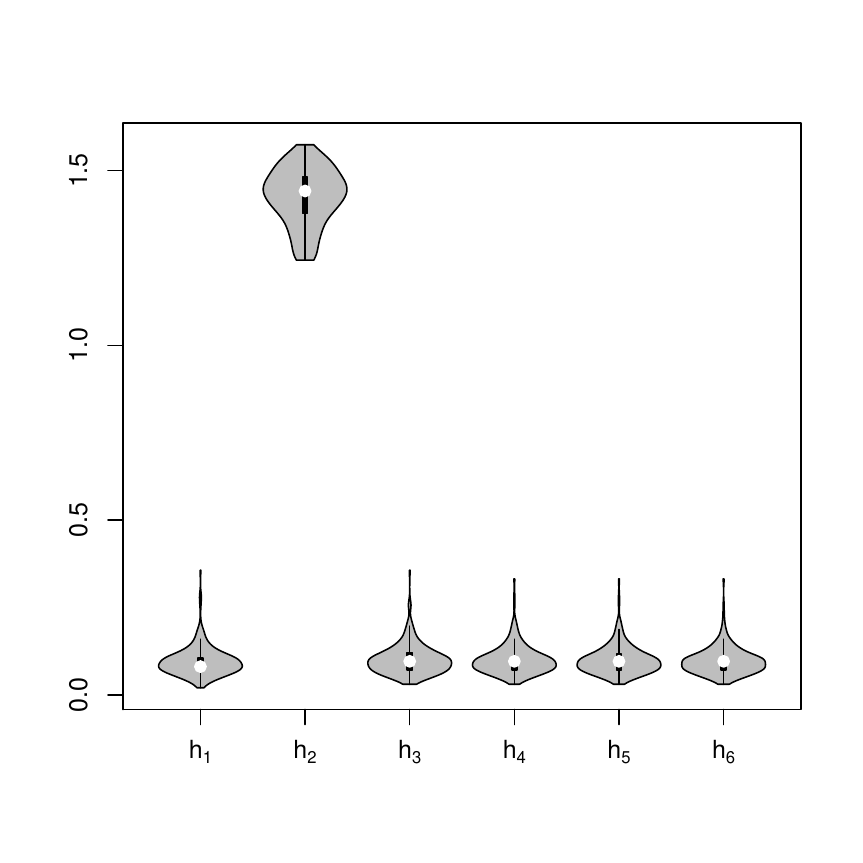}}
			\put(340,-110){\includegraphics[scale=0.35]{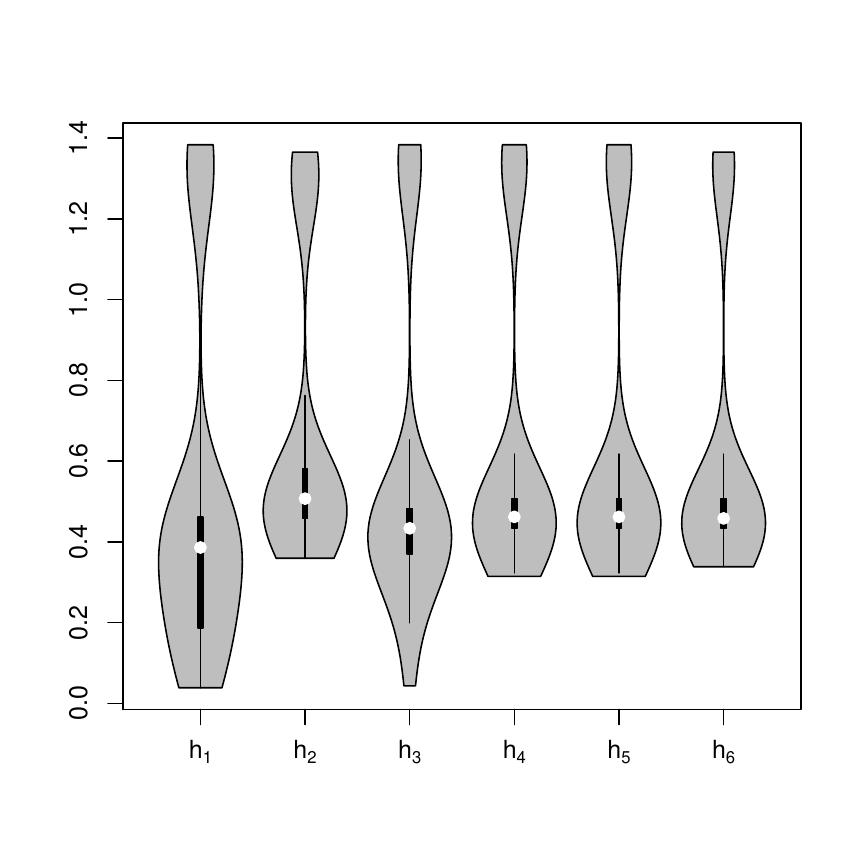}}
			\end{picture} \vspace{3.2cm}\\
			\caption{Violin plots of Hausdorff errors for models C1, C3, C6 and C8 when $\tau=0.5$ and $n=1000$. Note that due to the behaviour of $h_2$, the scale of these figures is different.}	\label{vioplot_cir_2}
		\end{figure}
		\restoregeometry
	\end{landscape}

	\newpage
	\begin{landscape}
		\newgeometry{left=2cm,right=2cm,top=5.5cm,bottom=1cm} 
		\thispagestyle{empty}
		\begin{table}[h!]\centering

			\caption{Means (M) and standard deviations (SD) of $250$ errors in Hausdorff distance for $\tau=0.8$, $n=500$ and $B=200$.}\label{meanstau08n500}
			$\hspace{-6cm}$\begin{tabular}{ccccccccccccccccccc}
				\hline
				&\multicolumn{2}{c}{\textbf{C1}}&\multicolumn{2}{c}{\textbf{C2}}&\multicolumn{2}{c}{\textbf{C3}}&\multicolumn{2}{c}{\textbf{C4}}&\multicolumn{2}{c}{\textbf{C5}}&\multicolumn{2}{c}{\textbf{C6}}&\multicolumn{2}{c}{\textbf{C7}}&\multicolumn{2}{c}{\textbf{C8}}&\multicolumn{2}{c}{\textbf{C9}}\\
				&M& SD&M& SD&M& SD&M& SD&M& SD&M& SD&M& SD&M& SD&M& SD\\

				\hline		
				
				$h_1$&0.022&0.014&0.151&0.108&\cellcolor{GGG}0.691&0.812&\cellcolor{GGG}0.610&0.679&0.079&0.075&\cellcolor{GGG}1.189&0.664&\cellcolor{GGG}1.076&0.333&0.353&0.243&\cellcolor{GGG}0.837&0.436\\
				$h_2$&0.020&0.013&0.184&0.063&1.849&0.282&0.905&0.756&0.058&0.036&1.747&0.059&1.820&0.068&0.356&0.060&1.809&0.188\\
				$h_3$&0.022&0.014&\cellcolor{GGG}0.141&0.106&0.705&0.835&0.635&0.701&0.088&0.073&1.246&0.644&1.143&0.236&\cellcolor{GGG}0.342&0.247&0.983&0.363\\
				$h_4$&\cellcolor{GGG}0.019&0.013&0.155&0.081&0.685&0.823&0.634&0.700&0.070&0.045&1.269&0.632&1.171&0.262&0.404&0.195&1.101&0.317\\
				$h_5$&\cellcolor{GGG}0.019&0.013&0.157&0.081&0.726&0.844&0.650&0.709&0.061&0.041&1.246&0.644&1.171&0.262&0.407&0.186&1.101&0.317\\
				$h_6$&0.022&0.014&0.184&0.060&0.784&0.873&0.673&0.721&\cellcolor{GGG}0.051&0.031&1.310&0.606&1.162&0.251&0.396&0.096&1.752&0.251\\
				
				\hline	
			\end{tabular}
		\end{table}  
		\begin{table}[h!]\centering
			\caption{Means (M) and standard deviations (SD) of $250$ errors in Hausdorff distance for $\tau=0.8$, $n=1000$ and $B=200$.}\label{meanstau08n1000}
			$\hspace{-6cm}$\begin{tabular}{ccccccccccccccccccc}
				\hline
				
				&\multicolumn{2}{c}{\textbf{C1}}&\multicolumn{2}{c}{\textbf{C2}}&\multicolumn{2}{c}{\textbf{C3}}&\multicolumn{2}{c}{\textbf{C4}}&\multicolumn{2}{c}{\textbf{C5}}&\multicolumn{2}{c}{\textbf{C6}}&\multicolumn{2}{c}{\textbf{C7}}&\multicolumn{2}{c}{\textbf{C8}}&\multicolumn{2}{c}{\textbf{C9}}\\
				&M& SD&M& SD&M& SD&M& SD&M& SD&M& SD&M& SD&M& SD&M& SD\\
				\hline		
				$h_1$&0.015&0.012&\cellcolor{GGG}0.127&0.075&0.472&0.678&0.441&0.594&0.056&0.038&\cellcolor{GGG}0.983&0.721&\cellcolor{GGG}1.048&0.30&\cellcolor{GGG}0.252&0.220&\cellcolor{GGG}0.708&0.416\\
				$h_2$&\cellcolor{GGG}0.013&0.011&0.157&0.042&1.878&0.231&0.561&0.707&0.045&0.029&1.746&0.058&1.825&0.063&0.341&0.051&1.810&0.188\\
				$h_3$&0.016&0.012&0.120&0.070&0.461&0.692&\cellcolor{GGG}0.410&0.588&0.064&0.041&1.011&0.724&1.086&0.242&0.257&0.219&0.897&0.334\\
				$h_4$&\cellcolor{GGG}0.013&0.011&0.146&0.046&0.467&0.699&0.412&0.593&0.050&0.032&1.004&0.727&1.112&0.241&0.360&0.176&1.038&0.199\\
				$h_5$&\cellcolor{GGG}0.013&0.011&0.146&0.046&\cellcolor{GGG}0.447&0.680&0.433&0.614&0.050&0.032&0.998&0.728&1.112&0.241&0.360&0.176&1.038&0.199\\
				$h_6$&0.016&0.012&0.136&0.050&0.460&0.702&0.431&0.613&\cellcolor{GGG}0.043&0.027&1.025&0.724&1.086&0.242&0.396&0.077&0.957&0.289\\
				
				\hline	
			\end{tabular}
		\end{table} 
		\begin{figure}[h!] 
			\begin{picture}(0,0)
			\put(60,-110){\includegraphics[scale=0.35]{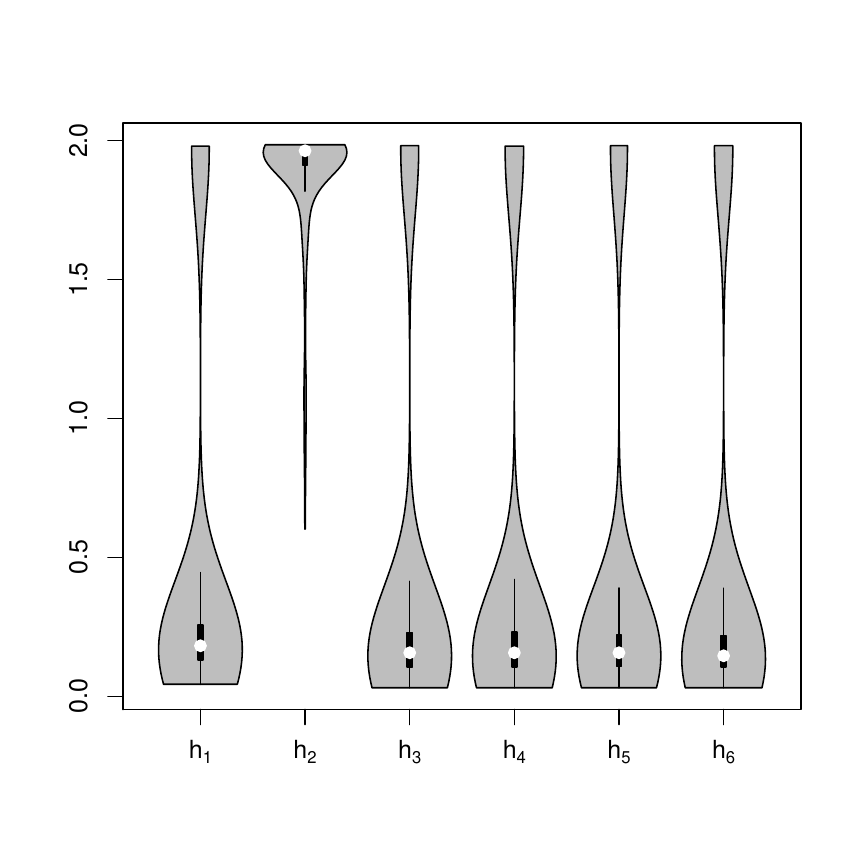}}
			\put(200,-110){\includegraphics[scale=0.35]{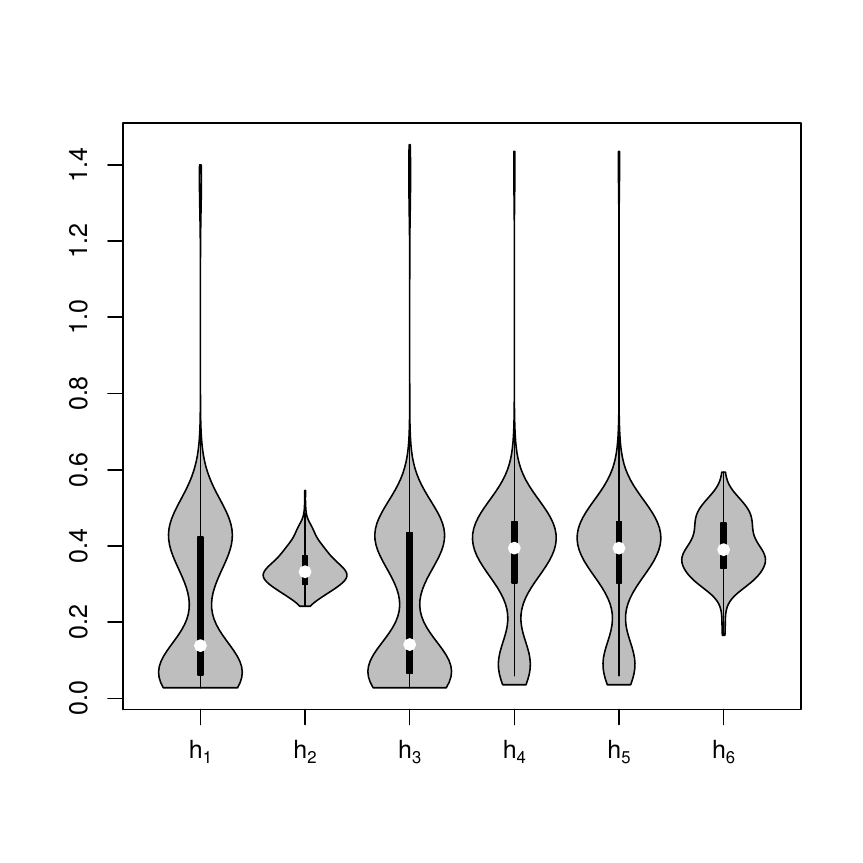}} 
			\put(340,-110){\includegraphics[scale=0.35]{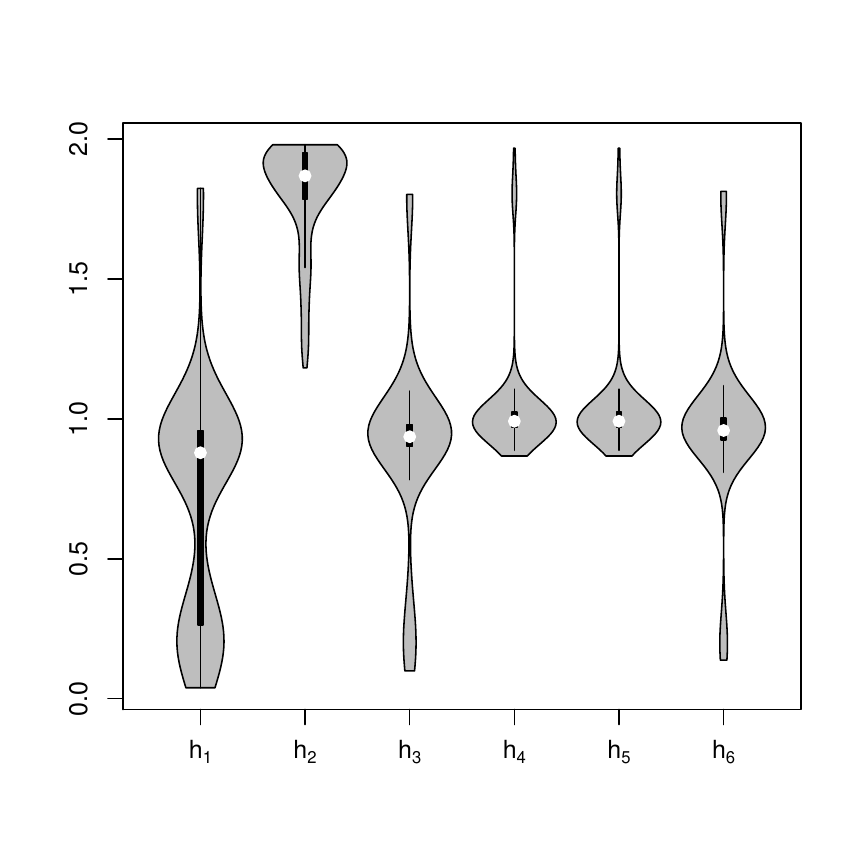}} 
			\end{picture} \vspace{3.2cm}\\
			\caption{Violin plots of Hausdorff errors for models C3, C8 and C9 when $\tau=0.8$ and $n=1000$.}
			\label{vioplot_cir_3}
		\end{figure}
		\restoregeometry
		
	\end{landscape}
	
	Figures \ref{vioplot_cir_1}, \ref{vioplot_cir_2} and \ref{vioplot_cir_3} show the violin plots of Hausdorff errors obtained for some of the simulation models when $\tau=0.2$, $\tau=0.5$ and $\tau=0.8$ ($n=1000$), respectively. According to Figure \ref{vioplot_cir_1}, $h_2$ is the selector that presents a worst behavior for the represented circular densities. If $\tau=0.5$, Figure \ref{vioplot_cir_2} shows that the same occurs for models C3 and C6. Furthermore, its variance is again specially large for model C3. Figure \ref{vioplot_cir_3} for models C3, C8 and C9 shows that the variance of the Hausdorff errors for $h_1$ can be large although this selector provides competitive mean errors.

	\subsection{Directional level set estimation}
	\label{sphericalsimus}
	As for the spherical scenario, 9 density models have been considered. These models, namely S1 to S9, are mixtures of von Mises-Fisher densities on the sphere, allowing to represent complex structures such as multimodality and/or asymetry. Parameters of mixtures are fully established in Table \ref{sphetable} in Appendix \ref{AppendixC}. Figure \ref{sphericalmodels} shows these densities and the corresponding HDRs for $\tau=0.2$, $\tau=0.5$ and $\tau=0.8$.

	\begin{figure}[h!]  
		\vspace{-7.8cm}
		$ \hspace{1.3cm}$\begin{picture}(0,0)
		\put(-25,-350){\includegraphics[scale=.5]{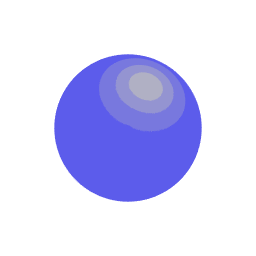}}
		\put(100,-350){\includegraphics[scale=.5]{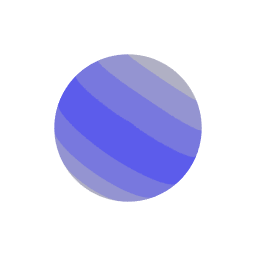}}
		\put(225,-350){\includegraphics[scale=.5]{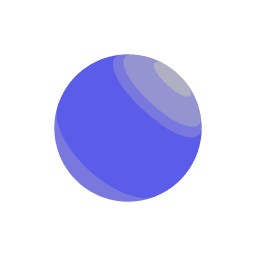}}

		\put(-25,-453){\includegraphics[scale=.5]{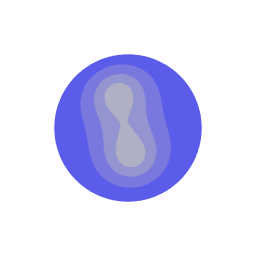}}
		\put(100,-453){\includegraphics[scale=.5]{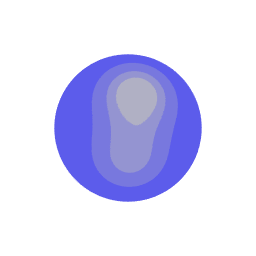}}
		\put(225,-453){\includegraphics[scale=.5]{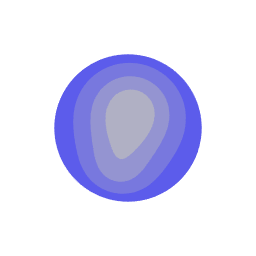}}

		\put(-25,-556){\includegraphics[scale=.5]{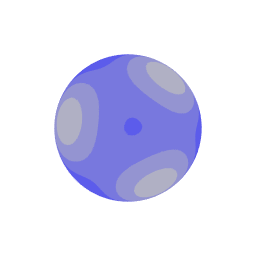}}
		\put(100,-556){\includegraphics[scale=.5]{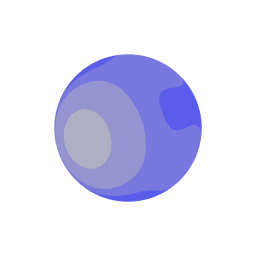}}
		\put(225,-556){\includegraphics[scale=.5]{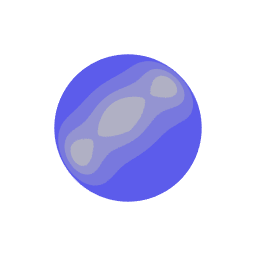}}
		\put(32,-235){ \textbf{S1}} 
		\put(160,-235){ \textbf{S2}} 	 
		\put(285,-235){ \textbf{S3}} 	 
		\put(32,-342){ \textbf{S4}} 
		\put(160,-342){ \textbf{S5}}  	 
		\put(285,-342){ \textbf{S6}}  	   
		\put(32,-445){ \textbf{S7}} 
		\put(160,-445){ \textbf{S8}}  	   
		\put(285,-445){ \textbf{S9}}  	
		\end{picture}\vspace{18.75cm}
		\caption{Finite mixtures of von Mises-Fisher spherical models for simulations. HDRs are represented for $\tau=0.2$, $\tau=0.5$ and $\tau=0.8$.}\label{sphericalmodels}
	\end{figure}
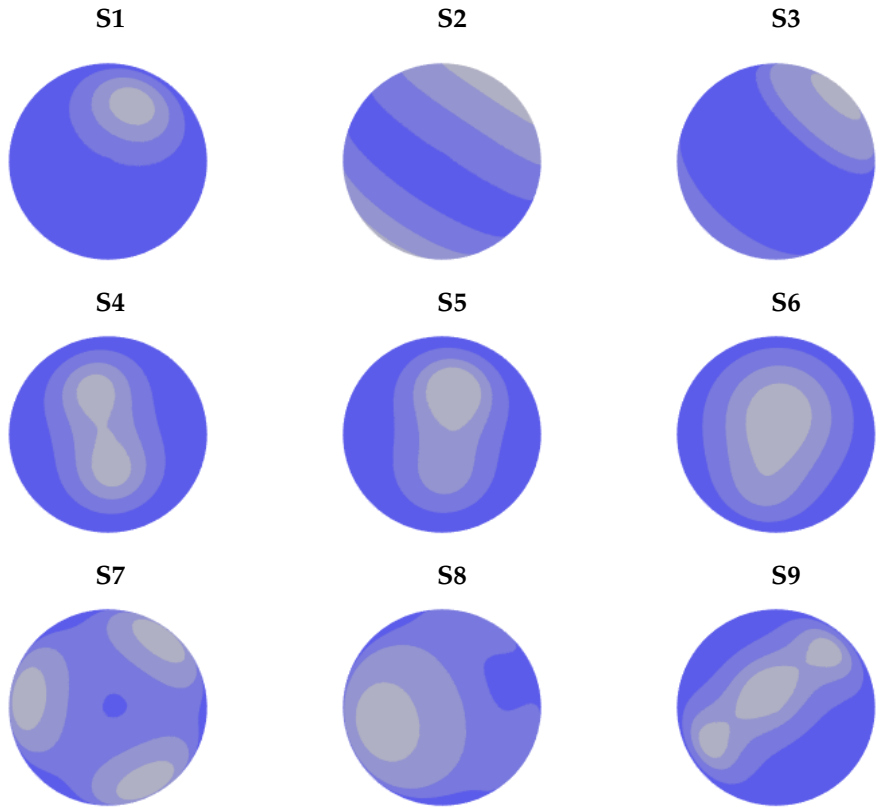
	
	For sample sizes $n=500$, $n=1500$ and $n=2500$, $200$ random samples were generated from models S1 to S9. From each sample, HDRs are reconstructed for $\tau=0.2$, $\tau=0.5$ and $\tau=0.8$. The performance of different plug-in methods that emerge from the consideration of different bandwidth parameters discussed in this work is checked. In this case, a total of $B=50$ resamples are established for estimating the proposed bootstrap bandwidth $h_1$, taking $h_5$ as a pilot bandwidth. 
	
	For each method and each sample, the estimation error is again measured calculating the Hausdorff distance between the boundaries of estimated level set and the frontier of theoretical set. As reference, note that the maximum value of both criteria $S^2$ is also $2$. In this case, this upper bound coincides exactly with the length of the diameter of the sphere. 
	
	Tables \ref{smeanstau02n1500} and \ref{smeanstau02n2500} show the means and the standard deviations of the $200$ estimation errors obtained when $\tau=0.2$ from samples of sizes $n=1500$ and $n=2500$, respectively. Again, blue cells corresponds to the smallest mean errors obtained for each density. Except for model S8, $h_1$ is the best or shows a competitive performance. For $\tau=0.5$ (see Tables \ref{smeanstau05n500}, \ref{smeanstau05n1500} and \ref{smeanstau05n2500}, for sample sizes  $n=500$, $n=1500$ and $n=2500$, respectively), the proposed selector is the best or second best in all cases. In fact, $h_1$ and $h_5$ usually behave similarly and $h_7$ is the worst selector for S3.

	Tables \ref{smeanstau08n1500} and \ref{smeanstau08n2500} show the means and the standard deviations of the $200$ estimation errors obtained when $\tau=0.8$ from samples of size $n=1500$ and $n=2500$, respectively.  Although results for S7 are not good when $h_1$ is considered, this selector is again the best or competitive with $h_5$. As for $h_7$, results are remarkably poor in S2 and S6.

	Figure \ref{vioplot_s_1} contains the violin plots of Hausdorff errors for models S3, S4, S6 and S9 when $\tau=0.2$ and $n=2500$. Note that he performance of selector $h_1$ is considerably good. Figure \ref{vioplot_s_2} for $\tau=0.5$ and $n=1500$ shows that $h_1$ and $h_5$ usually present similar results, see densities S5 and S9. However, $h_1$ is clearly more competitive for models S1 and S8.
	
	\begin{landscape}
		\thispagestyle{empty}
		
		\begin{table}[h!]\centering
			\caption{Means (M) and standard deviations (SD) of $200$ errors in Hausdorff distance for $\tau=0.2$, $n=1500$ and $B=50$.}\label{smeanstau02n1500}
			$\hspace{-3.5cm}$\begin{tabular}{ccccccccccccccccccc}
				\hline
				&\multicolumn{2}{c}{\textbf{S1}}&\multicolumn{2}{c}{\textbf{S2}}&\multicolumn{2}{c}{\textbf{S3}}&\multicolumn{2}{c}{\textbf{S4}}&\multicolumn{2}{c}{\textbf{S5}}&\multicolumn{2}{c}{\textbf{S6}}&\multicolumn{2}{c}{\textbf{S7}}&\multicolumn{2}{c}{\textbf{S8}}&\multicolumn{2}{c}{\textbf{S9}}\\
				&M& SD&M& SD&M& SD&M& SD&M& SD&M& SD&M& SD&M& SD&M& SD\\
				\hline
				$h_1$&\cellcolor{GGG}0.033&0.010&\cellcolor{GGG}0.757&0.261&0.445&0.168&\cellcolor{GGG}0.065&0.014&\cellcolor{GGG}0.074&0.018&\cellcolor{GGG}0.075&0.019&0.748&0.241& 0.302& 0.118     &0.081&0.017\\
				$h_5$&0.052&0.012&0.764&0.238&0.598&0.145&0.072&0.018&0.078&0.020&0.080&0.022&\cellcolor{GGG}0.651&0.335&0.275&0.082&0.088&0.018\\
				$h_7$&0.063&0.013&1.224&0.243&\cellcolor{GGG}0.371&0.072&0.073&0.017&0.078&0.019&0.089&0.023&0.921&0.199&\cellcolor{GGG}0.263&0.087&\cellcolor{GGG}0.079&0.016\\
				
				\hline		
			\end{tabular}\vspace{-.2cm}
		\end{table}

		\begin{table}[h!]\centering
			\caption{Means (M) and standard deviations (SD) of $200$ errors in Hausdorff distance for $\tau=0.2$, $n=2500$ and $B=50$.}\label{smeanstau02n2500}
			$\hspace{-3.5cm}$\begin{tabular}{ccccccccccccccccccc}
				\hline
				&\multicolumn{2}{c}{\textbf{S1}}&\multicolumn{2}{c}{\textbf{S2}}&\multicolumn{2}{c}{\textbf{S3}}&\multicolumn{2}{c}{\textbf{S4}}&\multicolumn{2}{c}{\textbf{S5}}&\multicolumn{2}{c}{\textbf{S6}}&\multicolumn{2}{c}{\textbf{S7}}&\multicolumn{2}{c}{\textbf{S8}}&\multicolumn{2}{c}{\textbf{S9}}\\
				&M& SD&M& SD&M& SD&M& SD&M& SD&M& SD&M& SD&M& SD&M& SD\\
				\hline
				$h_1$&\cellcolor{GGG}0.029&0.009&0.650&0.212&0.340&0.089&\cellcolor{GGG}0.059&0.013&\cellcolor{GGG}0.061&0.014&\cellcolor{GGG}0.064&0.016&0.667&0.272&0.247&0.099&0.070&0.015\\
				$h_5$&0.044&0.010&\cellcolor{GGG}0.648&0.199&0.484&0.097&0.063&0.014&0.067&0.015&0.071&0.018&\cellcolor{GGG}0.556&0.336&0.230&0.073&0.075&0.016\\
				$h_7$&0.053&0.011&1.137&0.245&\cellcolor{GGG}0.306&0.053&0.062&0.014&0.067&0.016&0.079&0.017&0.889&0.229&\cellcolor{GGG}0.218&0.073&\cellcolor{GGG}0.069&0.014\\
				
				\hline		
			\end{tabular}
		\end{table}

		\begin{figure}[h!] 
			\begin{picture}(0,0)
			\put(10,-110){\includegraphics[scale=0.35]{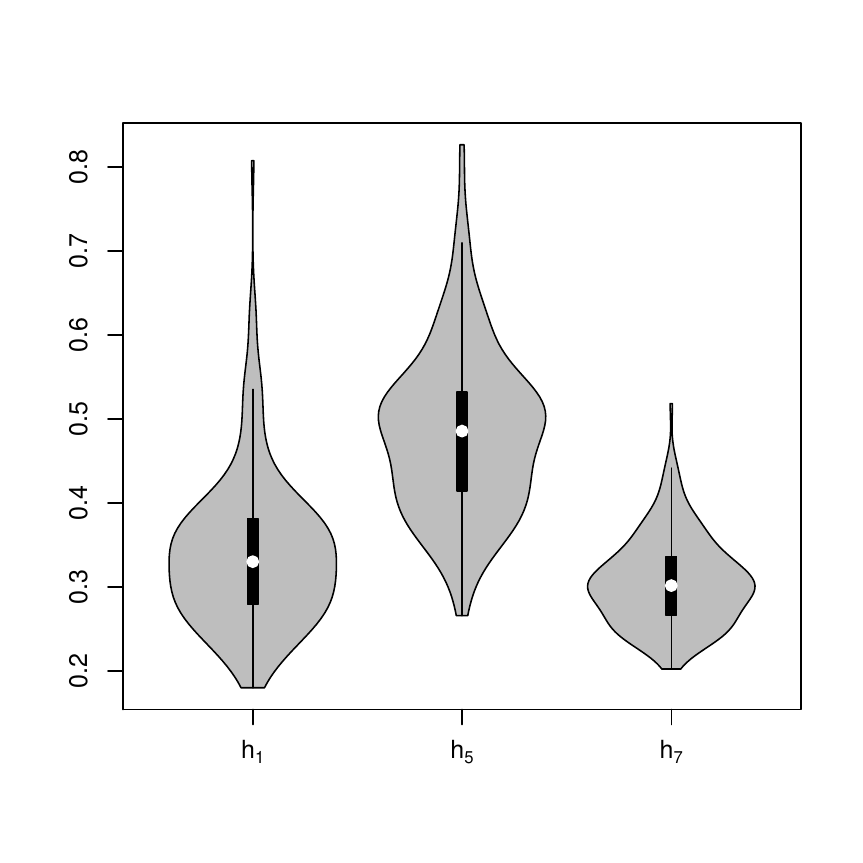}}
			\put(150,-110){\includegraphics[scale=0.35]{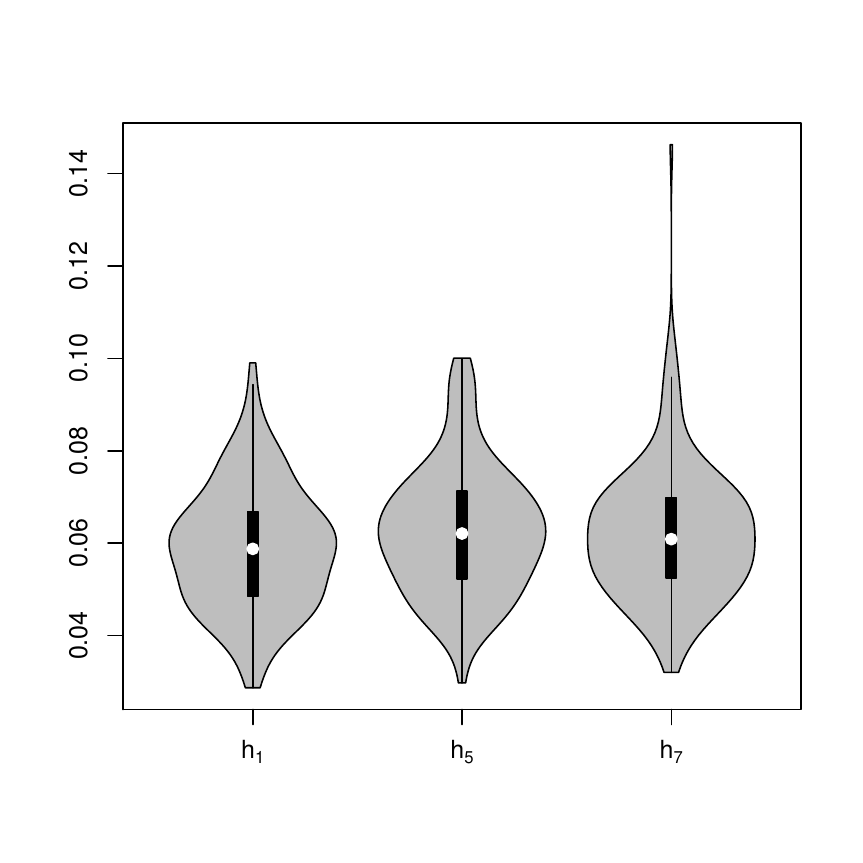}}
			\put(290,-110){\includegraphics[scale=0.35]{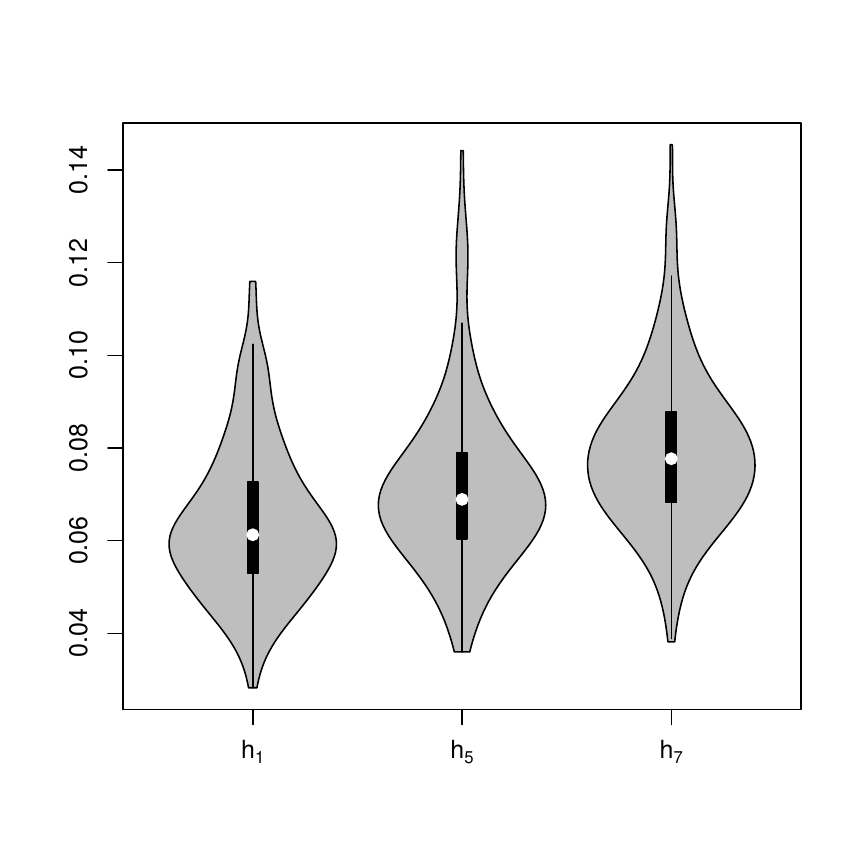}}
			\put(430,-110){\includegraphics[scale=0.35]{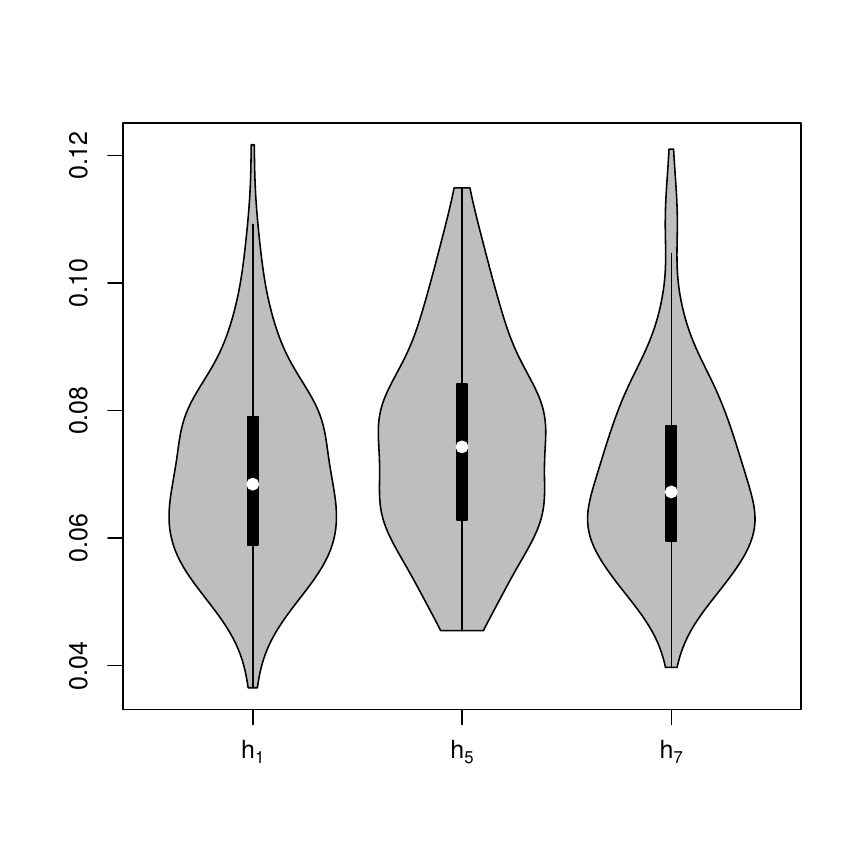}}
			\end{picture} \vspace{3.2cm}\\
			\caption{Violin plots of Hausdorff errors for models S3 and S4, S6 and S9  when $\tau=0.2$ and $n=2500$.}\label{vioplot_s_1}
		\end{figure}
		\restoregeometry

	\end{landscape}
	\newpage

	\begin{landscape}
		\thispagestyle{empty}

		\begin{table}[h!]\centering
			\caption{Means (M) and standard deviations (SD) of $200$ errors in Hausdorff distance for $\tau=0.5$, $n=500$ and $B=50$.}\label{smeanstau05n500}
			$\hspace{-3.5cm}$\begin{tabular}{ccccccccccccccccccc}
				\hline
				&\multicolumn{2}{c}{\textbf{S1}}&\multicolumn{2}{c}{\textbf{S2}}&\multicolumn{2}{c}{\textbf{S3}}&\multicolumn{2}{c}{\textbf{S4}}&\multicolumn{2}{c}{\textbf{S5}}&\multicolumn{2}{c}{\textbf{S6}}&\multicolumn{2}{c}{\textbf{S7}}&\multicolumn{2}{c}{\textbf{S8}}&\multicolumn{2}{c}{\textbf{S9}}\\			&M& SD&M& SD&M& SD&M& SD&M& SD&M& SD&M& SD&M& SD&M& SD\\
				\hline
				$h_1$&\cellcolor{GGG}0.044&0.018&0.843&0.249&0.924&0.567&\cellcolor{GGG}0.113&0.033&0.131&0.048&0.136&0.038&0.307&0.074&\cellcolor{GGG}0.140&0.070&0.172&0.061\\
				$h_5$&0.069&0.020&\cellcolor{GGG}0.796&0.245&\cellcolor{GGG}0.888&0.552&0.118&0.035&0.130&0.047&\cellcolor{GGG}0.135&0.043&\cellcolor{GGG}0.292&0.066&0.220&0.212&0.174&0.056\\
				$h_7$&0.082&0.022&0.880&0.181&1.497&0.514&0.115&0.031&\cellcolor{GGG}0.127&0.045&0.145&0.045&0.313&0.074&0.147&0.091&\cellcolor{GGG}0.149&0.049		\\	
				\hline		
				
			\end{tabular}
		\end{table}

		\begin{table}[h!]\centering
			\caption{Means (M) and standard deviations (SD) of $200$ errors in Hausdorff distance for $\tau=0.5$, $n=1500$ and $B=50$.}\label{smeanstau05n1500}
			$\hspace{-3.5cm}$\begin{tabular}{ccccccccccccccccccc}
				\hline
				&\multicolumn{2}{c}{\textbf{S1}}&\multicolumn{2}{c}{\textbf{S2}}&\multicolumn{2}{c}{\textbf{S3}}&\multicolumn{2}{c}{\textbf{S4}}&\multicolumn{2}{c}{\textbf{S5}}&\multicolumn{2}{c}{\textbf{S6}}&\multicolumn{2}{c}{\textbf{S7}}&\multicolumn{2}{c}{\textbf{S8}}&\multicolumn{2}{c}{\textbf{S9}}\\		&M& SD&M& SD&M& SD&M& SD&M& SD&M& SD&M& SD&M& SD&M& SD\\
				\hline
				$h_1$&\cellcolor{GGG}0.032&0.008&0.568&0.155&0.779&0.590&\cellcolor{GGG}0.077&0.024&\cellcolor{GGG}0.092&0.030&0.093&0.027&0.223&0.052&0.095&0.046&0.103&0.032\\
				$h_5$&0.048&0.013&\cellcolor{GGG}0.536&0.129&\cellcolor{GGG}0.591&0.392&0.080&0.021&\cellcolor{GGG}0.092&0.030&\cellcolor{GGG}0.086&0.023&\cellcolor{GGG}0.218&0.052&0.125&0.117&0.111&0.032\\
				$h_7$&0.057&0.014&0.648&0.144&1.473&0.505&0.079&0.020&\cellcolor{GGG}0.092&0.029&0.093&0.023&0.223&0.055&\cellcolor{GGG}0.093&0.027&\cellcolor{GGG}0.098&0.024\\
				\hline		
				
			\end{tabular}
		\end{table}

		\begin{figure}[h!] 
			\begin{picture}(0,0)
			\put(10,-110){\includegraphics[scale=0.35]{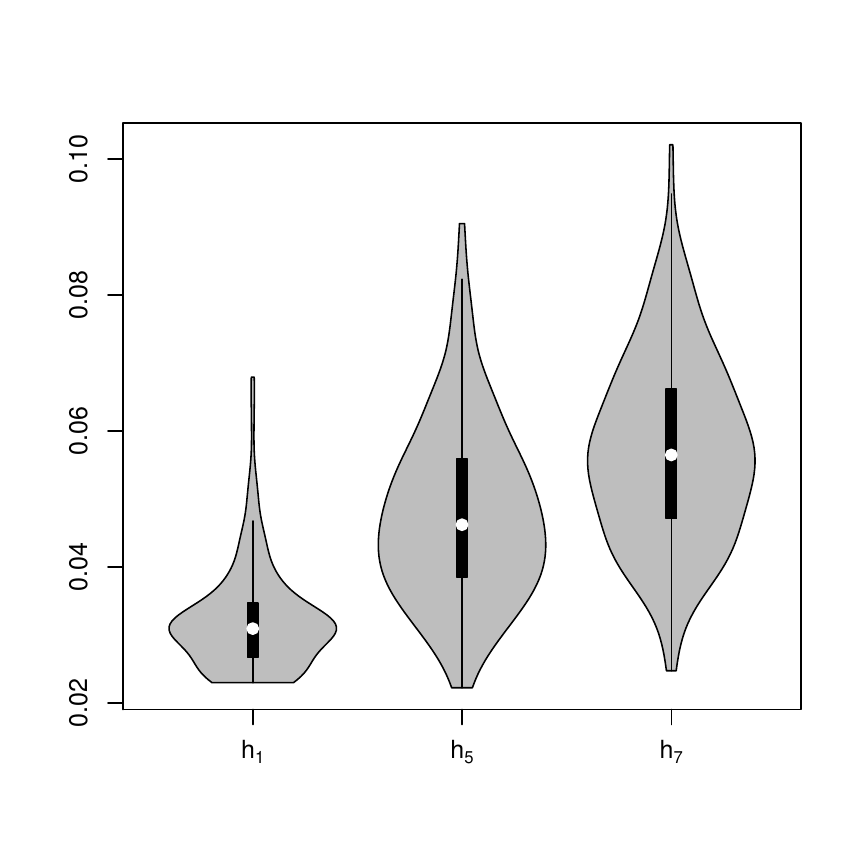}}
			\put(150,-110){\includegraphics[scale=0.35]{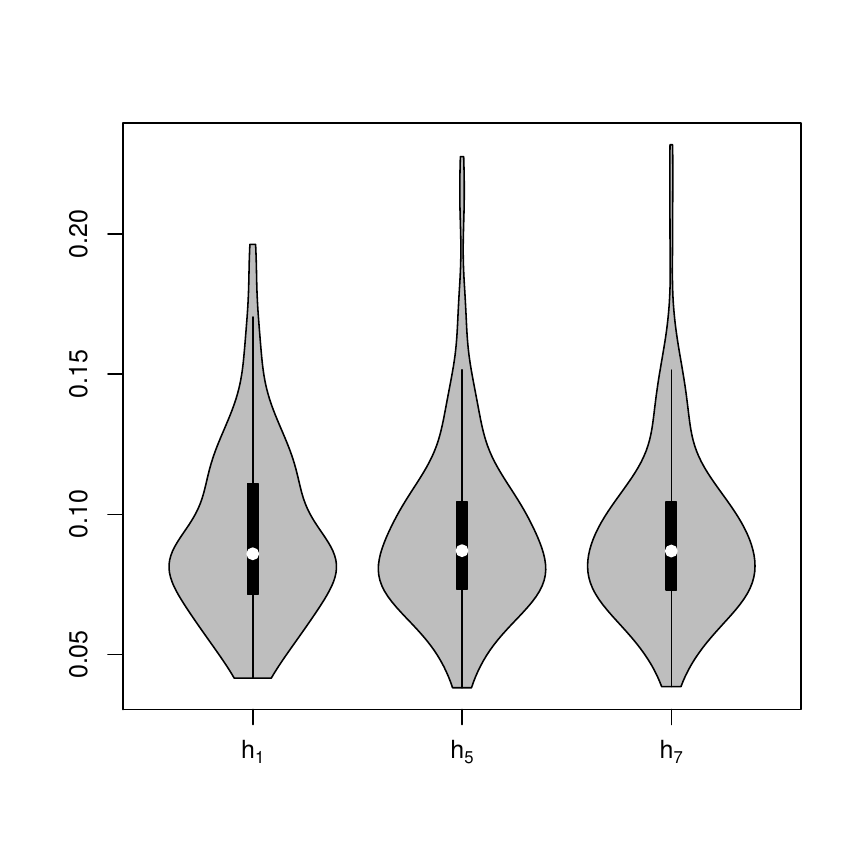}}
			\put(290,-110){\includegraphics[scale=0.35]{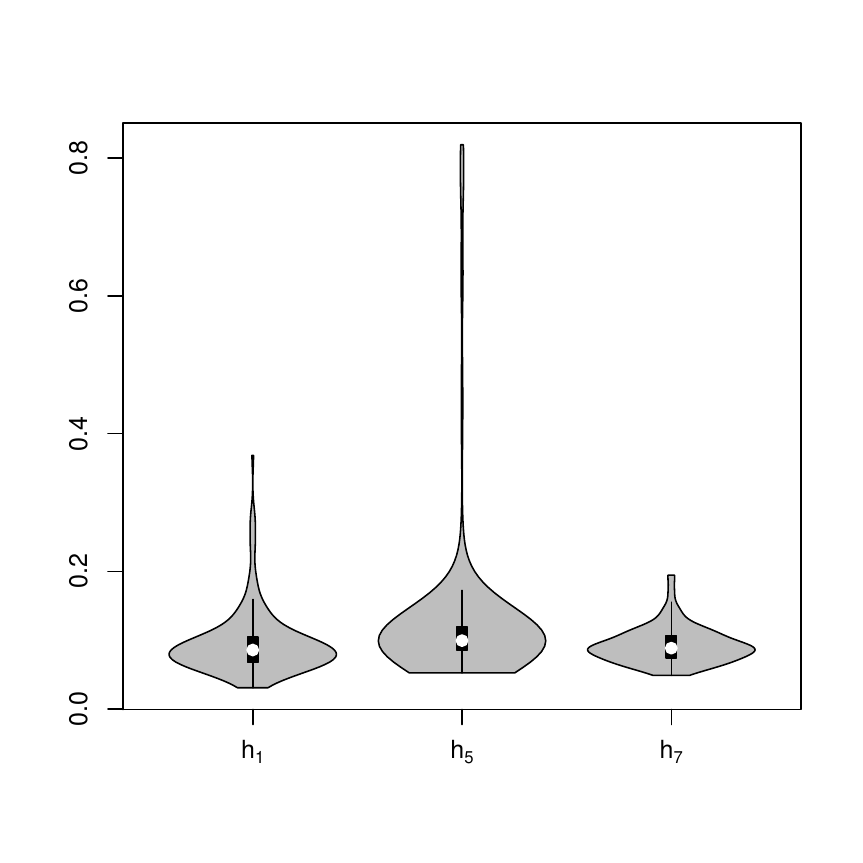}}
			\put(430,-110){\includegraphics[scale=0.35]{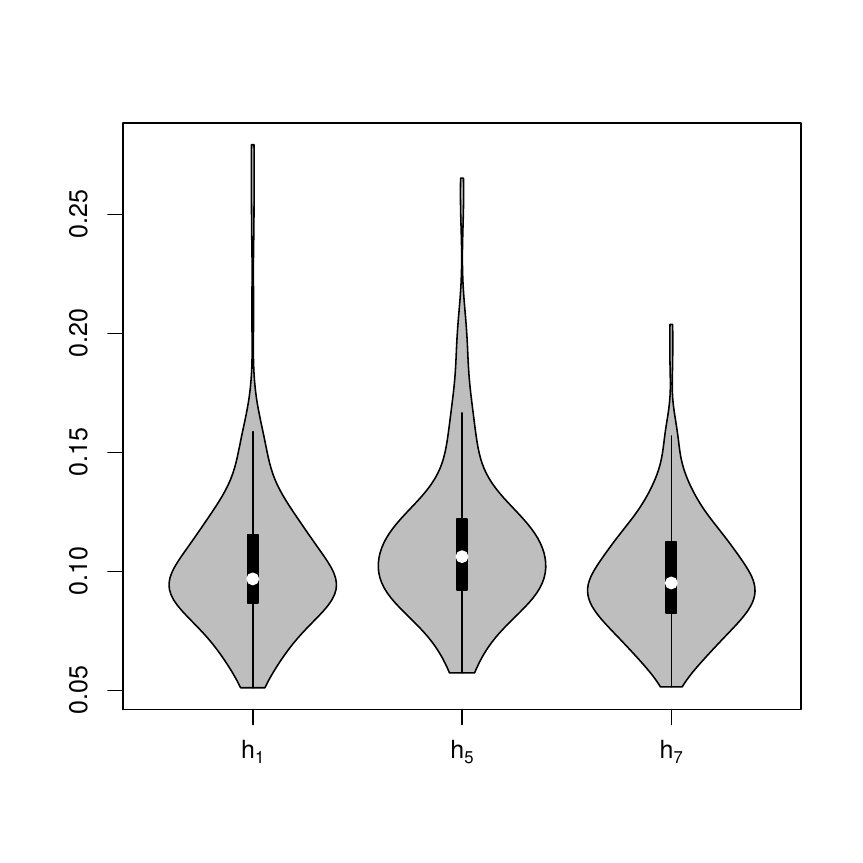}}
			\end{picture} \vspace{3.2cm}\\
			\caption{Violin plots of Hausdorff errors for models S1, S5, S8 and S9  when $\tau=0.5$ and $n=1500$.}\label{vioplot_s_2}
		\end{figure}
		\restoregeometry

	\end{landscape}
	\newpage

	\begin{landscape}
		\thispagestyle{empty}

		\begin{table}[h!]\centering
			\caption{Means (M) and standard deviations (SD) of $200$ errors in Hausdorff distance for $\tau=0.5$, $n=2500$ and $B=50$.}\label{smeanstau05n2500}
			$\hspace{-3.5cm}$\begin{tabular}{ccccccccccccccccccc}
				\hline
				&\multicolumn{2}{c}{\textbf{S1}}&\multicolumn{2}{c}{\textbf{S2}}&\multicolumn{2}{c}{\textbf{S3}}&\multicolumn{2}{c}{\textbf{S4}}&\multicolumn{2}{c}{\textbf{S5}}&\multicolumn{2}{c}{\textbf{S6}}&\multicolumn{2}{c}{\textbf{S7}}&\multicolumn{2}{c}{\textbf{S8}}&\multicolumn{2}{c}{\textbf{S9}}\\		&M& SD&M& SD&M& SD&M& SD&M& SD&M& SD&M& SD&M& SD&M& SD\\
				\hline
				$h_1$&\cellcolor{GGG}0.026&0.005&\cellcolor{GGG}0.437&0.131&0.760&0.595&\cellcolor{GGG}0.064&0.019&\cellcolor{GGG}0.074&0.022&0.088&0.027&0.181&0.052&\cellcolor{GGG}0.076&0.048&0.085&0.028\\
				$h_5$&0.042&0.009&0.458&0.123&\cellcolor{GGG}0.458&0.243&0.066&0.017&0.076&0.024&\cellcolor{GGG}0.076&0.023&\cellcolor{GGG}0.178&0.048&0.113&0.136&0.088&0.025\\
				$h_7$&0.050&0.012&0.523&0.124&1.495&0.508&0.066&0.017&0.076&0.024&0.082&0.023&0.180&0.049&0.081&0.053&\cellcolor{GGG}0.081&0.020\\
				
				\hline		
				
			\end{tabular}
		\end{table}

		\begin{table}[h!]\centering
			\caption{Means (M) and standard deviations (SD) of $200$ errors in Hausdorff distance for $\tau=0.8$, $n=1500$ and $B=50$.}\label{smeanstau08n1500}
			$\hspace{-3.5cm}$\begin{tabular}{ccccccccccccccccccc}
				\hline
				&\multicolumn{2}{c}{\textbf{S1}}&\multicolumn{2}{c}{\textbf{S2}}&\multicolumn{2}{c}{\textbf{S3}}&\multicolumn{2}{c}{\textbf{S4}}&\multicolumn{2}{c}{\textbf{S5}}&\multicolumn{2}{c}{\textbf{S6}}&\multicolumn{2}{c}{\textbf{S7}}&\multicolumn{2}{c}{\textbf{S8}}&\multicolumn{2}{c}{\textbf{S9}}\\
				&M& SD&M& SD&M& SD&M& SD&M& SD&M& SD&M& SD&M& SD&M& SD\\
				\hline
				$h_1$&\cellcolor{GGG}0.030&0.012&0.629&0.359&\cellcolor{GGG}0.040&0.014&0.138&0.042&\cellcolor{GGG}0.132&0.078&\cellcolor{GGG}0.089&0.023&0.462&0.231&\cellcolor{GGG}0.063&0.021&0.243&\cellcolor{GGG}0.109\\
				$h_5$&0.054&0.018&\cellcolor{GGG}0.537&0.295&0.053&0.018&\cellcolor{GGG}0.134&0.041&0.154&0.101&0.110&0.033&0.212&0.143&0.083&0.025&\cellcolor{GGG}0.233&0.123\\
				$h_7$&0.068&0.023&0.915&0.533&\cellcolor{GGG}0.040&0.012&\cellcolor{GGG}0.134&0.041&0.155&0.102&0.128&0.039&\cellcolor{GGG}0.195&0.160&0.073&0.022&\cellcolor{GGG}0.233&0.124\\
				
				\hline		
			\end{tabular}
		\end{table}

		\begin{table}[h!]\centering
			\caption{Means (M) and standard deviations (SD) of $200$ errors in Hausdorff distance for $\tau=0.8$, $n=2500$ and $B=50$.}\label{smeanstau08n2500}
			$\hspace{-3.5cm}$\begin{tabular}{ccccccccccccccccccc}
				\hline
				&\multicolumn{2}{c}{\textbf{S1}}&\multicolumn{2}{c}{\textbf{S2}}&\multicolumn{2}{c}{\textbf{S3}}&\multicolumn{2}{c}{\textbf{S4}}&\multicolumn{2}{c}{\textbf{S5}}&\multicolumn{2}{c}{\textbf{S6}}&\multicolumn{2}{c}{\textbf{S7}}&\multicolumn{2}{c}{\textbf{S8}}&\multicolumn{2}{c}{\textbf{S9}}\\			&M& SD&M& SD&M& SD&M& SD&M& SD&M& SD&M& SD&M& SD&M& SD\\
				\hline
				$h_1$&\cellcolor{GGG}0.023&0.008&0.403&0.171&0.031&0.008&0.122&0.033&\cellcolor{GGG}0.115&0.073&\cellcolor{GGG}0.081&0.023&0.234&0.200&\cellcolor{GGG}0.051&0.016&0.201&0.086\\
				$h_5$&0.043&0.013&\cellcolor{GGG}0.399&0.128&0.047&0.011&\cellcolor{GGG}0.117&0.033&0.133&0.097&0.096&0.032&0.166&0.056&0.066&0.018&\cellcolor{GGG}0.186&0.090\\
				$h_7$&0.054&0.017&0.670&0.475&\cellcolor{GGG}0.030&0.009&\cellcolor{GGG}0.117&0.032&0.135&0.098&0.112&0.037&\cellcolor{GGG}0.140&0.047&0.059&0.017&0.194&0.110\\
				\hline		
			\end{tabular}
		\end{table}

	\end{landscape}
	
	\newpage

	\section{Real data analysis}
	\label{sec:real_data}
	The proposed methodology is now applied to the two real datasets presented in the Introduction exemplifying the aplicability of the method for circular and spherical data.
	
	\subsection{Behavioral plasticity of sandhoppers}
	\label{sec:sandhoppers}
	Adaptation to changing beach environments for the real example on sandhoppers introduced in Section \ref{intro:data} is analyzed from density level set estimation perspective.
	
	HDRs are estimated for $\tau=0.8$ disaggregating the sandhoppers data taking into account the categories of variables specie, sex, time of day and month of year. As consequence, a total of 24 set estimators are determined, numbered E1 to E24. Variables combinations yielding this group classification are presented in Table \ref{tab:multicol} in Appendix \ref{AppendixB}.
	
	Note that the estimated HDRs correspond to the largest modes of the orientation distributions. Distances between these 24 sets are able to establish the degree of dissimilarity of HDRs. Apart from $d_H$, the Euclidean distance $d_E$ between $A$ and $B$ defined as
	$$d_E(A,B)=\inf\left\{d_E(x,y),\mbox{ }x\in A,\mbox{ }y\in B\right\}$$is also used. Although $d_E(A,B)=0$ does not imply that $A=B$, this distance is useful to determine large differences between orientations. In general, large distances between the boundaries of two sets indicate the existence of modes in different directions. If the categories of all variables with the exception of one are fixed, it is possible to check if the different values of the non-fixing variable has some influence in sandhoppers orientation through the comparison of the estimated level sets. As a reference, note that the maximum value of Hausdorff and Euclidean distances between two points in $S^1$ is $2$. The upper triangular matrix in Table \ref{24tab} contains the Hausdorff distances between boundaries and the lower triangular matrix, the Euclidean ones. The largest distances are represented in blue color for both criteria. Grey color is used in order to depict the next largest values. Furthermore, Table \ref{24tab} (top) contains some of the estimated HDRs that present the largest distances.  
	
	In particular, Hausdorff distance between regions 5 and 11 is equal to $1.91$. According to Table \ref{names}, the variable configuration 5 corresponds to the largest orientation modes for females of the specie Talitrus saltator when the orientation is measure in noon during October. Region 11 refers to same measurements taken in April. Therefore, the month can be seen as variable that has influence on the orientation for sandhoppers. 
	
	Euclidean distance between regions 5 and 6 is equal to $1.57$. According to Table \ref{names}, set 6 also corresponds to the level set for females of the specie Talitrus saltator but, in this case, when the orientation is registered in morning during October. Then, the moment of the day also seems a factor with influence on the sandhoppers behavior.

	\begin{landscape}
		\thispagestyle{empty}
		
		\begin{table}[h!]\vspace{-2cm}
			$\hspace{-14.1cm}$\caption{Upper triangular matrix contains the Hausdorff distances between boundaries of sets from $1$ to $24$. Lower triangular matrix of Euclidean distances between boundaries of sets from $1$ to $24$ (bottom). HDRs representations for $\tau=0.8$ for some sets between 1 and 24 (top).}\label{24tab}\vspace{-.05cm}
			$\hspace{.5cm}$\begin{picture}(10,40)
			\put(-100,-110){\includegraphics[scale=.42]{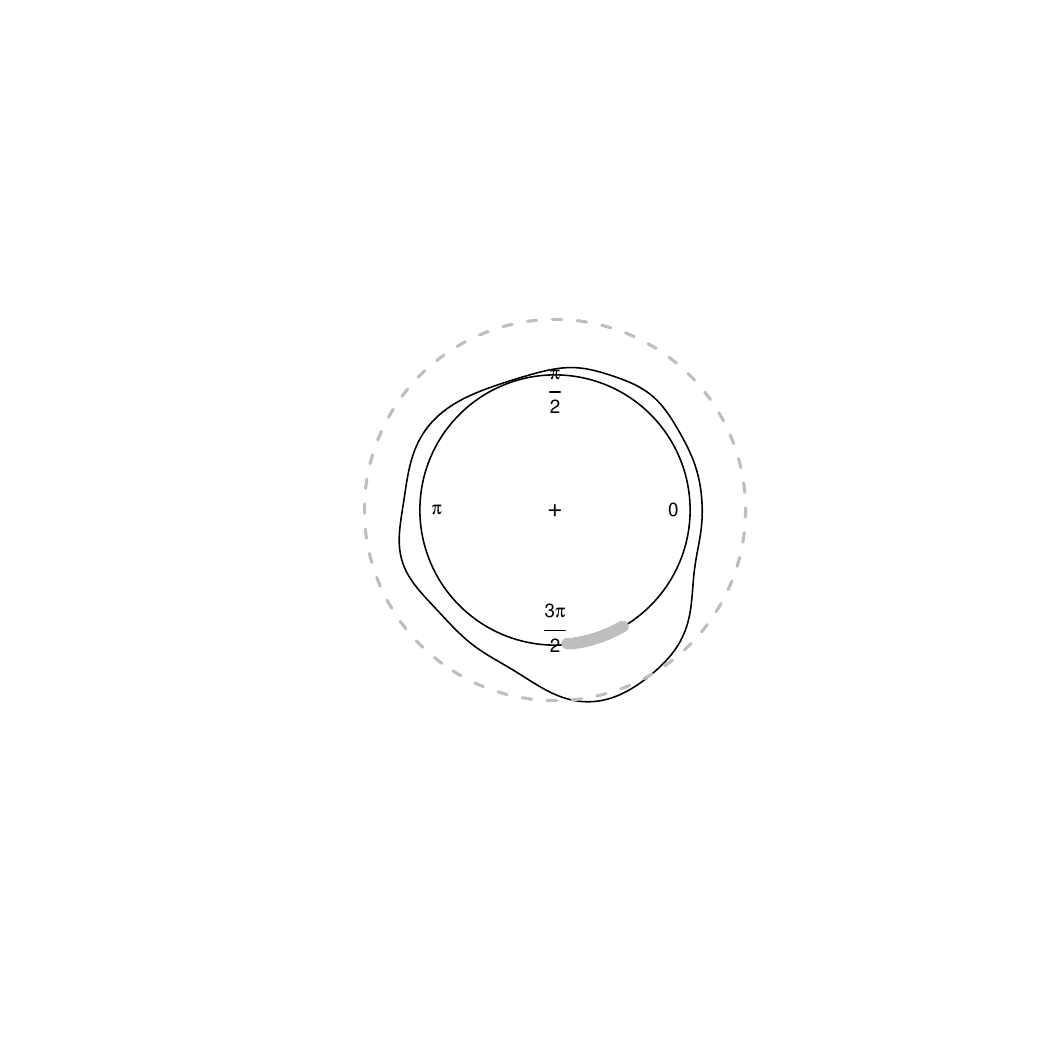}} 
			\put(0,-110){\includegraphics[scale=.42]{saltOctobernoonM.pdf}}
			\put(100,-110){\includegraphics[scale=.42]{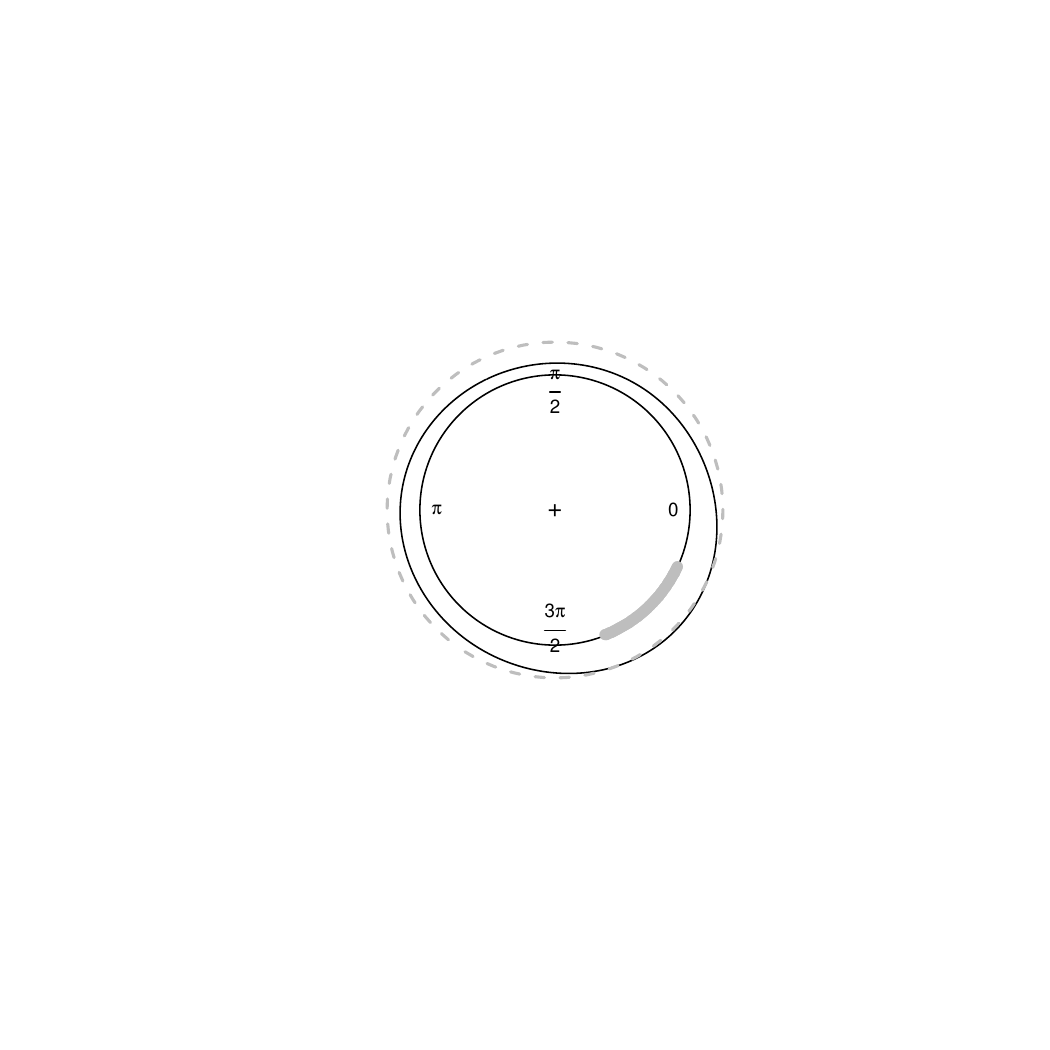}}
			\put(195,-110){\includegraphics[scale=0.42]{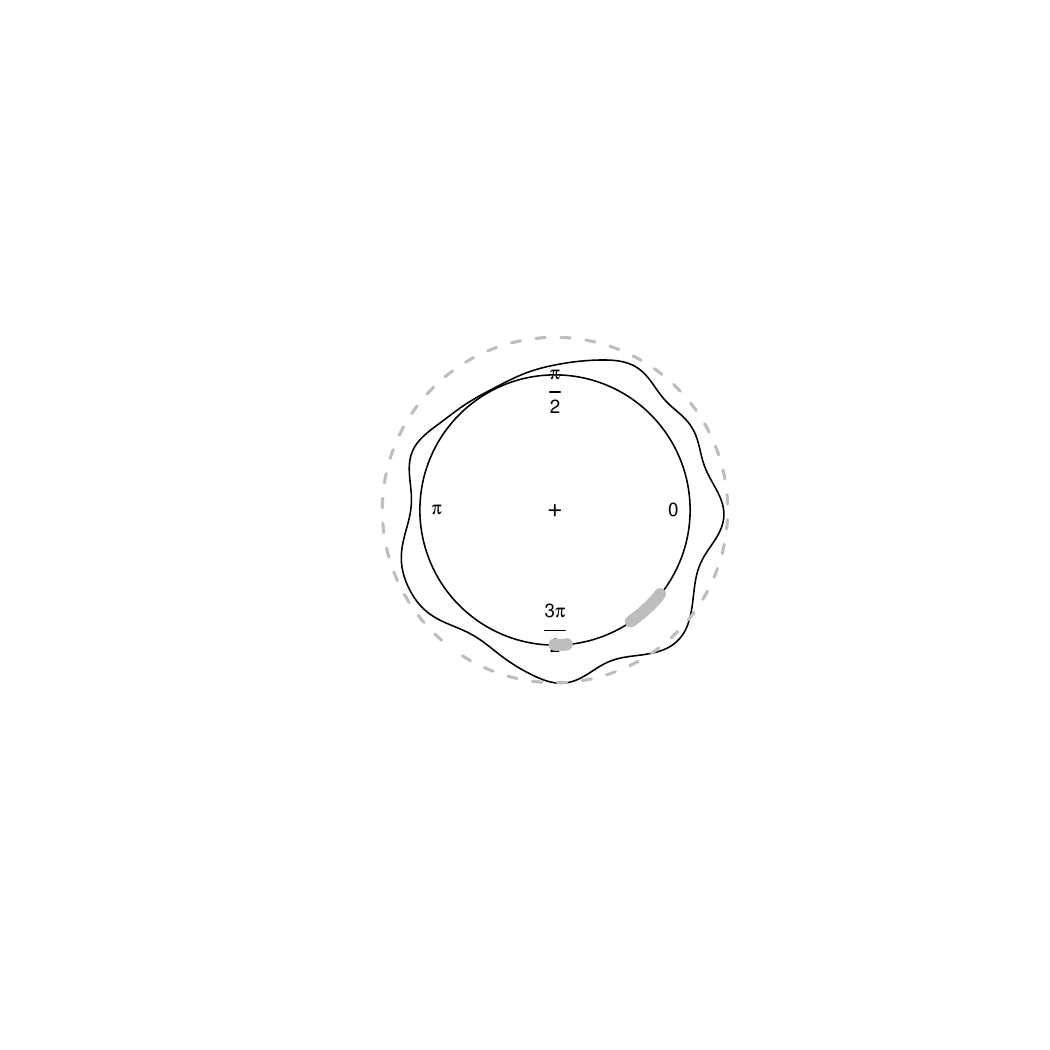}}
			\put(290,-110){\includegraphics[scale=.42]{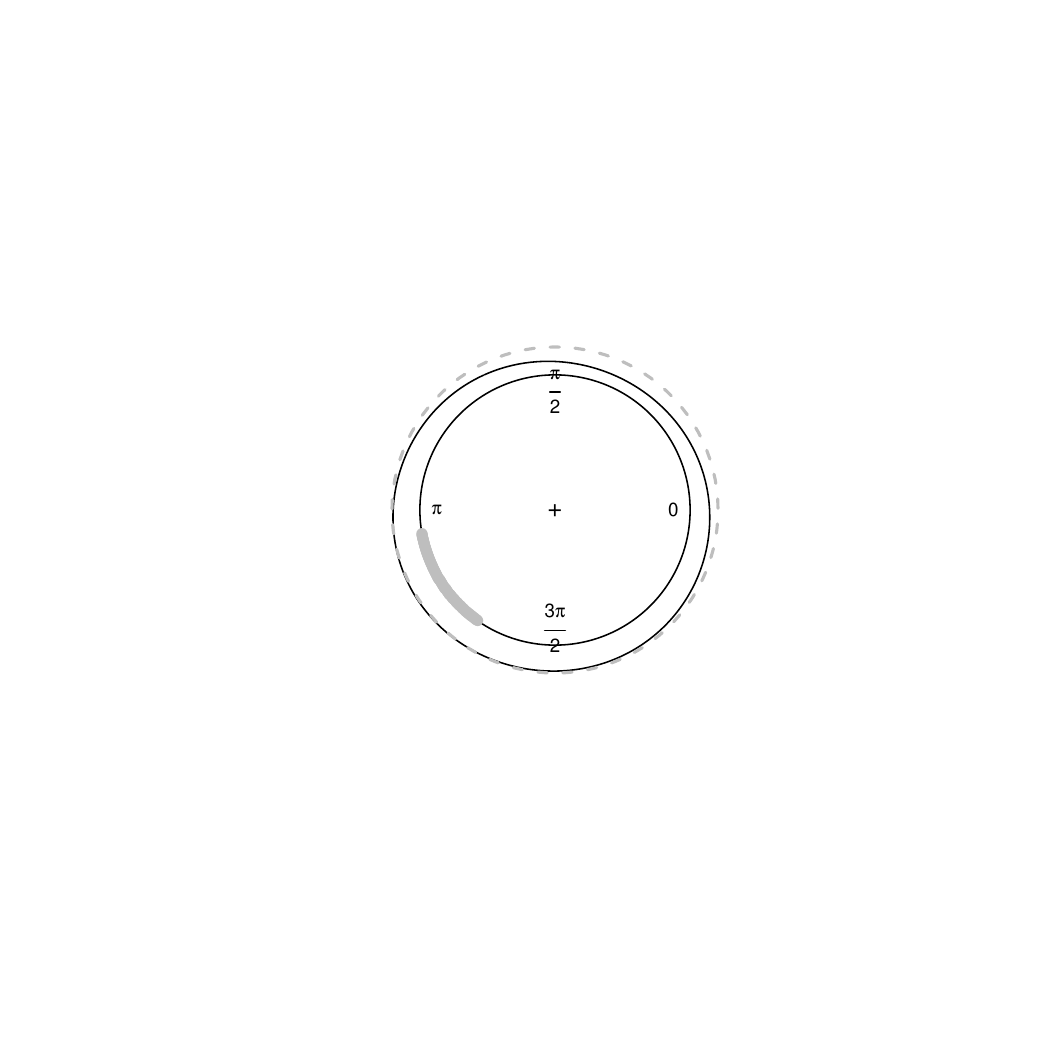}}
			\put(400,-110){\includegraphics[scale=0.42]{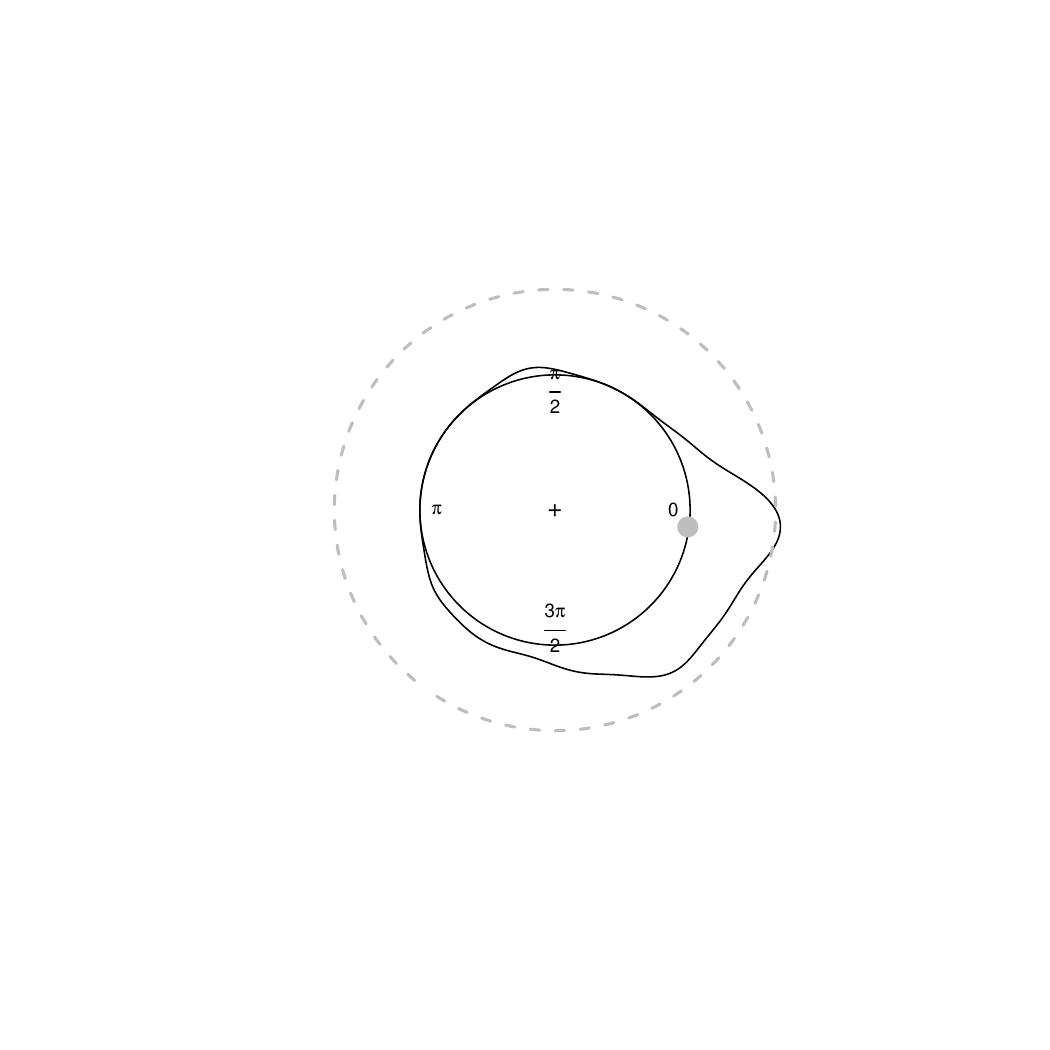}}
			\put(258,-20){\textbf{4}}
			\put(460,-20){\textbf{7}}

			\put(350,-20){\textbf{5}}
			\put(158,-20){\textbf{3}}
			\put(60,-20){\textbf{2}}
			\put(-40,-20){\textbf{1}}

			\put(-100,-190){\includegraphics[scale=0.42]{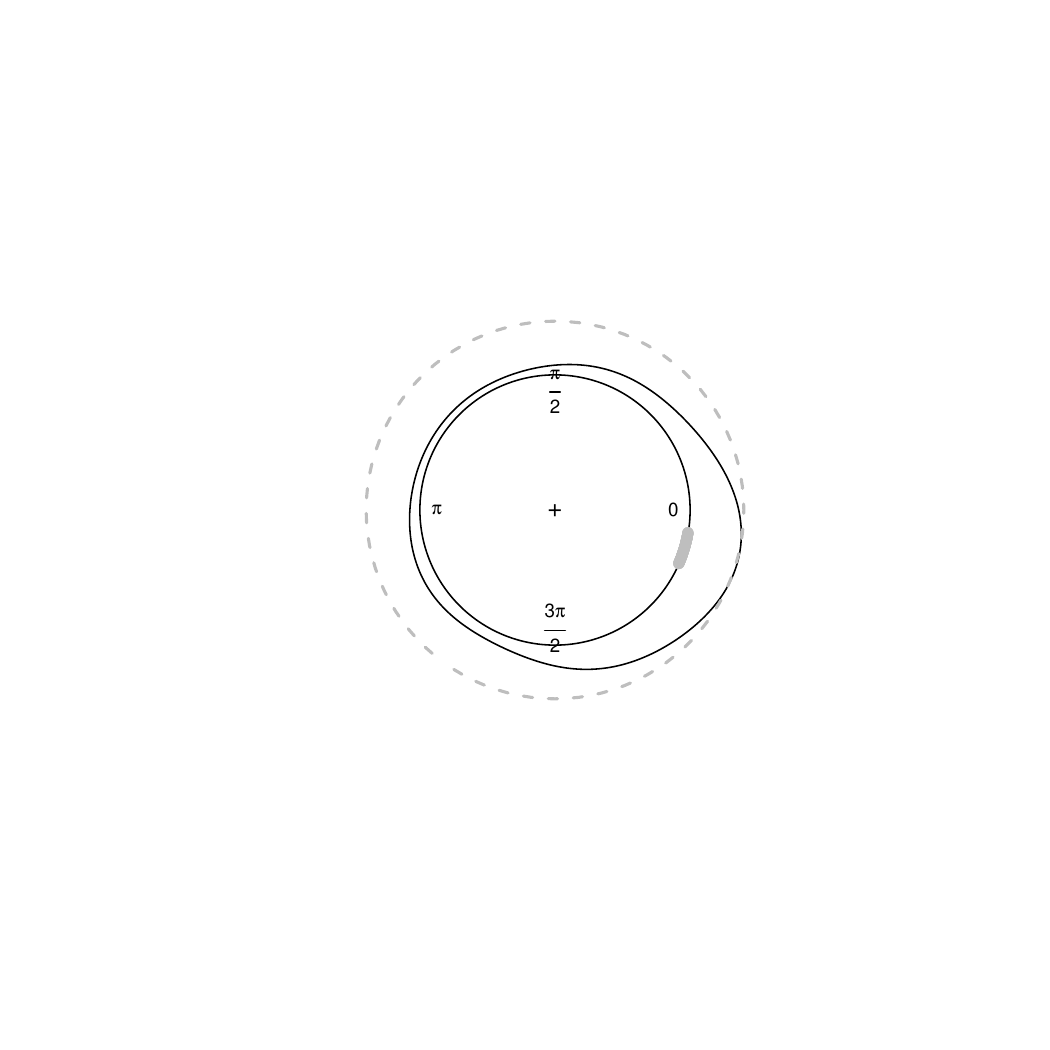}}
			\put(0,-190){\includegraphics[scale=0.42]{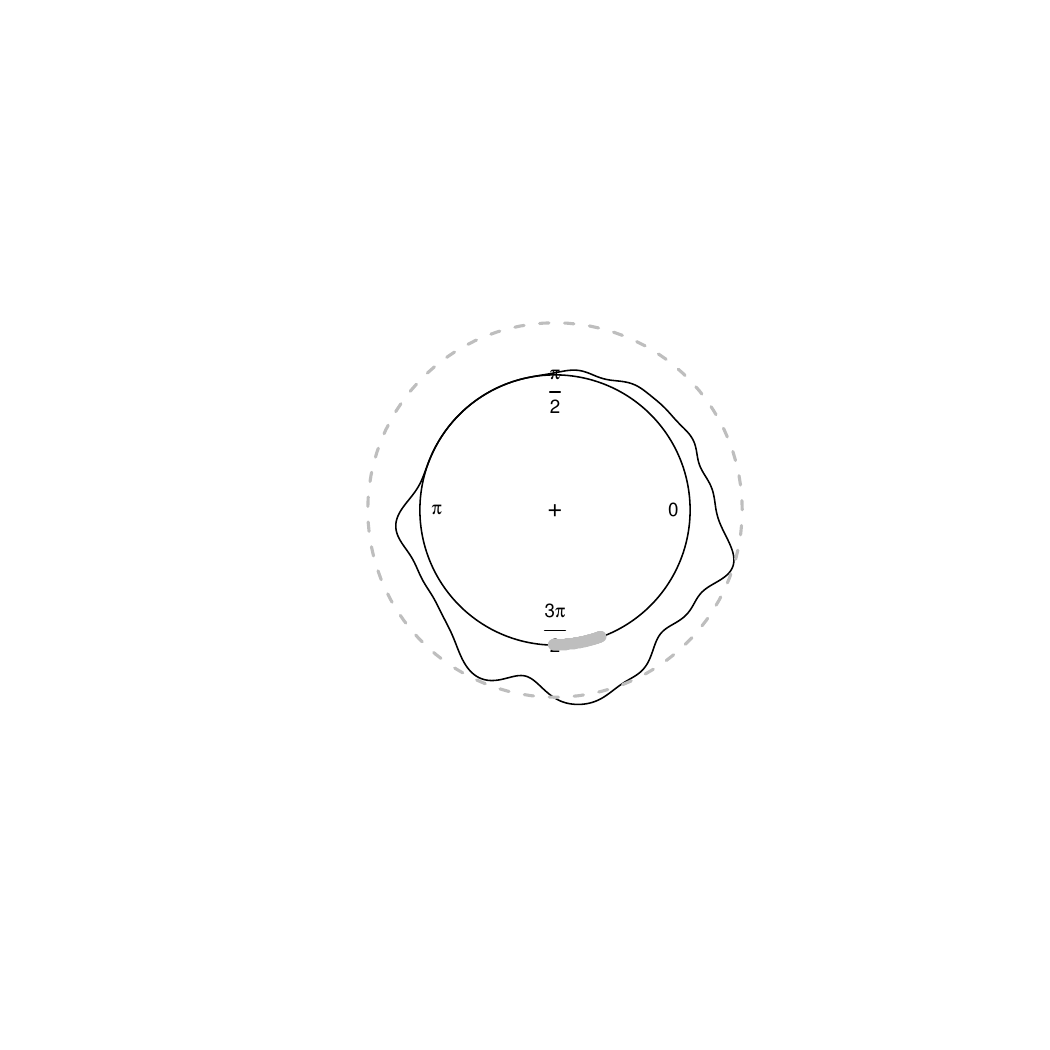}}
			\put(100,-190){\includegraphics[scale=0.42]{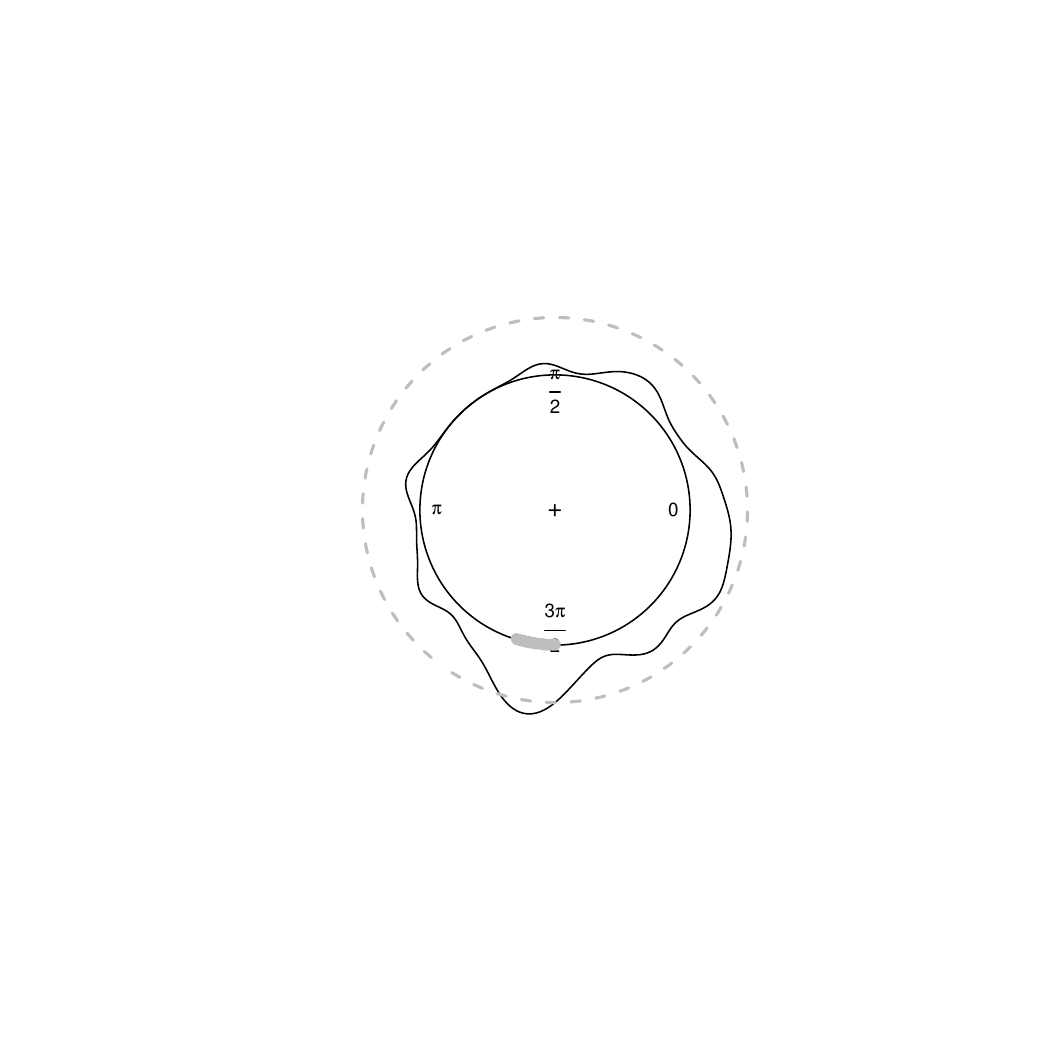}}
			\put(195,-190){\includegraphics[scale=0.42]{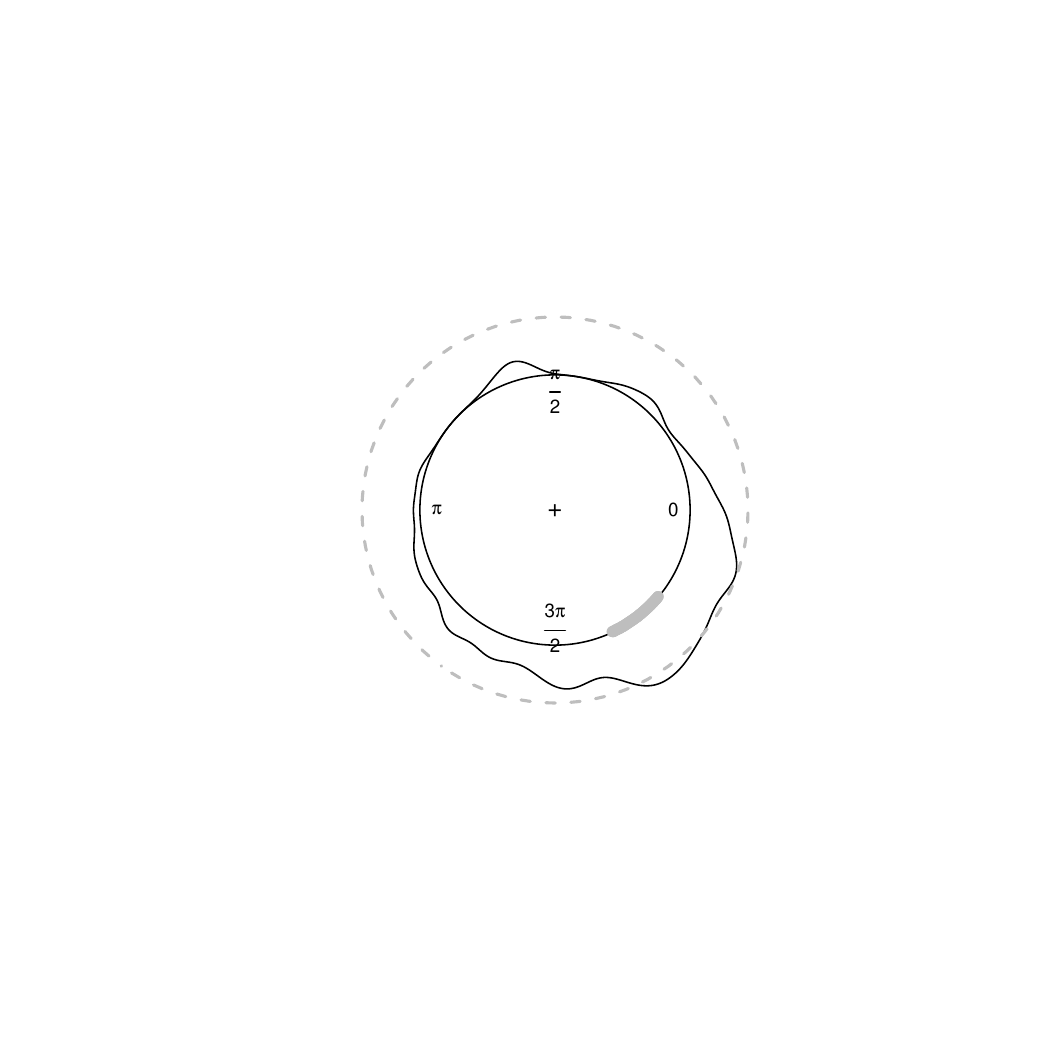}} 
			\put(290,-190){\includegraphics[scale=0.42]{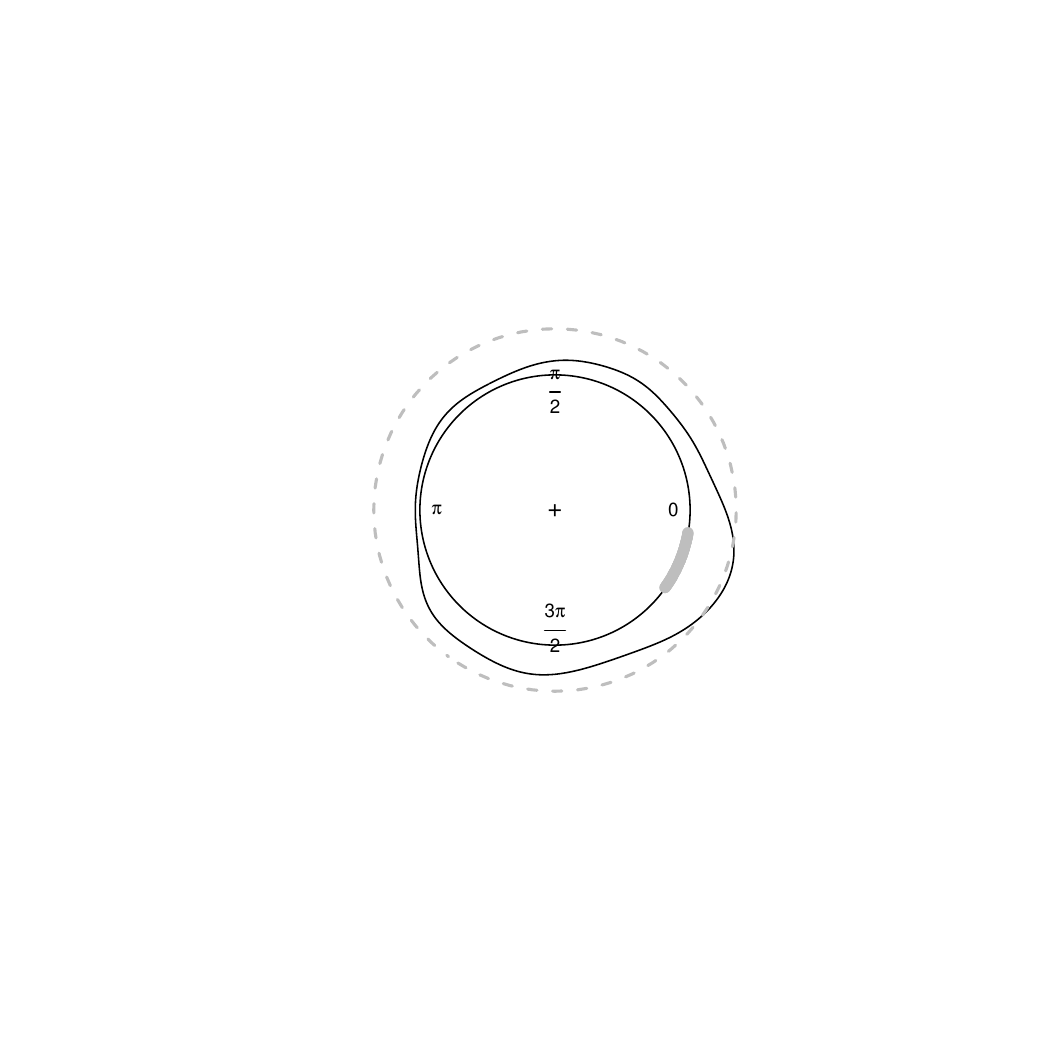}}
			\put(400,-190){\includegraphics[scale=0.4]{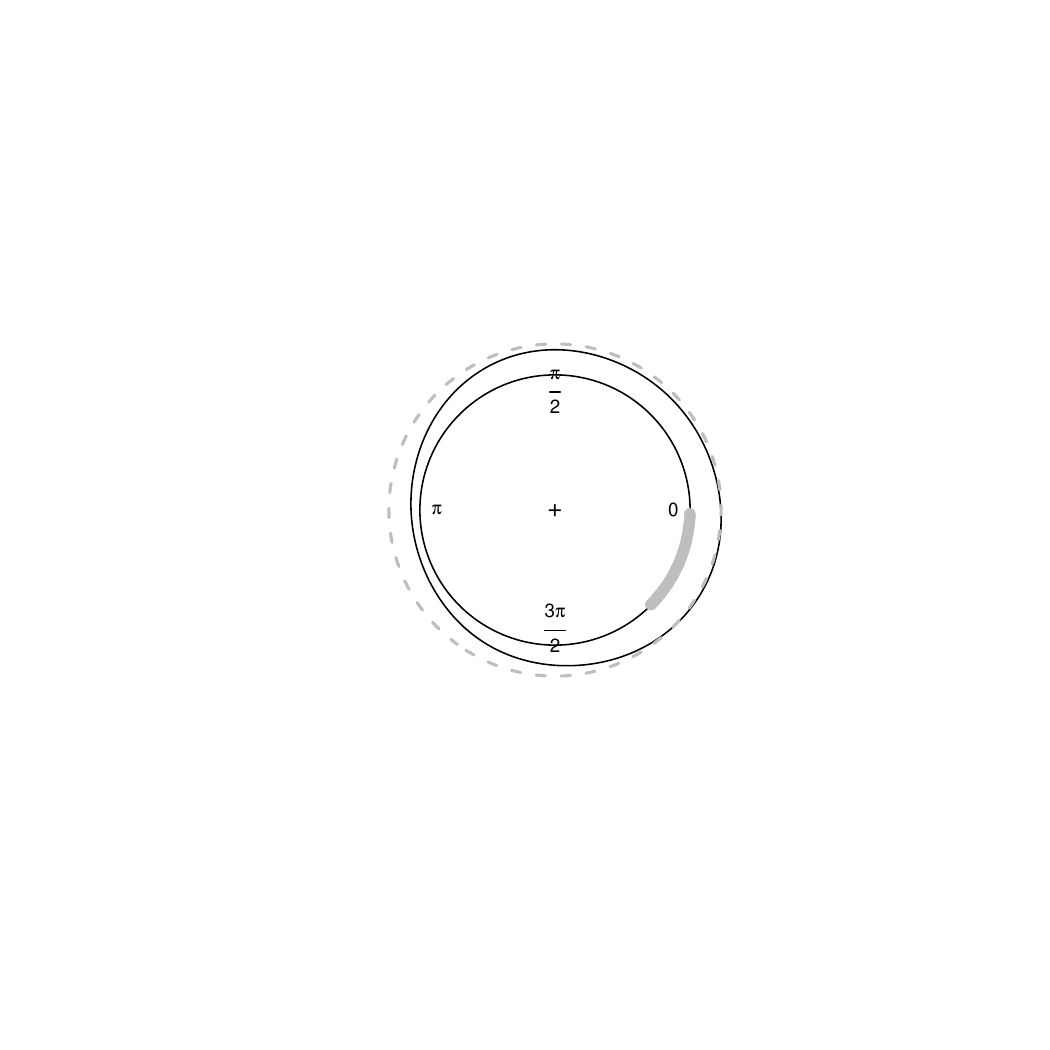}}
			
			\put(-50,-110){\textbf{11}}
			\put(55,-110){\textbf{14}}
			\put(158,-110){\textbf{15}}
			\put(253,-110){\textbf{17}}
			\put(350,-110){\textbf{18}}
			\put(460,-110){\textbf{20}}
			

			
			\end{picture}
			$$ $$
			
			$$ $$

			$$ $$

			$$ $$

			$$ $$
			
			$\hspace{-3.5cm}$\begin{tabular}{c|cccccccccccccccccccccccc}
				&1&2&3&4&5&6&7&8&9&10&11&12&13&14&15&16&17&18&19&20&21&22&23&24\\\hline\\
				1&\cellcolor{diag}&0.35&0.60&0.36&1.35&1.53&\cellcolor{var}1.21&0.68&0.51&0.42&1.02&0.60&0.30&0.19&0.52&0.52&0.35&0.84&0.51&0.97&0.30&0.67&1.34&0.75\\
				2&0.18&\cellcolor{diag}&0.76&0.53&\cellcolor{var}1.07&1.63&1.47&1.00&0.84&0.76&1.31&0.92&0.60&0.25&0.35&0.85&0.69&1.14&0.84&1.12&0.64&0.83&1.47&0.91\\
				3&0.15&0.03&\cellcolor{diag}&0.38&1.55&\cellcolor{var}1.07&0.97&0.40&0.33&0.29&0.76&0.31&0.30&0.77&\cellcolor{var}1.07&0.24&0.26&0.56&0.23&0.40&0.38&0.14&0.84&0.36\\
				4&0.00&0.24&0.21&\cellcolor{diag}&\cellcolor{var}1.37&\cellcolor{var}1.27&1.28&0.76&0.60&0.51&1.09&0.68&0.34&0.55&0.86&0.61&0.44&0.91&0.60&0.76&0.39&0.31&1.14&0.40\\
				5&0.69&0.35&0.95&0.61&\cellcolor{diag}&\cellcolor{GGG}1.93&\cellcolor{GGG}1.97&1.77&1.68&1.63&\cellcolor{GGG}1.91&1.73&1.53&1.28&1.05&1.69&\cellcolor{var}1.59&\cellcolor{GG}1.84&1.68&1.77&1.56&1.58&\cellcolor{GGG}1.93&1.63\\
				6&0.66&0.82&0.06&0.30&\cellcolor{GG}1.57&\cellcolor{diag}&0.85&1.00&1.33&1.31&0.84&\cellcolor{var}1.15&1.32&1.64&\cellcolor{GG}1.81&1.15&1.29&0.84&1.18&0.71&1.37&1.01&0.26&0.93\\
				7&0.83&0.99&0.25&0.49&\cellcolor{GGG}1.68&0.19&\cellcolor{diag}&0.60&0.76&0.84&0.23&0.68&1.00&1.29&1.49&0.75&0.91&0.44&0.76&0.60&0.95&1.02&0.61&1.06\\
				8&0.25&0.42&0.09&0.11&1.29&0.03&0.16&\cellcolor{diag}&0.41&0.38&0.38&0.18&0.44&0.85&1.15&0.18&0.35&0.18&0.21&0.31&0.46&0.46&0.76&0.51\\
				9&0.08&0.25&0.23&0.01&1.15&0.39&0.57&0.03&\cellcolor{diag}&0.09&0.58&0.24&0.26&0.61&0.86&0.24&0.16&0.58&0.20&0.71&0.21&0.40&1.13&0.49\\
				10&0.01&0.16&0.14&0.05&1.07&0.35&0.53&0.06&0.04&\cellcolor{diag}&0.63&0.20&0.18&0.52&0.82&0.20&0.08&0.55&0.16&0.68&0.13&0.36&1.09&0.45\\
				11&0.62&0.78&0.03&0.26&\cellcolor{GG}1.55&0.04&0.01&0.06&0.35&0.31&\cellcolor{diag}&0.46&0.80&1.11&1.32&0.53&0.70&0.21&0.55&0.38&0.75&0.82&0.60&0.86\\
				12&0.16&0.34&0.09&0.10&1.22&0.15&0.34&0.09&0.09&0.15&0.11&\cellcolor{diag}&0.35&0.69&1.00&0.08&0.25&0.35&0.09&0.49&0.30&0.38&0.92&0.42\\
				13&0.19&0.01&0.04&0.06&0.92&0.36&0.55&0.05&0.03&0.01&0.33&0.14&\cellcolor{diag}&0.49&0.81&0.28&0.10&0.60&0.26&0.69&0.08&0.38&1.11&0.46\\
				14&0.10&0.01&0.04&0.01&0.60&0.83&1.00&0.44&0.26&0.18&0.80&0.35&0.00&\cellcolor{diag}&0.34&0.69&0.52&1.01&0.66&\cellcolor{var}1.13&0.41&0.84&1.48&0.92\\
				15&0.09&0.03&0.38&0.00&0.33&1.13&1.28&0.76&0.60&0.51&1.09&0.68&0.34&0.01&\cellcolor{diag}&1.00&0.84&\cellcolor{var}1.29&0.97&1.39&0.74&1.14&1.68&1.21\\
				16&0.09&0.26&0.09&0.03&1.16&0.15&0.34&0.16&0.01&0.10&0.11&0.00&0.21&0.28&0.61&\cellcolor{diag}&0.18&0.35&0.04&0.49&0.29&0.30&0.92&0.35\\
				17&0.09&0.09&0.06&0.03&\cellcolor{var}1.01&0.33&0.51&0.09&0.06&0.03&0.29&0.18&0.04&0.10&0.44&0.18&\cellcolor{diag}&0.52&0.16&0.66&0.11&0.34&1.07&0.42\\
				18&0.41&0.58&0.19&0.05&1.41&0.20&0.01&0.16&0.14&0.10&0.00&0.10&0.11&0.60&0.91&0.10&0.08&\cellcolor{diag}&0.39&0.16&0.63&0.62&0.60&0.67\\
				19&0.08&0.25&0.13&0.01&1.15&0.19&0.38&0.18&0.00&0.09&0.15&0.04&0.18&0.26&0.60&0.01&0.14&0.06&\cellcolor{diag}&0.52&0.25&0.29&0.95&0.34\\
				20&0.25&0.42&0.35&0.11&1.29&0.34&0.15&0.00&0.03&0.06&0.14&0.09&0.05&0.44&0.76&0.16&0.09&0.14&0.18&\cellcolor{diag}&0.76&0.46&0.46&0.51\\
				21&0.14&0.04&0.01&0.14&0.97&0.44&0.62&0.03&0.05&0.09&0.40&0.06&0.05&0.05&0.39&0.14&0.05&0.19&0.15&0.03&\cellcolor{diag}&0.45&1.17&0.53\\
				22&0.01&0.04&0.06&0.08&0.90&0.01&0.18&0.01&0.09&0.00&0.05&0.05&0.03&0.03&0.31&0.05&0.08&0.05&0.01&0.21&0.08&\cellcolor{diag}&0.86&0.25\\
				23&0.67&0.83&0.08&0.31&\cellcolor{GG}1.58&0.01&0.18&0.01&0.40&0.36&0.05&0.16&0.38&0.84&1.14&0.16&0.34&0.19&0.20&0.33&0.45&0.00&\cellcolor{diag}&0.91\\
				24&0.18&0.09&0.11&0.15&0.85&0.06&0.09&0.04&0.06&0.10&0.10&0.05&0.08&0.08&0.26&0.13&0.13&0.10&0.14&0.04&0.01&0.05&0.05&\cellcolor{diag}\\
			\end{tabular} 
			
		\end{table}
		
	\end{landscape}

	Several cells in Table \ref{24tab} are represented in rose color. All of them corresponds to considerable large values of distances (larger than $1.00$) and they are used to analyze briefly the influence of each of the variables in the dataset. Under the same values of the rest of variables Talitrus saltator and Talorchestia brito present different behaviors. For instance, distances between sets 5 and 17 or 3 and 15 correspond to this situation. Sets 5 and 17 can be compared using their representations in Table \ref{24tab} (top). The importance of the sex variable for the specie Talitrus saltator can be also seen considering the Hausdorff distances of the sets 2 and 5, 3 and 6 or 18 and 15. According to images in Table \ref{24tab}, these sets present their largest modes in completely different directions. Note that the role of the variable month is clearly remarkable. The relatively high values of the distances between sets 1 and 7 and 6 and 12 or 14 and 20 for the species Talitrus saltator and Talorchestia brito also corresponds to the existence of modes in different directions. Finally, the importance of the moment of the day for the Talitrus saltator can be studied through the distances between sets 4 and 5 or 4 and 6. Remark that set 4 has two connected components while set 5 only presents one.\vspace{-.8cm}

	\begin{figure}[h!]\centering
		\includegraphics[scale=.4]{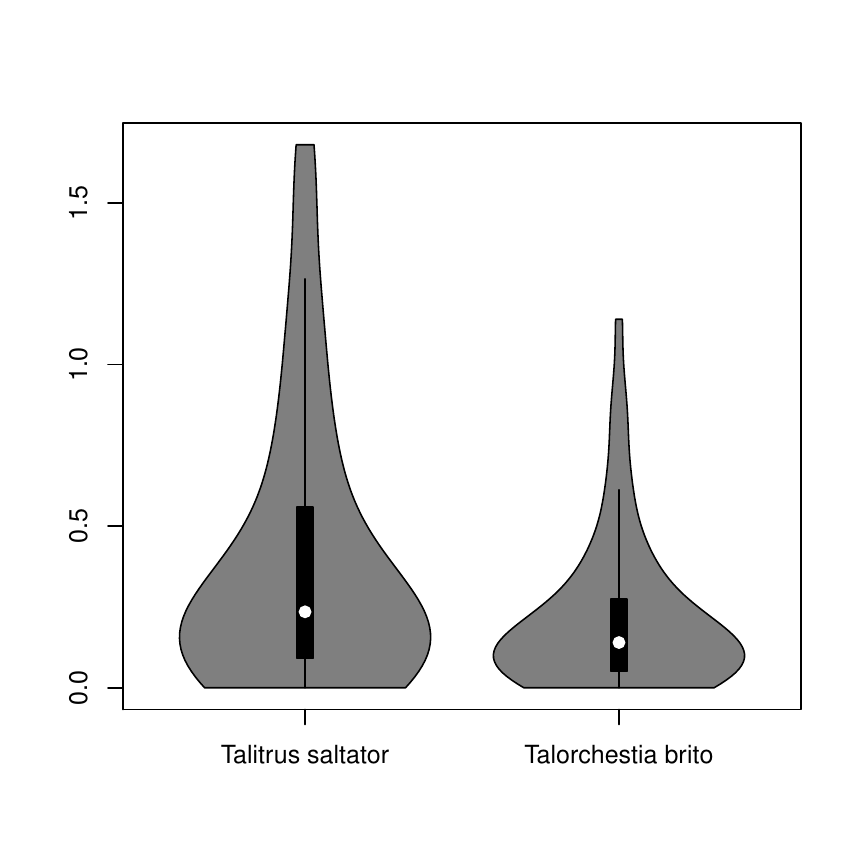}
		\includegraphics[scale=.4]{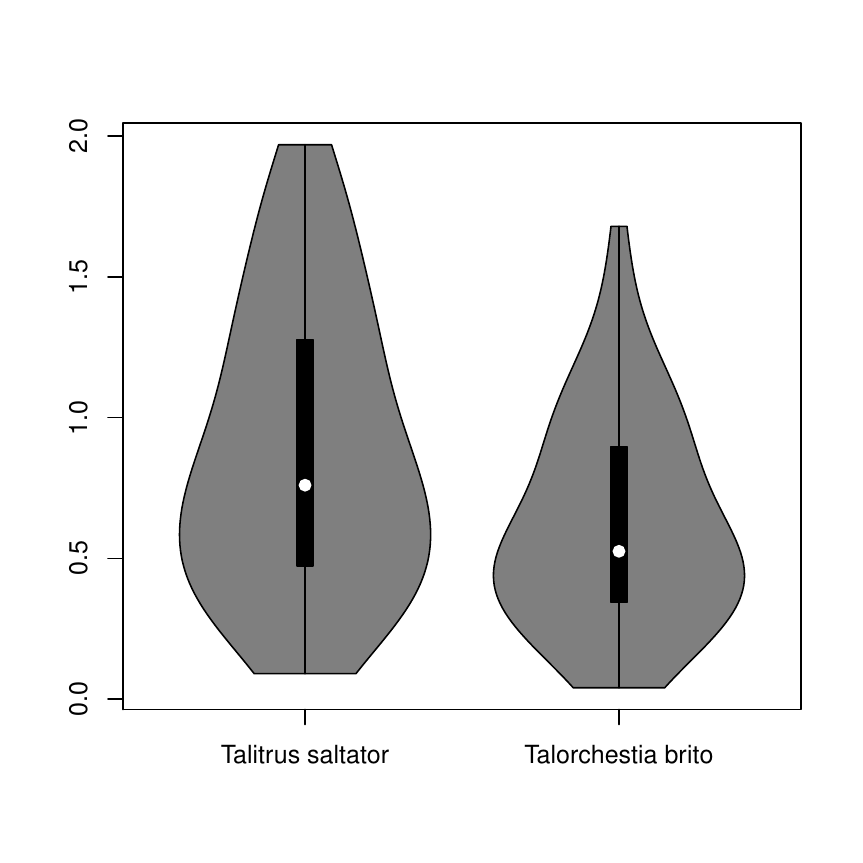}\vspace{-.7cm}\\
		\caption{Violin plots of Euclidean (left) and Hausdorff (right) distances for species Talitrus saltator and Talorchestia brito.}\label{violin}
	\end{figure}

	Finally, Figure \ref{violin} shows the violin plots for Hausdorff and Euclidean distances for the two species of sandhoppers. Note that the median of the Talitrus saltator in Hausdorff (Euclidean) distance is $0.76$ ($0.23$), clearly bigger than the median of Talorchestia brito that is equal to $ 0.52$ ($0.14$). This shows that Talitrus saltator presents more differentiated orientations, depending on the time of day, period of year and sex, with respect to Talorchestia brito. Therefore, conclusions in \cite{scapini2002} are corroborated from this perspective.

	\subsection{Earthquakes distribution on Earth}
	\label{sec:earthquakes}
	According to the theory of plate tectonics, Earth is an active planet. Its surface is composed of about 15 individual plates that move and interact, constantly changing and reshaping Earth's outer layer. These movements are usually the main cause of volcanoes and earthquakes. In fact, seismologists have related these natural phenomena to the boundaries of tectonic plates because they tend to occur there. In fact, the concentration of earthquake epicenters traces the filamentary network of fault lines and, consequently, they could be analyzed alternatively from the perspective of nonparametric filamentary structure estimation (see, for instance,\cite{genovese}). Moreover, tectonic hazards can provoque important damages (destroy buildings, infrastructures or even cause deaths). Therefore, it is important to detect which areas are specially risky. As an illustration, the recent world earthquakes distribution is analyzed next through HDRs estimation.

	\begin{figure}[h!]
		\begin{picture}(-200,430)
		\put(0,270){\includegraphics[scale=0.7]{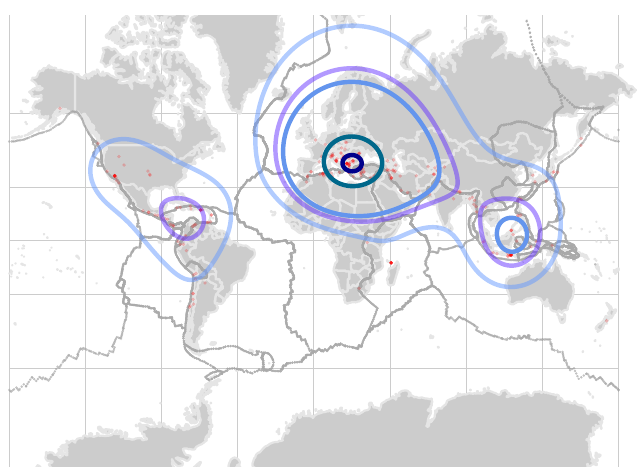}}
		\put(205,215){\includegraphics[scale=0.5]{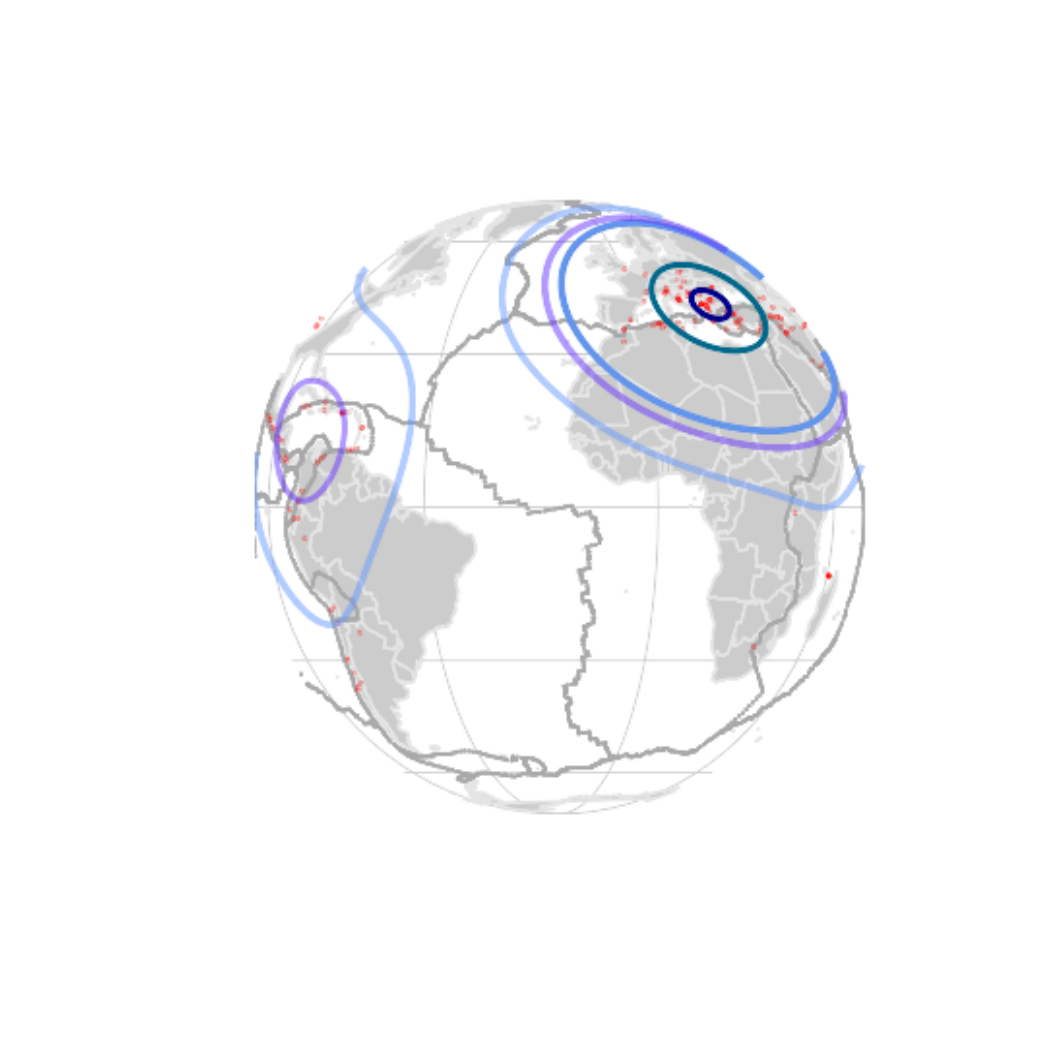}}
		\end{picture}  \vspace{-9.2cm}
		\caption{Contours of HDRs for $\tau_1=0.1$, $\tau_2=0.3$, $\tau_3=0.5$, $\tau_4=0.7$ and $\tau_5=0.9$ obtained from the sample of world earthquakes registered between October 2004 and April 2020.}\label{earthquakes}
	\end{figure}

	Figure \ref{earthquakes} shows the margins of the tectonic plates (grey color) and the geographical coordinates (red points) of a total of $272$ medium and strong earthquakes registered between 1$^{th}$ October 2004 and 9$^{th}$ April 2020 already introduced in Section \ref{intro:data}. Note that most of events are exactly located on the plates boundaries.

	Our main goal is to detect which areas or countries are really problematic nowadays. In Section \ref{intro:data}, we show that the largest mode is located on the Southeast Europe considering a value of $\tau=0.8$. However, a more general view on earthquakes distribution could be obtained if more HDRs are reconstructed for a range of values of $\tau$. Specifically, they were estimated choosing $\tau_1=0.1$, $\tau_2=0.3$, $\tau_3=0.5$, $\tau_4=0.7$ and $\tau_5=0.9$. The bandwidth parameter used is the proposed in \cite{eduardo2013}. The corresponding contours are also represented in Figure \ref{earthquakes} using blue colors. An interactive representation of these HDRs can be seen in Appendix \ref{AppendixA}.
	
	The two smallest contours (dark blue colors) corresponds to density regions with probability at least $1-\tau_5=0.1$ and $1-\tau_4=0.3$, respectively. Therefore, they match with the greatest modes of earthquakes world distribution and they identify the more risky parts of the world. They are located on Europe. Concretely, on the boundaries intersection for the Eurasian and African Plates. Note that the second of these regions even includes the frontier of the Arabian Plate. Contours for $\tau_2=0.3$ and $\tau_3=0.5$ are related to Indo-Australian Plate and margins of Philippine Sea and Pacific Plates appears when $\tau_1=0.1$. 
	
	As for America, the most problematic area is detected in Central America. Concretely, it is mainly located on the frontiers of Cocos, Nazca and Caribbean Plates. According to the contours shown, this region belongs to the zone of the world where the 70\% ($1-\tau_2$\%) of earthquakes are registered. If $\tau_1=0.1$ is considered then Pacific, North and South American plates appears as risky areas.

	

	\section{Conclusions}
The main goals of this work are to extend the definition of HDRs for directional data and propose a plug-in estimator based on a new bootstrap bandwidth selector that is focused on the problem of level set reconstruction. The route designed to reach this goal can be summarized
as follows: (1) Extending the definition of HDRs for directional data, (2)
	proposing the plug-in estimator, (3) introducing a suitable bootstrap selector of the bandwidth parameter, (3) studying
	the behavior of the plug-in estimators (using the new selector and other classical directional bandwidths) and (4) applying the plug-in reconstruction of HDRs to the real data on sandhoppers orientation and earthquakes.
	  
	Finally, natural extensions of this work are discussed. First, other suitable bootstrap bandwidth selectors could be proposed in order to estimate HDRs by using, for instance, the Lebesgue measure instead of the Hausdorff distance. Secondly, an estimator for the number of population clusters can be proposed in the directional setting as the number of connected components of the HDRs plug-in estimators. Theoretical results on its consistency might probably be proved under certain regularity conditions. Another important achievement
	would be to introduce a nonparametric test for comparing two or more populations in general dimension using distances between sets. The test statistic could measure the discrepancy
	(for example, boundary distances) among the directional level set estimators of these
	populations. This test procedure could use explicitly the distance between
	boundaries of the estimated level sets. The simple geometric structure of
	estimators could be used to compute the procedure and calibrate the test
	using re-sampling schemes. Finally, earthquakes on earth could be analyzed following a different approach of set estimation theory. Since the concentration of earthquake epicenters traces the filamentary network of fault lines, the performance of nonparametric filament estimators could be analyzed.

	\paragraph{Acknowledgements} R.M. Crujeiras and P. Saavedra-Nieves acknowledge the financial support of Ministerio de Economía y Competitividad of the Spanish government under grants MTM2016-76969P and MTM2017-089422-P and ERDF. Authors also thank Prof. Felicita Scapini for providing the sandhoppers data (collected under the support of the European Project ERB ICI8-CT98-0270) and the computational resources of the CESGA Supercomputing Center.


\begin{thebibliography}{9}

 
		
		
		\bibitem{Ame1}	Ameijeiras-Alonso, J., Crujeiras, R.M. and Rodríguez-Casal, A., Mode testing, critical bandwidth and excess mass, Test, 28, 900-919 (2019)
		
			\bibitem{Ame2}Ameijeiras-Alonso, J., Benali, A., Crujeiras, R.M., Rodríguez-Casal, A. and Pereira, J. M. C., Fire seasonality identification with multimodality tests, Annals of Applied Statistics, 13, 2120-2139 (2019)
		
		
			\bibitem{Anderberg}Anderberg, M. R., Cluster Analysis for Applications, Academic Press, New York (1973)
		
		
		
		\bibitem{azza}	Azzalini, A. and Torelli, N., Clustering via nonparametric density estimation, Statistics and Computing, 17(1), 71-80, (2007)
		
		\bibitem{bai}Bai, Z. D., Rao, C. R. and Zhao, L. C., Kernel estimators of density function of directional data, Multivariate Statistics and Probability, 24-39  (1989)
		
		
		\bibitem{bai3}		
		Ba\'illo, A., Total error in a plug-in estimator of level sets, Statist. Probab. Lett., 65, 411-417 (2003)
		
		
		\bibitem{bai2}		
		Ba\'illo, A. and Cuevas, A., Parametric versus nonparametric tolerance regions in detection problems,
		Comput. Statist., 21, 523-536 (2006)  
		
		\bibitem{bai1}		
		Ba\'illo, A., Cuevas, A. and Justel, A., Set estimation and nonparametric detection, Canad. J. Statist., 28, 765-782 (2000) 
		
		
		\bibitem{banerjee}Banerjee, A., Dhillon, I. S., Ghosh, J., and Sra, S., Clustering on the unit hypersphere	using von Mises-Fisher distributions, J. Mach. Learn. Res., 6, 1345-1382 (2005)
		
		\bibitem{biau}Biau, G., Cadre, B. and Pelletier, B., A graph-based estimator of the number of clusters. ESAIM: Probability and Statistics, 11, 272-280 (2007)
		
		\bibitem{box}Box, G. E. P., and Tiao, G. C., Bayesian Inference in Statistical
		Analysis, Reading, MA: Addison-Wesley (1973)
		
		
		
		
		
		\bibitem{burman}Burman, P. and Polonik, W., Multivariate mode hunting: Data analytic tools with measures of significance, Journal of Multivariate Analysis, 100(6), 1198-1218 (2009)
		
		\bibitem{chen}Chen, Y. C., Genovese, C. R. and Wasserman, L., Density level sets: Asymptotics, inference, and visualization, J. Am. Stat. Assoc., 112, 1684-1696 (2017) 
		
		
		
		
		
		\bibitem{cholaq}Cholaquidis, A., Fraiman, R. and Moreno, L., Level set and density estimation on manifolds, arXiv preprint arXiv:2003.05814 (2020)
		
		\bibitem{cue}Cuevas, A., Febrero, M. and Fraiman, R., Estimating the number of clusters, Canad. J. Statist., 28, 367-382 (2000) 
		
		
		
		\bibitem{cuefebr} Cuevas, A., Febrero, M. and Fraiman, R., Estimating the number of clusters. Canadian Journal of Statistics, 28, 367-382 (2000)
		
		
		\bibitem{cueff}Cuevas, A., Febrero, M. and Fraiman, R., Cluster analysis: a further approach based on density estimation, Computational Statistics and Data Analysis, 36(4), 441-459 (2001)
		
		\bibitem{cuefra}Cuevas, A. and Fraiman, R.,  A plug-in approach to support estimation, Ann. Statist, 25, 2300-2312 (1997) 
		
		
		\bibitem{cuevasManteiga}Cuevas, A., Gonz\'alez-Manteiga, W. and Rodr\'iguez-Casal, A., Plug-in estimation of general level sets, Australian and New Zealand Journal of Statistics, 48(1), 7-19 (2006)
		
		
		\bibitem{dekker1978}Dekker, W., Strandvlooien (Talitridae). Tabellenserie van de Strandwerkgemeenschap, 24 (1978)
		
	 
	 \bibitem{dev}Devroye, L. and Wise, G., Detection of abnormal behavior via nonparametric estimation of the support, SIAM J. Appl. Math., 38, 480-488 (1980)
		 
		\bibitem{dimarzio}Di Marzio, M., Panzera, A., and Taylor, C. C., Local polynomial regression for circular
		predictors, Statistics \& Probability Letters, 79, 2066-2075 (2009)
		
		\bibitem{dimarzio2011}Di Marzio, M., Panzera, A., Taylor, C.C., Kernel density estimation on the torus, Journal of Statistical Planning and Inference, 141, 2156-2173 (2011)
		
		\bibitem{doss}Doss, C. R. and Weng, G., Bandwidth selection for kernel density estimators of multivariate level sets and highest density regions,  Electron. J. Stat., 12. 4313-4376 (2018)
		
		
		
			\bibitem{fraiman}Fraiman, R. and Meloche, J., Counting bumps. Ann. Inst. Stat. Math. (1997)
		
		\bibitem{gar}Gardner, A.B., Krieger, A.M., Vachtsevanos, G and Litt, B., One-class novelty detection for seizure
		analysis from intracranial EEG, J. Mach. Learn. Res., 7, 1025-1044 (2006) 
		
		
		\bibitem{gar}Garcia, J. N., Kutalik, Z., Cho, K. H. and Wolkenhauer, O., Level sets and minimum volume sets of probability density functions, Int. J. Approx. Reason., 34, 25-47 (2003)
		
		\bibitem{eduardo2013} Garc\'ia-Portugu\'es, E., Exact risk improvement of bandwidth selectors for kernel density estimation with directional data, Electronic Journal of Statistics, 7, 1655-1685 (2013)
		
		\bibitem{genovese}Genovese, C. R., Perone-Pacifico, M., Verdinelli, I., and Wasserman, L., The geometry of nonparametric filament estimation,  Journal of the American Statistical Association, 107(498), 788-799 (2012)
		
		\bibitem{everitt}	Everitt, B. S., Cluster Analysis. Arnold-Halsted, New York (1993)
		
		\bibitem{hall84}Hall, P., Central limit theorem for integrated square error of multivariate nonparametric
		density estimators, J. Multivariate Anal., 14(1), 1-16 (1984)
		
		
		
		\bibitem{hall87}Hall, P., Watson, G. S., and Cabrera, J., Kernel density estimation with spherical data,
		Biometrika, 74(4), 751-762 (1987)
		
			\bibitem{hall}Hall, P. and Wood, A. T., Approximations to distributions of statistics used for testing hypotheses about the number of modes of a population, J. Statist. Plann.
		Inf., 55, 299-317 (1996)
		
		\bibitem{har}Hartigan, J., Clustering algorithms, Wiley Series in Probability and Mathematical Statistics, John Wiley \& Sons, New York-London-Sydney (1975) 
		
		
		
		
		\bibitem{huo}Huo, X. and Lu, J.C., A network flow approach in finding maximum likelihood estimate of high
		concentration regions, Comput. Statist. Data Anal., 46, 33-56 (2004)  
		
		\bibitem{hyn}Hyndman, R.J., Computing and graphing highest density regions, Am. Stat., 50, 120-126 (1996)
		
		
		\bibitem{jan}Jang, W., Nonparametric density estimation and clustering in astronomical sky surveys, Comput. Statist. Data Anal., 50, 760-774 (2006)  
		
		
		
		
			\bibitem{kle3}Klemel\"a, J., Estimation of densities and derivatives of densities with directional data, J.
		Multivariate Anal., 73(1), 18-40 (2000)
		
		
		\bibitem{mammen1995}Mammen, E., On qualitative smoothness of kernel density estimates, Statistics,	26, 253-267 (1995) 
		
		\bibitem{mammen1992}Mammen, E., Marron, J. S. and Fisher, N. I., Some asymptotics for multimodality tests based on kernel density estimates, Prob. Theory Relat. Fields, 91, 115-132 (1992)
		
		
		
		\bibitem{rig2}Mammen, E. and Polonik, W., Confidence regions for level sets, J. Multivariate Anal., 122, 202-214 (2013)
		
		
		
		\bibitem{mam}Mammen, E. and Tsybakov, A. B., Asymptotical minimax recovery of sets with smooth boundaries, Ann. Stat., 23, 502-524 (1995)  
		
		
		
		
		\bibitem{Mardia} Mardia, K.V. and Jupp, P., Directional Statistics, Wiley (2009).
		
		\bibitem{mar}Markou, M. and Singh, S., Novelty detection: a reviewpart 1: statistical approaches,   Signal Processing, 83, 2481-2497  (2003)
		
		
		
		\bibitem{scapini2003}Marchetti, G. M. and Scapini, F., Use of multiple regression models in the study of sandhopper
		orientation under natural conditions, Estuarine, Coastal and Shelf Science, 58, 207-215 (2003)
		
		\bibitem{mas}Mason, D.M. and Polonik, W., Asymptotic normality of plug-in level set estimates,  Ann. Appl. Probab., 19, 1108-1142 (2009)  
		
		
		
		\bibitem{minnote}Minnotte, C. M. and Scott, D. W., The mode tree: A tool for visualization of nonparametric density features, J. Comp. Graph. Stat., 2, 51-68 (1993)
		
		
		\bibitem{oli1}Oliveira, M., Crujeiras, R.M. and Rodr\'iguez-Casal, A.,  A plug-in rule for bandwidth selection in circular density estimation, Computational Statistics and Data Analysis, 56, 3898-3908 (2012)
		
		\bibitem{oli2013} Oliveira, M., Crujeiras R.M. and Rodr\'iguez-Casal, A.,  Nonparametric circular methods for exploring environmental data, Environmental and Ecological Statistics, 20, 1-17 (2013) 
		
		
		\bibitem{oli2}Oliveira, M., Crujeiras R.M. and Rodr\'iguez-Casal, A.,  NPCirc: an R package for nonparametric circular methods, Journal of Statistical Software, 61(9), 1-26 (2014)
		
		
		\bibitem{parzen}Parzen, E., On estimation of a probability density function and mode, The Annals of Mathematical Statistics, 33, 1065-1076 (1962)
		
		
		\bibitem{pol3}Polonik, W., Minimum volume sets and generalized quantile processes, Stoch. Proc. Appl., 69, 1-24  (1997)  
		
		
		\bibitem{pol2} Polonik, W., Confidence regions for level sets, J. Multivar. Anal., 122, 202-214  (2013)  
		
		
		\bibitem{Qiao}Qiao, W., Asymptotics and optimal bandwidth selection for nonparametric estimation of density level sets, arXiv preprint arXiv:1707.09697 (2017)
		
		\bibitem{rig}Rigollet, P. and Vert, R., Optimal rates for plug-in estimators of density level sets, Bernoulli, 15, 1154-1178 (2009)  
		
		
		\bibitem{rin}Rinaldo, A. and Wasserman, L., Generalized density clustering, Ann. Stat., 38, 2678-2722 (2010)
		
		
		
		\bibitem{roe}Roederer, M. and Hardy, R.R., Frequency difference gating: a multivariate method for identifying
		subsets that differ between samples, Cytometry,  45, 56-64 (2001)
		
		
		
		
		
		
		
		\bibitem{rosaa}Rodríguez-Casal, A. and Saavedra-Nieves, P., Minimax Hausdorff estimation of density level sets. arXiv preprint arXiv:1905.02897 (2019)
		
		\bibitem{rosenblatt}Rosenblatt, M., Remarks on some nonparametric estimate of a density function, The
		Annals of Mathematical Statistics, 27, 832-837 (1956)
		
		\bibitem{sam}Samworth, R. and Wand, M., Asymptotics and optimal bandwidth selection for highest density
		region estimation, Ann. Statist., 38, 1767-1792 (2010)   
		
		
		\bibitem{scapini2002} Scapini, F., Aloia, A., Bouslama, M. F., Chelazzi, L., Colombini, I., ElGtari, M., Fallaci, M. and
		 Marchetti, G. M. Multiple regression analysis of the sources of variation in orientation of two sympatric sandhoppers, Talitrus saltator and Talorchestia brito, from an exposed Mediterranean beach, Behavioral Ecology and Sociobiology, 51(5), 403-414 (2002)
		
		
		
		\bibitem{sin}Singh, A., Scott, C. and Nowak, R., Adaptive Hausdorff estimation of density level sets, Ann. Statist., 37, 2760-2782 (2009)
		
		
		\bibitem{silverman1981}	Silverman, B. W., Using kernel density estimates to investigate multimodality, J. Roy. Statist. Soc. B, 43, 97-99 (1981)
		
		\bibitem{silverman}Silverman, B. W., Density Estimation for Statistics and Data Analysis, Chapman and
		Hall, London (1986)
		
		\bibitem{ste}Steinwart, I., Fully adaptive density-based clustering, Ann. Statist., 43, 2132-2167 (2015)
		
		
		\bibitem{stue}Stuetzle, W. and Nugent, R., A generalized single linkage method for estimating the cluster tree of a density, Journal of Computational and Graphical Statistics, 19(2), 397-418 (2010) 
		
		\bibitem{Taylor} Taylor, C. C., Automatic bandwidth selection for circular density estimation, Computational
		Statistics \& Data Analysis, 52, 3493-3500 (2008)
		
		\bibitem{Tibshirani}Tibshirani, R., Walther, G., and Hastie, T., Estimating the number of clusters in a data set via the gap statistic, Journal of the Royal Statistical Society: Series B (Statistical Methodology), 63(2), 411-423 (2001) 
		
		\bibitem{r4}Tsybakov, A. B., On nonparametric estimation of density level sets, Ann. Statist., 25, 948-969 (1997)  
		
		
		
		
		
		\bibitem{wand1995}	Wand, M. P., and Jones, M. C., Kernel Smoothing, London: Chapman
		and Hall (1995)
		
		\bibitem{zhao}Zhao, L. and Wu, C., Central limit theorem for integrated square error of kernel estimators of spherical density, Sci. China Ser. A, 44(4), 474-483 (2001)
		
		\end{thebibliography}

	\newpage
	\appendix

	\section{Further details on the datasets}\label{AppendixB}

	\subsection{Levels to the estimated HDRs disaggregating the sandhoppers variables}
	
	\begin{table}[h!]  \centering
		\caption{Associated levels to the 24 estimated HDRs.}\label{tab:multicol}\label{names}
	\begin{tabular}{|cc|ccc|ccc|}
			\hline 
			\multicolumn{2}{|c|}{Variables}  	&	\multicolumn{3}{c|}{Males}&\multicolumn{3}{c|}{Females}\\ 
			\multicolumn{2}{|c|}{levels}    &	Afternoon &Noon&Morning&Afternoon&Noon&Morning\\ \hline                      
			Talitrus saltator	&October&	E1&E2&E3&E4&E5&E6\\
			&April&	E7&E8&E9&E10&E11&E12\\
			\hline
			Talorchestia brito	&October		&E13&E14&E15&E16&E17&E18\\
			&April          &E19&E20&E21&E22&E23&E24\\
			\hline
		\end{tabular}
		\vspace{-.3cm}
	\end{table}

	\subsection{Interactive representation of HDRs for eathquakes on Earth}
	
	\begin{figure}[h!]
		\begin{picture}(-200,430)
		\put(60,150){\animategraphics[scale=0.65]{6}{file3_}{1}{40}}
		\put(60,228){\mediabutton[
			jsaction={
				if(anim[?taylor?].isPlaying)
				anim[?taylor?].pause();
				else
				anim[?taylor?].playFwd();
			}
			]{\fbox{Play/Pause}}}  
		
		\end{picture}  \vspace{-7cm}
		\caption{Distribution of earthquakes around the world between October 2004 and April 2020  (red color). Contours of HDRs for $\tau_1=0.1$, $\tau_2=0.3$, $\tau_3=0.5$, $\tau_4=0.7$ and $\tau_5=0.9$  (bluish colors).}\label{earthquakes00}
	\end{figure}

	\newpage
	 
	\section{Simulated spherical models}\label{AppendixC}

	\begin{table}[h!]\centering
		\begin{tabular}{cccc}
			\hline
			Model&$\mu$ & $\kappa$ & Mixture probabilities \\
			\hline
			S1& $(0, 0, 1)$ & $10$ &$1$\\
			S2& $( 0  ,  0,    1)$; $(0   , 0,   -1)$&$1$; $1$&$1/2$; $1/2$\\
			S3 & $(0  ,  0 ,   1)$; $( 0   , 0,   -1)$&$10$; $1$&$1/2$; $1/2$\\
			S4&$( 0,0, 1)$; $(0, 1/\sqrt{2}, 1/\sqrt{2})$&$10$; $10$&$1/2$; $1/2$\\
			S5&$( 0, 0, 1)$; $(0, 1/\sqrt{2},1/\sqrt{2})$&$10$; $10$&$2/5$; $3/5$\\
			S6&$( 0,0, 1)$; $(0 , 1/\sqrt{2},  1/\sqrt{2})$ &$10$; $5$&$1/5$; $4/5$\\
			S7&$(  0   , 0 ,   1)$; $(0,    1,    0)$; $(1,    0,    0)$&$5$; $5$; $5$&$1/3$; $1/3$; $1/3$\\
			S8&$( 0 ,   0,    1)$; $( 0   , 1 ,   0)$; $(1    ,0   , 0)$&$5$; $5$; $5$&$2/3$; $1/6$; $1/6$\\
			S9&$( 0 ,0,1)$; $(0,1/\sqrt{2},1/\sqrt{2})$; $(  0, 1,0)$&$10$; $10$; $10$&$1/3$; $1/3$; $1/3$\\
			\hline
		\end{tabular}\caption{Finite mixtures of von Mises-Fisher spherical distributions considered as models for simulations.}\label{sphetable}
	\end{table}
	
	
		\newpage
	 
		\section{Some details on the directional bandwidth selectors}\label{AppendixA}

	We briefly revise in this section some bandwidth selection methods designed for kernel density estimation. Although these methods do not focus on HDRs, but on the reconstruction of the whole density curve, it may be argued that they could also be used for constructing the proposed plug-in estimator. The performance of our proposal is compared in all the simulated scenarios with different bandwidth selectors for circular and spherical data.
	
	As in the Euclidean setting, most used techniques for selecting $h$ are based on the minimization of some error criteria that quantify the accuracy of the kernel density estimator. One of the most simple errors to be considered is the mean integrated squared error that can be written as follows:
	\begin{equation}\label{MISE}
	MISE(h)=\mathbb{E}\left[ \int_{S^{d-1}}(f_n(x)-f(x))^2 \omega_d(dx)\right],
	\end{equation}where $\omega_d$ denotes the Lebesgue in $S^{d-1}$. Then, a possibility is to search for the bandwidth that minimizes (\ref{MISE}). However, the asymptotic version of $MISE$, $AMISE$, is more commonly used in literature. A rule of thumb proposed in \cite{Taylor} adapts the idea in \cite{silverman} in kernel linear density estimation to the circular setting. The resulting plug-in selector assumes that the data follow a von Mises distribution to determine the $AMISE$. The bandwidth is chosen by first obtaining an estimation $\hat{\kappa}$ of the concentration parameter $\kappa$ in the reference density (for example, by maximum likelihood) through the formula
	$$h_{2}=\left[\frac{4\pi^{1/2}\mathcal{I}_0(\hat{\kappa})^2}{3\hat{\kappa}^2\mathcal{I}_2(2\hat{\kappa}) n}\right]^{1/5}.$$
	Remark that the parametrization in \cite{Taylor} has been adapted to the context of the estimator
	(\ref{estimacionnucleo}) by denoting by $h$ the inverse of the squared concentration parameter employed in his paper. The poor performance of this rule is sometimes due to the non robust estimation by maximum likelihood of the concentration parameter. An alternative and robustified estimation procedure is considered in \cite{oli2013}.
	
	A new selector also devoted to the circular case is established in \cite{oli1}. It improves the performance of the Taylor's proposal allowing for more flexibility in the reference density, considering a mixture of von Mises. This selector is mainly based on two elements. First, the
	$AMISE$ expansion derived in \cite{dimarzio} for the circular kernel density estimator
	by the use of Fourier expansions of the circular kernels. This expression has the following form
	when the kernel is a circular von Mises (the estimator is equivalent to consider $L(r) = e^{-r}$
	and $h$ as the inverse of the squared concentration parameter in (\ref{estimacionnucleo}):
	\begin{equation}\label{oli}
	AMISE(h)=\frac{1}{16}\left[1-\frac{\mathcal{I}_2(h^{-1/2}) }{\mathcal{I}_0(h^{-1/2})}   \right]^2\int_{0}^{2\pi}f^{''}(\theta)^2 d\theta+\frac{\mathcal{I}_0(2h^{-1/2}) }{2n\pi\mathcal{I}_0(h^{-1/2})^2 }.
	\end{equation} 
	The second element is the Expectation-Maximization (EM) algorithm in \cite{banerjee}
	for fitting mixtures of directional von Mises. The selector, that is denoted by $h_{3}$, proceeds
	as follows: first, apply the EM algorithm to fit mixtures with different number of components; then, choose the fitted mixture with the lowest AIC. Finally, compute the curvature term in (\ref{oli}) using the fitted mixture and seek for the $h$ that	minimizes this expression. This value of $h$ is denoted by $h_{3}$. 	
	
	Of course, plug-in rules are not the only alternative to smoothing parameter selection. Some other data-driven directional procedures were already proposed in \cite{hall87} using cross-validation ideas. Specifically, Least Squares Cross-Validation
	(LSCV) and Likelihood Cross-Validation (LCV) bandwidth are introduced, arising as the minimizers
	of the cross-validated estimates of the squared error loss and the Kullback-Leibler loss,
	respectively. The selectors have the following expressions:
	$$h_{4}=\arg\max_{h>0} 2n^{-1}\sum_{i=1}^n f_n^{-i}(X_i)-\int_{S^{d-1}} f_n(x)^2 \omega_q(dx)$$
	and
	$$h_{5}=\arg\max_{h>0} \sum_{i=1}^n\log{f_n^{-i}(X_i)},$$
	where $f_n^{-i}$ represents the kernel estimator computed without the $i-$th observation. 
	
	A bootstrap bandwidth selection procedure for data lying on a $d-$dimensional torus is proposed in \cite{dimarzio2011}. If a von Mises kernel is used, then the bootstrap MISE has a closed expression. Then, $h_{6}$ is selected as the value that minimizes 
	$$\int_{S^{1}} \mathbb{E}_B\left[{f_{n}}^*(X)-f_n(X)\right]^2 \omega_d(dx)$$
	where $\mathbb{E}_B$ denotes the bootstrap expectation with respect to random samples $\{X_1^*,\cdots,X_n^*\}$ generated from $f_n(X)$. A common problem for small samples is that a local minimum may be chosen, as pointed out by \cite{oli1}.
	
	Apart from existing cross-validation procedures in the directional setting, \cite{eduardo2013} derives a plug-in directional analogue to the rule of thumb in \cite{silverman} using the properties of the von Mises density.  Moreover, it is the optimal $AMISE$ bandwidth for normal reference density and normal kernel. Concretely, if the von Mises kernel is considered and $\kappa$ is estimated by maximum likelihood, 
	$$h_{7}=\left\{\begin{matrix}  \left[\frac{4\pi^{1/2}\mathcal{I}_0(\hat{k})^2}{\hat{k}[ \mathcal{I}_1(2\hat{k})+3\hat{k}  \mathcal{I}_2(2\hat{k})    ]n}\right]^{1/5} \mbox{ in} & S^1 \\     
	\left[\frac{ 8\sinh^2(\hat{k})}{\hat{k}[ (1+4\hat{k}^2)\sinh(2\hat{k})-2\hat{k}\cosh{2\hat{k}} ]n}\right]^{1/6}        \mbox{ in}&S^2.
	\end{matrix}\right\}$$

\end{document}